\documentclass[a4paper,10pt]{article}

\usepackage{amsmath,amsfonts,amssymb,graphicx,multirow,overcite,url}

\usepackage{subcaption}

\newcommand{\boma}[1]{\mbox{\boldmath ${#1}$}}
\newcommand{\argmax}{\mathop{\mbox{argmax}}}

\newcommand{\sgn}{\mathop{\mbox{sgn}}}

\begin{document}

%opening
\title{Robust multivariate methods in Chemometrics\thanks{Article has appeared as: {\em Comprehensive Chemometrics}, 2nd Edition, Steven Brown, Rom\'{a} Tauler and Beata Walczak (Eds.), Elsevier, 26 May 2020, Section 3.19, pages 393--430. ISBN: 9780444641656, \url{https://doi.org/10.1016/B978-0-12-409547-2.14642-6}} \thanks{\textcopyright\  2020. This manuscript version is made available under the CC-BY-NC-ND 4.0 license \url{http://creativecommons.org/licenses/by-nc-nd/4.0/}} \thanks{This is an update of P. Filzmoser, S. Serneels, R. Maronna, P.J. Van Espen, 3.24 - Robust Multivariate Methods in Chemometrics, in: {\em Comprehensive Chemometrics}, 1st Edition, Steven D. Brown, Rom\'{a} Tauler and Beata Walczak (Eds.), Elsevier, 2009, \url{https://doi.org/10.1016/B978-044452701-1.00113-7}}}
\author{Peter Filzmoser$^1$ \and Sven Serneels$^2$ \and Ricardo Maronna$^3$ \and Christophe Croux$^4$\\
$^1$ Institute of Statistics and Mathematical Methods in Economics\\
TU Wien, Vienna, Austria\\
$^2$ Aspen Technology, Bedford, Massachusetts, USA\\
$^3$ Department of Mathematics, National University of La Plata, Argentina\\
$^4$ EDHEC Business School, Lille, France}
\date{}
\maketitle

\begin{abstract}
This chapter presents an introduction to robust statistics with applications of a chemometric nature. Following a description of the basic ideas and concepts behind robust statistics, including how robust estimators can be conceived, the chapter builds up to the construction (and use) of robust alternatives for some methods for multivariate analysis frequently used in chemometrics, such as principal component analysis and partial least squares. The chapter then provides an insight into how these robust methods can be used or extended to classification. To conclude, the issue of validation of the results is being addressed: it is shown how uncertainty statements associated with robust estimates, can be obtained.    
\end{abstract}

\newpage

\tableofcontents

\section*{List of Abbreviations}
\begin{tabular}{ll}
CV & cross-validation\\
D-PLS & discrimination partial least squares\\
EIF & empirical influence function\\
EPXMA & electron probe x-ray micro-analysis\\
IF & influence function\\
IRPLS & iteratively re-weighted partial least squares\\
IRWLS & iteratively re-weighted least squares\\
LAD & least absolute deviation\\
LASSO & Least absolute shrinkage and selection operator\\
LDA & linear discriminant analysis\\
LIBRA & Library for Robust Analysis\\
LMS & least median of squares\\
LS & least squares\\
LTS & least trimmed squares\\
MAD & median absolute deviation\\
MATLAB & Matrix Laboratory\\
MCD & minimum covariance determinant\\
MSE & mean squared error\\
MSPE & mean squared prediction error\\
NIPALS & Nonlinear iterative partial least squares\\
NIR & near-infrared\\
OLS & ordinary least squares\\
PARAFAC & parallel factor analysis\\
PC & principal component\\
PCA & principal component analysis\\
PCR & principal component regression\\
PLAD & partial least absolute deviations\\
PLS & (uni- or multivariate) partial least squares\\
PLS2 & multivariate partial least squares\\
PM & partial M\\
PP & projection pursuit\\
PP-PLS & projection pursuit partial least squares\\
PRM & partial robust M\\
QDA & quadratic discriminant analysis\\
RAPCA & reflection-based algorithm for principal component analysis\\
RCR & robust continuum regression\\ 
RMSE & root mean squared error\\
RMSECV & root mean squared error of cross validation\\
RMSEP & root mean squared error of prediction\\
ROBPCA & robust PCA (one specific method by Hubert {\em et al.}\cite{Hubert2})\\
RSIMPLS & robust SIMPLS (one specific method by Hubert {\em et al}\cite{Hubert})\\
SPC & spherical principal components\\
SNIPLS & Sparse NIPALS\\
SPLS & Sparse Partial Least Squares\\
SPRM & Sparse Partial Robust M\\
SPRM-DA & Sparse Partial Robust M DA\\
SVD & singular value decomposition\\
TOMCAT & Toolbox for Multivariate Calibration Techniques\\
tri-PLS & trilinear partial least squares\\
TMSPE & trimmed mean squared prediction error\\
\end{tabular}

\newpage

\section{Introduction}

\subsection{The concept of robustness} 

Many statistical methods are based on distributional assumptions. Especially the normal distribution takes on a primordial role: well-known optimality properties of the most frequently applied estimators do only hold at the normal model. For instance, the least squares estimator for regression is known to be the maximum likelihood estimator at the normal model. Nearly all regression methods which are common in chemometrics, are to some extent derived from least squares. This implies that the normal distribution assumes a key position in multivariate chemometric methods. 

In practice data do never exactly follow the normal distribution. In most cases the normality assumption is satisfactory and methods based on it will produce reliable results. However, sometimes the normality approximation to the data is rather poor or even completely wrong. Data may intrinsically follow a different distribution than the normal (e.g. think of counting statistics such as X-ray counts which are Poisson distributed). The data may also show bi- or multimodality because the individual cases have been drawn from different populations. Here one can think of a data set containing cases which are known to appertain to different groups, but for which a joint calibration model is desired. E.g. different types of wines need to be analysed for their ester concentration. From the offset it is known that samples of different years and soils have different properties and thus belong to different populations. Nevertheless a model which predicts the ester concentration reliably independently of their origin or year may be required. For such a model probably a regression technique will be used, albeit it is clear that the data were not generated by a single normal distribution. Alternatively, the data may have been generated by the same model but have been influenced by different processes. Samples may have been generated in a similar manner but have undergone exposition to different effects (temperature, light, etc.), changing their behaviour, such that the assumption of a single distribution becomes invalid. 

The normality assumption may also be violated by an entirely different process. Outliers may occur which have atypical properties compared to the majority of the data. Outliers can be generated in several ways: they can be objects which intrinsically have different properties or they can be artifacts produced by the data generation process. A typical example in chemometrics would be that some cases have been measured with a different light source or detector such that the spectra cannot be included in a single model with the regular cases. When outliers are present in the data it does not make sense to model the data distribution including the outliers. The true model according to which the non-outlying data points have been generated will differ significantly from a model estimated by data containing outliers.  

All the above situations (multimodality, outliers) are examples of situations where the data do not follow a normal model. Whereas they are all examples of nonnormality, robust methods have explicitly been developed for the last mentioned situation. Robust estimators are estimators derived for a given model, including slight deviations from this model. More precisely, if the main group of data points is assumed to come from a distribution $G$, then a robust estimator for such data is designed for the distribution $G_{\varepsilon}=(1-\varepsilon)G+\varepsilon H$, where $H$ is another distribution and $\varepsilon\in[0,1)$. Because robust estimators are usually especially designed for such $\varepsilon$ contaminated distributions, they should resist any type of moderate deviation from $G$. This implies that robust estimators can in practice also perform well at distributions which are close to $G$. For instance, if $G$ is the normal distribution, then robust estimators are designed for a normal contaminated with outliers coming from a given outlier generating distribution $H$, but they may perform well as well for heavier tailed ``close to normal" distributions such as the Cauchy and Student's $t$ distributions. 

\subsection{Visualising multivariate data for outlier identification}

\begin{figure}
\begin{center}
\includegraphics[width=0.5\textwidth]{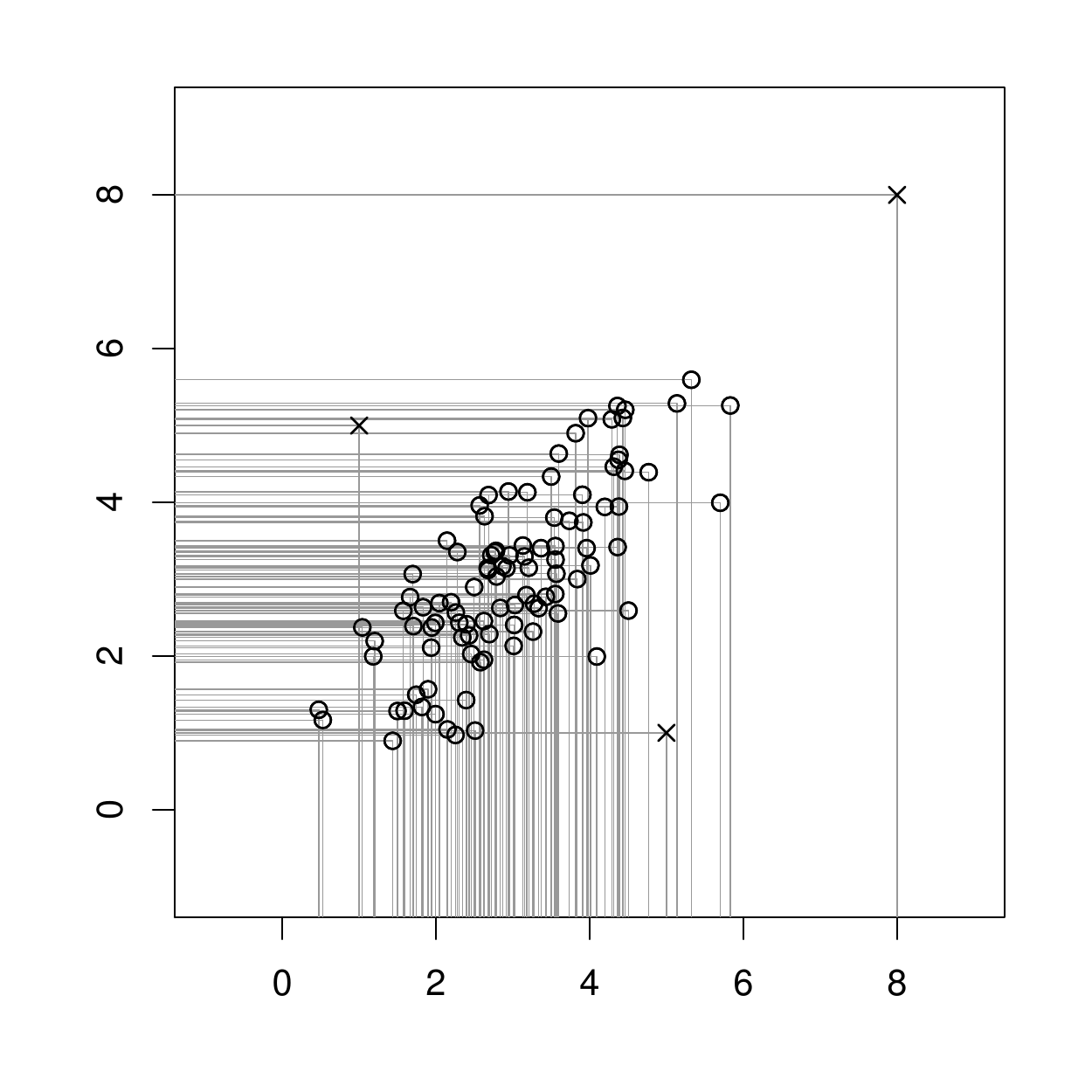}
\caption{\label{BiOut}Bivariate normally distributed data cloud ($\circ$) and three outliers at different positions ($\times$) as well as the projection of each point onto both axes.}      
\end{center}
\end{figure}

Detection of either multimodality or outliers is straightforward for univariate and bivariate data. For multivariate data clouds it will be difficult or even impossible to visualise the data and graphically detect the outliers. In particular chemometric calibration and classification problems are usually of a high dimensional nature: e.g. in spectrophotometry, the spectra are commonly measured at $p>1000$ variables. Visual inspection by simply plotting the data is practically impossible as one would need to inspect plots of all possible pairs of two variables. Even then the outliers can be of a multivariate nature such that they will not be detected by inspecting only two dimensions. Consider the reduction of a bivariate problem to one variable. In Figure \ref{BiOut} a bivariate distribution is plotted. Three outliers are added; it can be seen that a projection of the data onto either one of both axes does only reveal one of the three outliers. It is straightforward to imagine that a similar effect occurs when one projects multivariate data on two dimensions. 
  
\subsection{Masking effect}\label{Sec:Mask}

\begin{figure}
\begin{tabular}{cc}
\resizebox{0.5\textwidth}{!}{\includegraphics{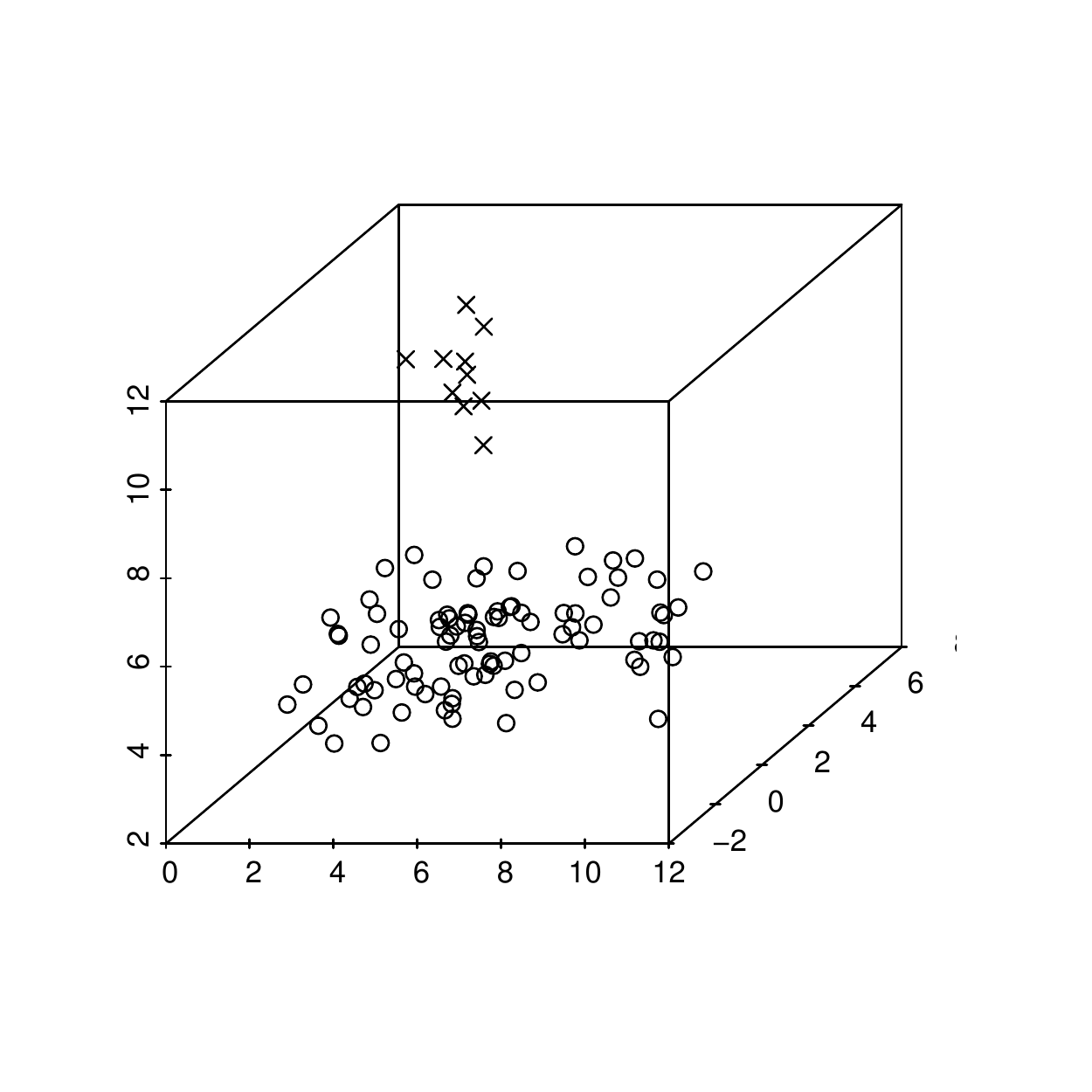}} & 
\resizebox{0.5\textwidth}{!}{\includegraphics{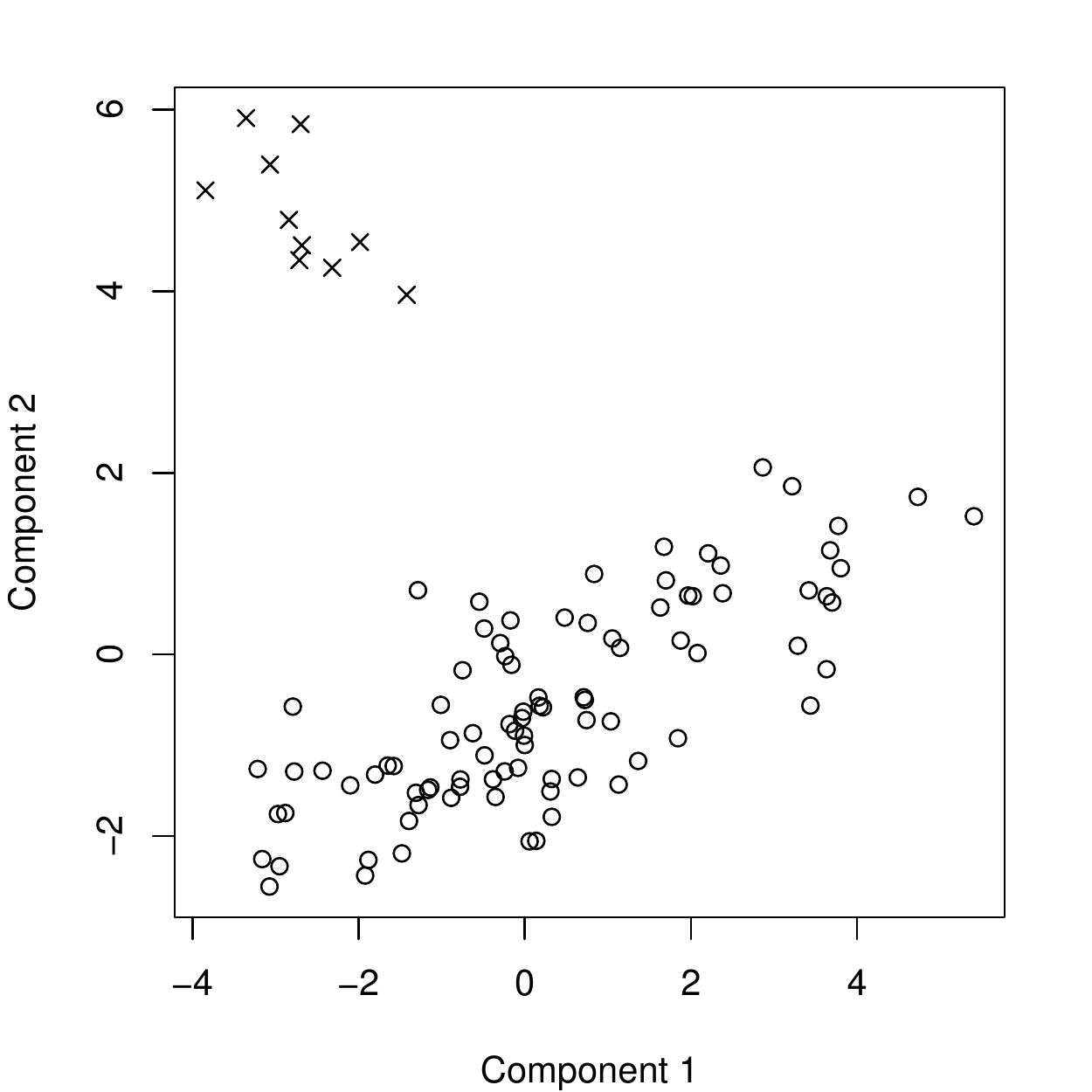}}\\
(a) & (b)
\end{tabular}
\caption{\label{TriOutNoMask} Trivariate normally distributed data cloud ($\circ$) and a group of outliers ($\times$); plotted are (a) the data and (b) a scatterplot of PC1 vs. PC2}      
\end{figure}

\begin{figure}
\begin{tabular}{cc}
\resizebox{0.5\textwidth}{!}{\includegraphics{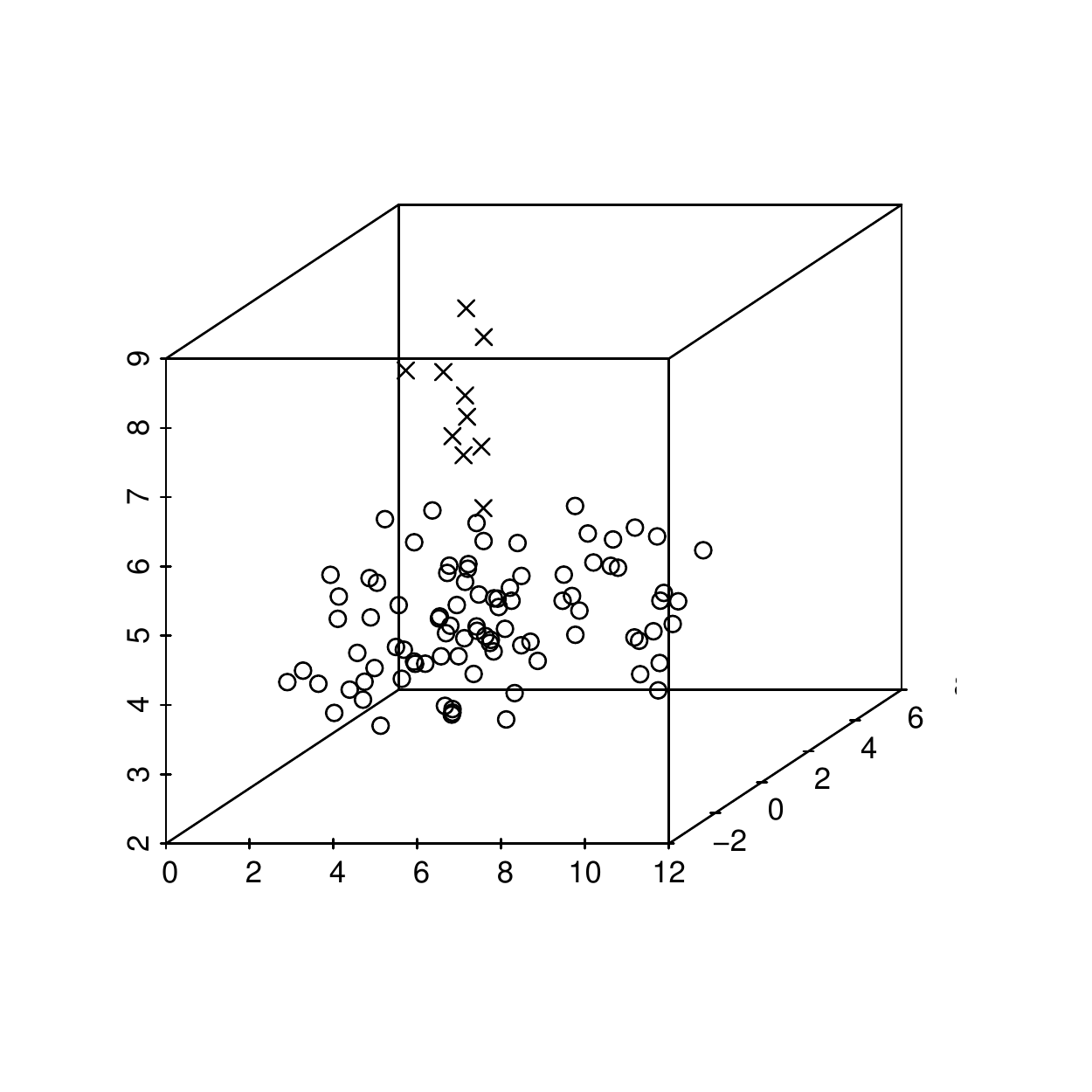}} & 
\resizebox{0.5\textwidth}{!}{\includegraphics{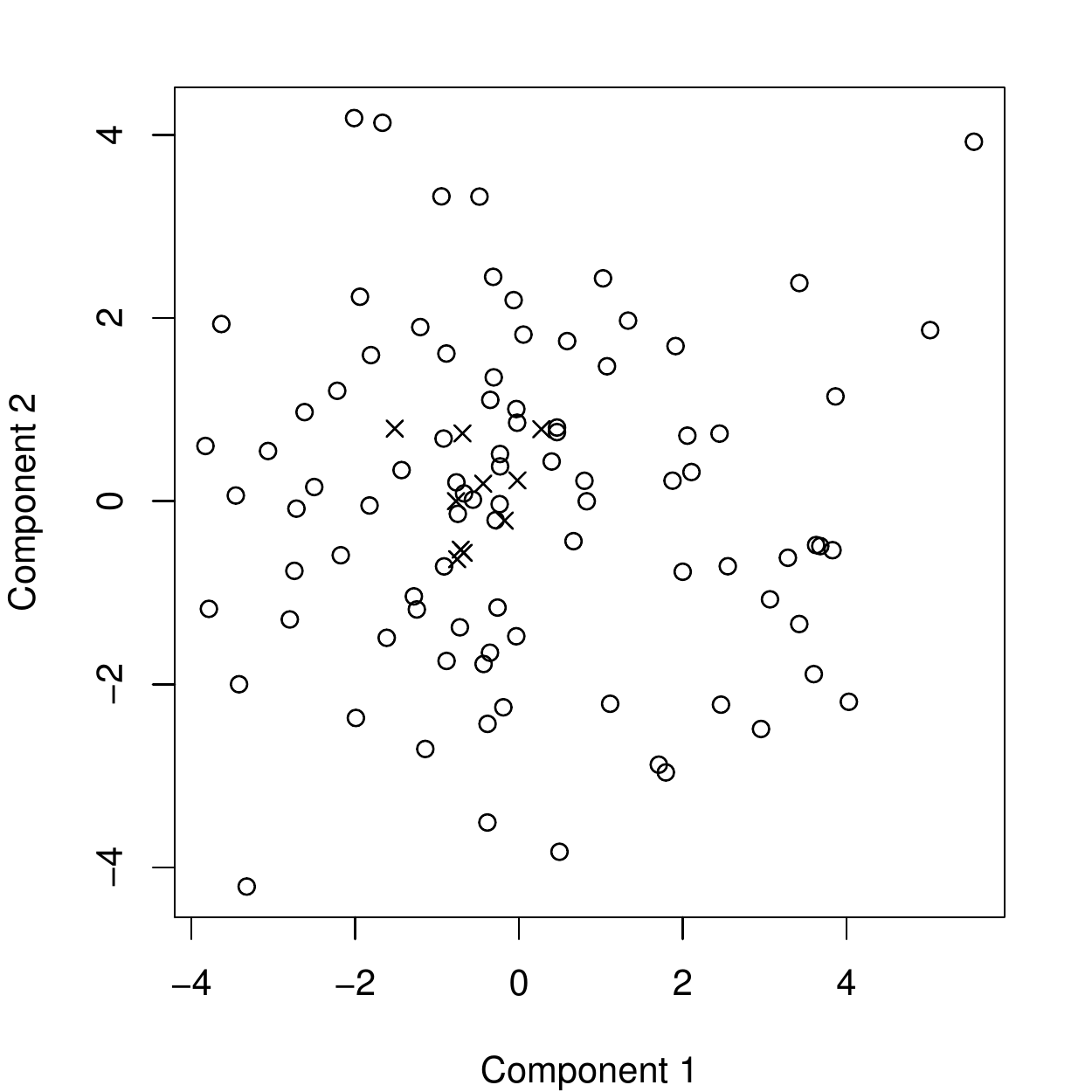}}\\
(a) & (b)
\end{tabular}
\caption{\label{TriOutMask} Trivariate normally distributed data cloud ($\circ$) and a group of outliers ($\times$); plotted are (a) the data and (b) a scatterplot of PC1 vs. PC2}      
\end{figure}

Instead of simply projecting the data onto a pair of its variables, classical statistical methods can be used for data reduction. A straightforward approach is summarise the data into principal components and then making pairwise plots of these principal components. Alternatively, if a dependent variable exists as well, a data reduction technique can be used which takes into account the relation to the predictand, such summarising the data into latent variables by partial least squares or canonical correlation analysis. However, methods like principal component analysis or partial least squares are classical (i.e. nonrobust) estimators, which implies that the outliers do also have an effect on the estimates of the latent variables obtained by those methods. The estimated latent variables can be biased in such a way that in fact no outliers are detected. The effect that due to the outliers' presence one is not able to detect them is referred to as the {\em masking effect}. In Figure \ref{TriOutNoMask} we show a trivariate data distribution to which a cluster (ten percent) of outliers has been added. In the left subplot the data are shown as a three dimensional plot; the right subplot shows a  biplot of the first two principal components (PCs) of this data set. In Figure \ref{TriOutMask} similar plots are shown for the same data cloud where the outliers have been added at a different position. 
From Figures \ref{TriOutNoMask} and \ref{TriOutMask} we see that it depends on the position in space where the outliers are situated whether principal components reveal them or not. A good robust data reduction method (discussed more into detail later in this chapter) would reveal the outliers in both configurations, whereas it would also still yield reasonable results if the data are not contaminated by any type of outliers.
     
\subsection{Swamping effect}

Apart from the masking effect, outliers can also cause typical points to be identified as outliers. This is easily understood by taking regression as an example. 
\begin{figure}
\begin{center}
\includegraphics[width=0.5\textwidth]{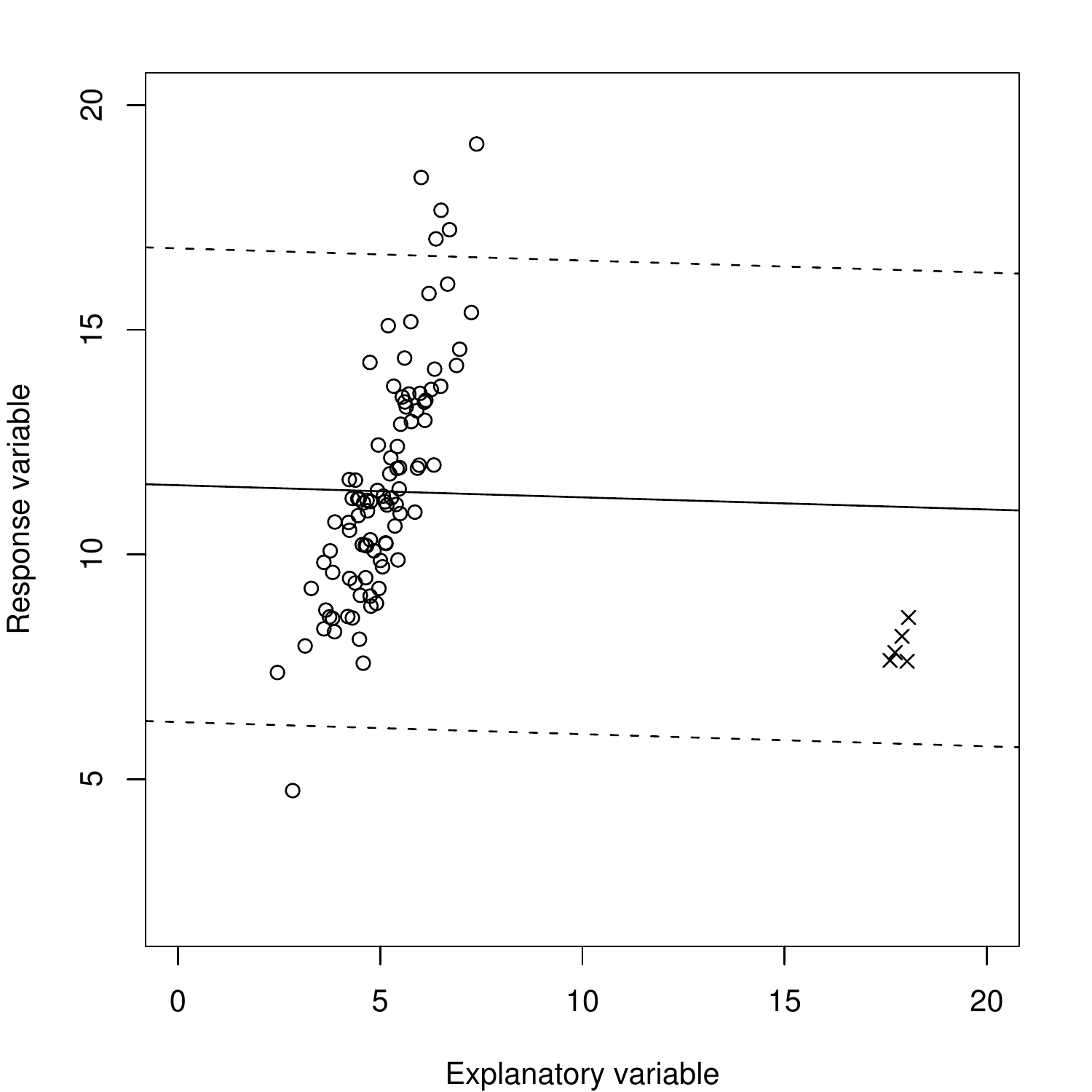}
\caption{\label{Swamp} A typical regression problem with normally distributed errors; a cloud of outliers has been added. The solid line shows the least squares fit to the data. The dotted lines show the limits beyond which points are considered to be outlying.}      
\end{center}
\end{figure}
Figure \ref{Swamp} shows a typical regression problem with normally distributed data to which a group of outliers is added. The least squares regression tries to fit all data points, i.e. also the outliers, and is thus attracted towards the outliers. Normally one would assign outliers to a least squares fit as those points which have the highest residuals to the model. In Figure \ref{Swamp} the regression line is shown together with two dashed lines indicating a residual distance of two standard errors. Potential outliers will fall outside these bands. In the example shown, the points having the largest residuals, which are thus detected ``outliers" are in this case not the true outliers, but typical points which are far away from the regression line just because the latter badly fits the majority of the data. The effect that regular data points are identified as outliers is denominated the {\em swamping effect}.

\begin{figure}
\begin{center}
\includegraphics[width=0.5\textwidth]{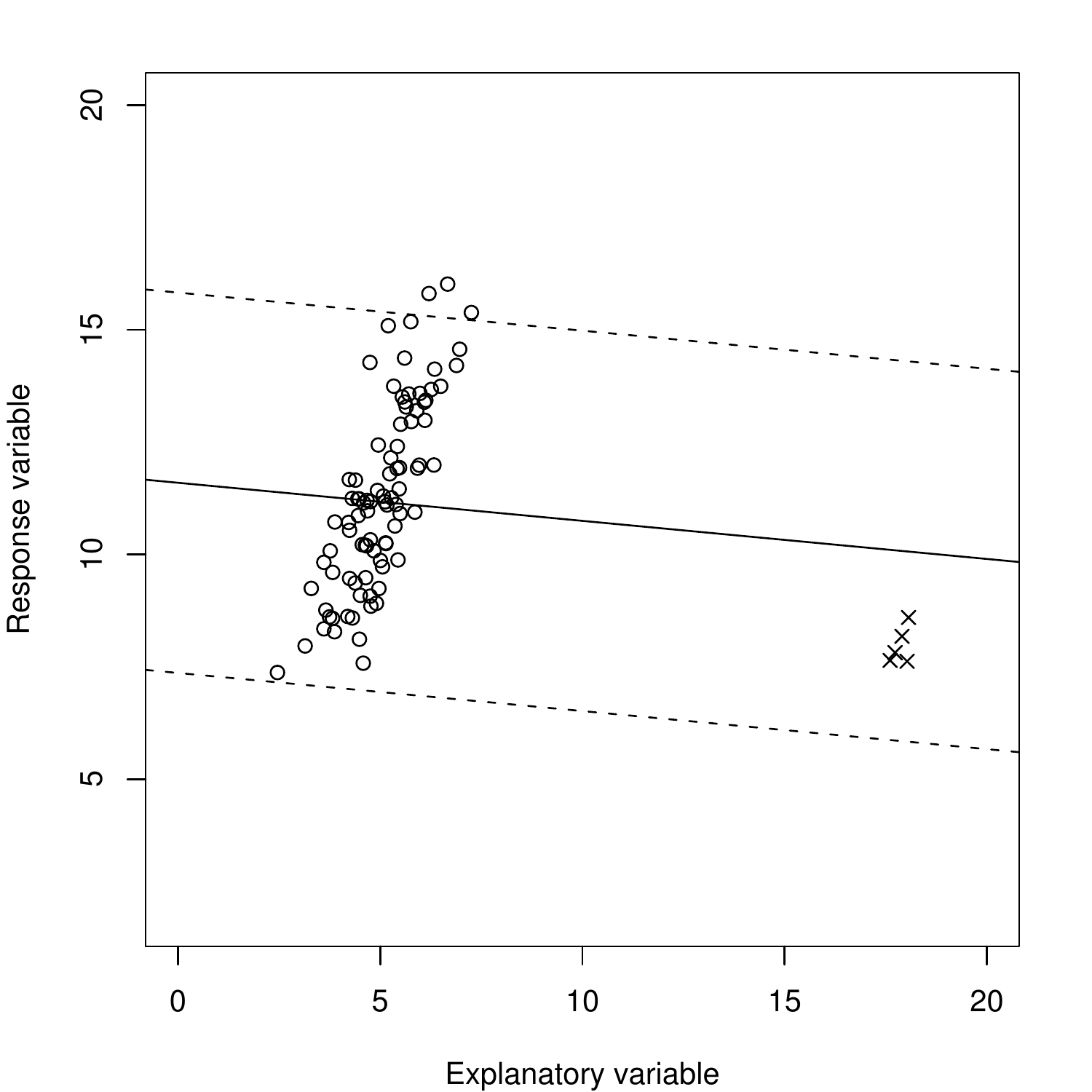}
\caption{\label{Swamp2} The regression problem from Figure \ref{Swamp} after removal of the outliers detected there.}      
\end{center}
\end{figure}

The above figures show two main effects of outliers. At first the outliers themselves are hard to detect (remember that in a multivariate setting, a plot which shows the entire data distribution cannot be constructed). Secondly, the outliers cause the regression line to be severely distorted. Predictions of future responses made according to this line will be unreliable. At this point one would intuitively proceed by detection of the outliers after which the regression analysis is performed on the remaining points. But exactly due to the masking and swamping effects the wrong data points may be identified as outliers such that eventually the regression analysis on the assumed ``clean" data remains a jeopardy. In Figure \ref{Swamp2} we show what happens if one would proceed by omitting the identified outliers in Figure \ref{Swamp}. 
Some of the good data points have been removed; the true outliers are still present. The slope of the regression line is still erroneous due to the outliers and the latter are still not detected. 

\subsection{Majority fit}

The goal of robust methods is to estimate parameters under conditions of slight deviations
to the model. Once the parameters are estimated in a robust way it is possible to
identify outliers which are considered to be deviating data points according to the
underlying statistical model. It should be observed that outlier detection always
involves some subjectivity, while parameter estimation does not.
Gross outliers can be detected and henceforth omitted from the analysis. Apart from these, various other deviations from the underlying model may be present such as a small groups of points which are known to have slightly different properties but cannot be omitted. In this case it is not possible to delete the deviating data points prior to a classical analysis. Nonetheless the classical analysis may probably accord too great an importance to the few differing points. A well chosen robust estimator will provide a reliable fit for the whole range spanned by the data points without being influenced by deviating points, regardless the type of deviation. Even gross outliers may be present in the data without influencing the robust fit. 

Most robust methods may be described as classical methods where the data are weighted,
with weights depending on the data.
The majority of the data will receive a (quasi) uniform weight while the more atypical individual cases are, the lower the weight they will get. Summarising, a robust fit can be considered to be a majority fit, where the fraction of data which makes up the majority depends on the points in space of the individual cases.

\subsection{Is robustness a synonym to wasting information?}

Many early robust estimators were based on trimming (e.g. the well known least trimmed squares estimator for regression
where a pre-determined percentage of the largest squared residuals is trimmed and thus
not considered for finding the regression parameters). 
On the other hand, trimming is not done for {\it any} data points, but especially for
the largest and smallest values. If any data points would be trimmed, the precision
(or efficiency) of the resulting estimator would be much lower than that where
the extremes are trimmed. This fact underlines already that even trimming does not
``waste'' information.
Ideally, robust estimation techniques should only discard data points which are extremely distinct from the bulk of data and are thus very likely to be gross outliers. All other data points should to some extent be taken into account. The amount of information taken from each data point is then regulated by weights between 0 and 1 given to them.
This procedure will in general lead to estimators with higher efficiency.

\section{Designing robust multivariate estimators}

\subsection{Which properties should a robust estimator have?}
\label{properties}
 
Robust estimators should be resistant to a sizeable proportion of outliers or deviation from assumptions. They should also still yield reasonable results if these ideal assumptions are valid. In this section some tools are introduced to assess an estimator's robustness properties: the influence function, the maxbias curve and the statistical efficiency.

\subsubsection{Empirical influence function and influence function}

\begin{figure}
\begin{tabular}{cc}
\resizebox{0.5\textwidth}{!}{\includegraphics{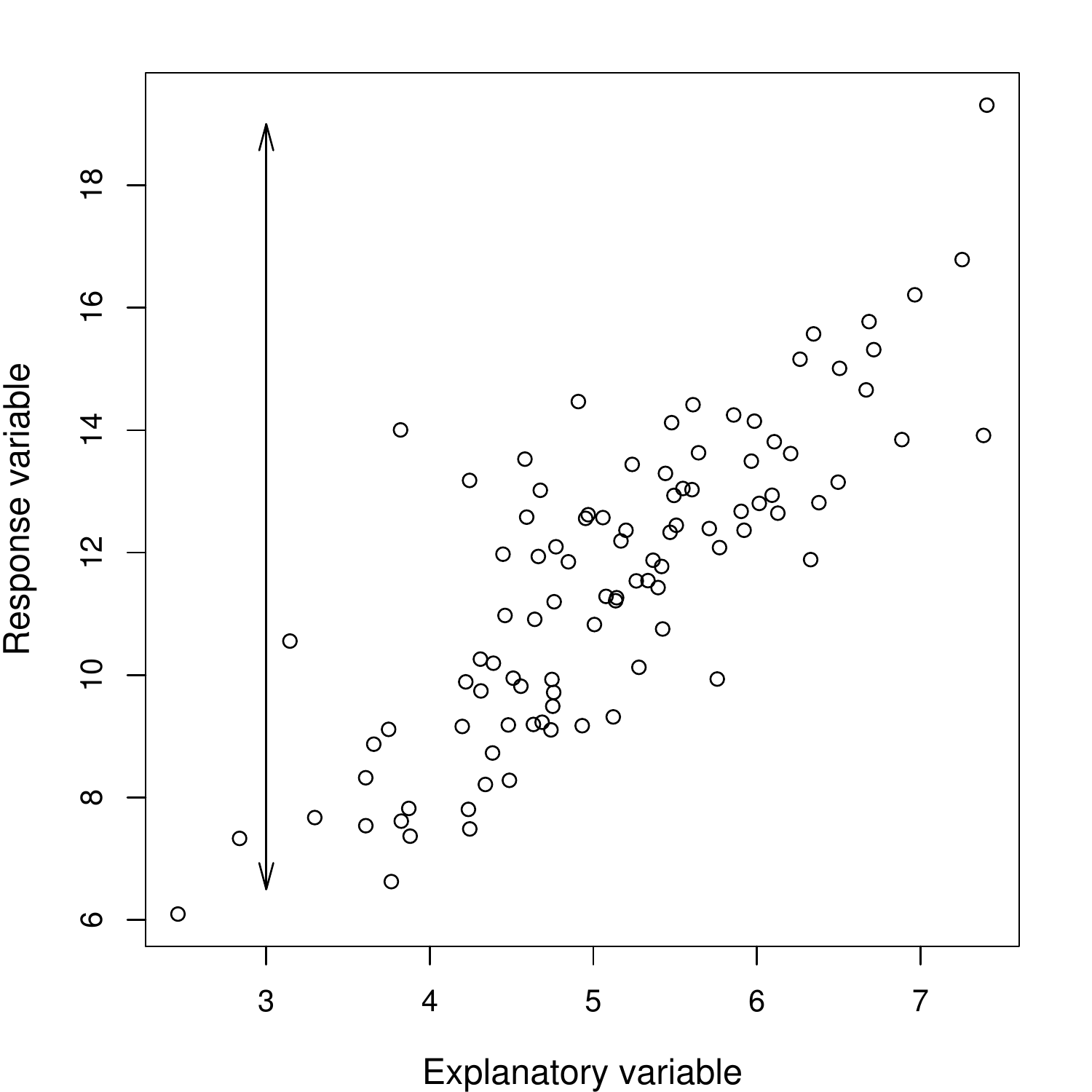}} &
\resizebox{0.5\textwidth}{!}{\includegraphics{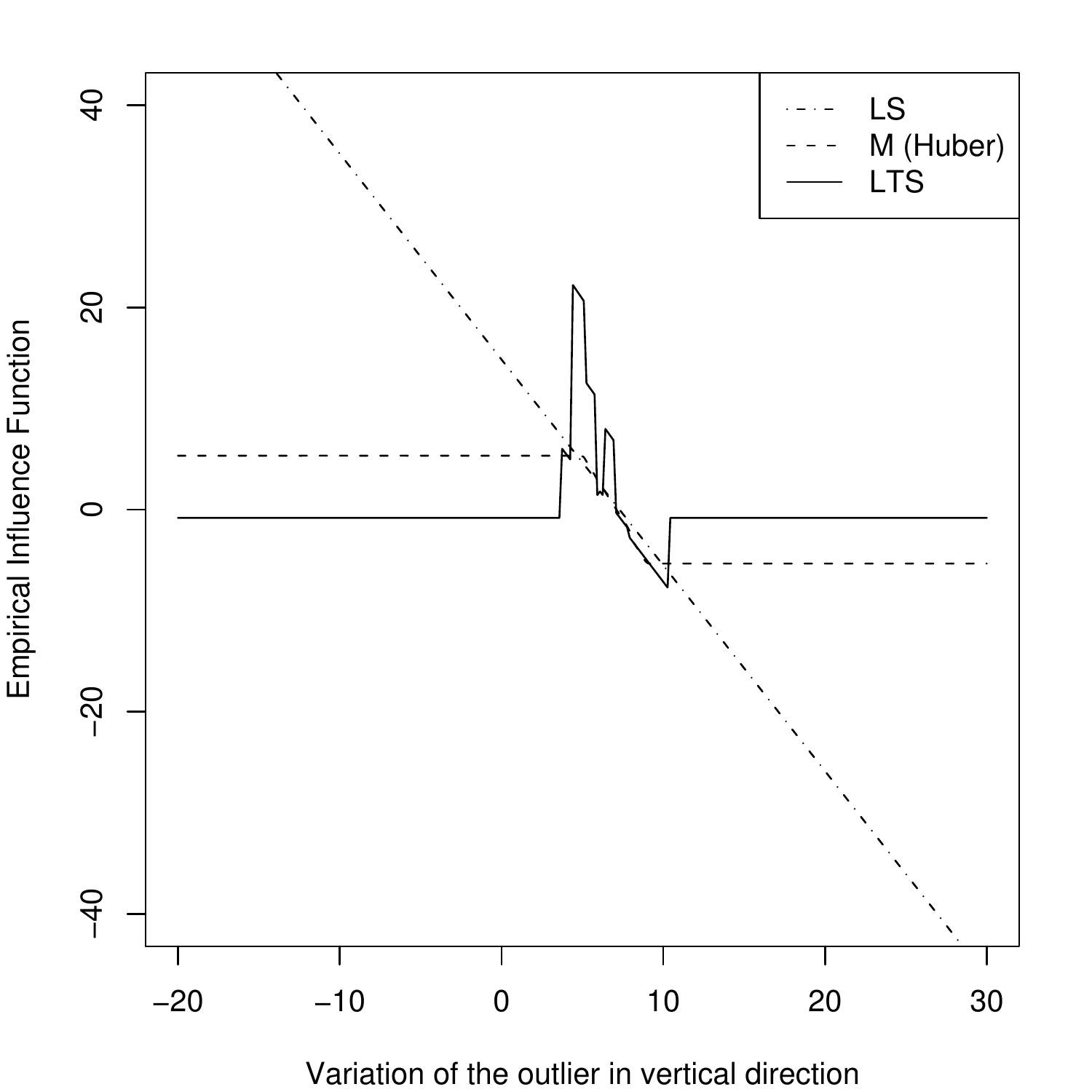}}\\
(a) & (b)
\end{tabular}
\caption{\label{EIFreg} Empirical influence functions for regression estimators. Subplots 
show (a) the data with varying positions of the outlier and (b) the empirical influence 
functions of the slope parameter estimated by least squares (LS), M regression and least 
trimmed squares (LTS). }
\end{figure}

One of the basic ideas of robustness is that a robust estimator should not be influenced by a limited amount of contamination, regardless where this contamination is situated. 
A simple way to check the behaviour of an estimator under small contamination is
to vary a single data point. As an example, Figure 6 shows the effect of an observation
which is varied in space to different regression estimators.
The data were taken from a normal distribution. The position of one data point was changed 
as is shown in Figure \ref{EIFreg}a. For each position of the data point, we are interested
in the change of the slope parameter of the following regression estimators:
least squares regression and two 
robust regression estimators: Huber M regression \cite{Huber} and LTS regression 
estimators (see later in this chapter). 
The result is known as the empirical influence function (EIF), but 
in order to make the results of the different estimators comparable, we compute the
difference of the slopes for the contaminated and uncontaminated data, and divide
by the amount $1/n$ of contamination. 
The right subplot of Figure \ref{EIFreg} shows the results.
It can be seen that the least squares estimator has an unbounded EIF.
This means that a single gross outlier can have an arbitrarily large effect on the estimator.
Both robust estimators possess a bounded EIF. 
However, not only the bound but also the shape of the EIF is of importance 
in order to understand how a robust estimator deals with contamination. Ideally the EIF should
be smooth: it should not show local spikes or should not be a step function. In practice, the
effect of placing a data point at one location and then shifting it to a very close position
should be very small. One observes that indeed the M estimator has a smooth EIF
and is thus virtually insensitive to local shifts in the data. On the contrary, the response of the least trimmed squares estimator to small data perturbations is far from smooth (in the literature it is said to be prone to a high {\em local shift 
sensitivity}).

The concept of the EIF can be formalised by the so-called influence function (IF).
The IF measures the influence an infinitesimal amount of contamination has on an estimator 
with respect to its position in space \cite{HRRS}. More precisely, the influence function of 
an estimator $T$ at a given distribution $G$ is defined as:
\begin{equation}\label{eq:ten}
\mathrm{IF}(\mathbf{z},T,G)=\lim_{\varepsilon\downarrow 0}\frac{T\left[(1-\varepsilon)G+\varepsilon \delta_{\mathbf{z}}\right]-T(G)}{\varepsilon},
\end{equation} 
where $\varepsilon$ is the fraction of contamination and $\delta_\mathbf{z}$ is a probability 
measure which puts all the mass at $\mathbf{z}$. The point $\mathbf{z}$ can be any point in the 
$p$ dimensional space but in practice it will often be a measured data point. Evaluating the 
influence function at the points of a data set reveals how each data point changes the 
estimator's behaviour. The influence of an outlier in the data set on the estimator can be 
measured by evaluating the influence function at the outlier. For nonrobust estimators evaluation 
of the influence function at the outlier will yield significantly different results compared to 
evaluating the IF at the typical data, whereas for a robust estimator the effect will be limited.

\subsubsection{Maxbias curve}

Hitherto we have considered the influence of a limited amount of contamination at varying positions in space. An interesting question is what happens if instead of the position in space one changes the proportion of contamination. What one expects is that a robust estimator can withstand a certain fraction of contamination. The mathematical tool to examine to which extent an estimator is distorted with respect to the fraction of contamination in the data is the maxbias curve. The maxbias curve measures the bias an estimator has with respect to the percentage of the worst possible type of contamination. Let $Z$ be the original data set and $\check{Z}$ be a data set in which $m$ out of $n$ observations have been replaced with arbitrary values and let $\parallel\cdot\parallel$ denote the Euclidean norm, then the maxbias curve for an estimator $T$ is defined as: 
\begin{equation}\label{eq:maxbias}
\mathrm{maxbias}(m, T,Z)= \sup_{\check{Z}} \parallel T(\check{Z})-T(Z) \parallel.
\end{equation}
It is known that for 
some estimates of 
regression the worst possible type of outliers is found at points where $y$, $x$ and the fraction $y/x$ increase to infinity. In what follows a numerical example is shown which does not reflect the exact maxbias curve but illustrates what happens if bad (but not the worst type) of outliers are added to data for the regression problem discussed in the previous section. For the data set presented in Figure \ref{EIFreg}, we have added vertical outliers in the following manner: points were added in a range $(\mu_x + a, \mu_y + b)$ about the mean, where $a \in [0,10]$ and $b \in [10^4,10^5]$. We then computed the bias of the slope compared to the known regression slope. 
\begin{figure}
\begin{center}
\includegraphics[width=0.65\textwidth]{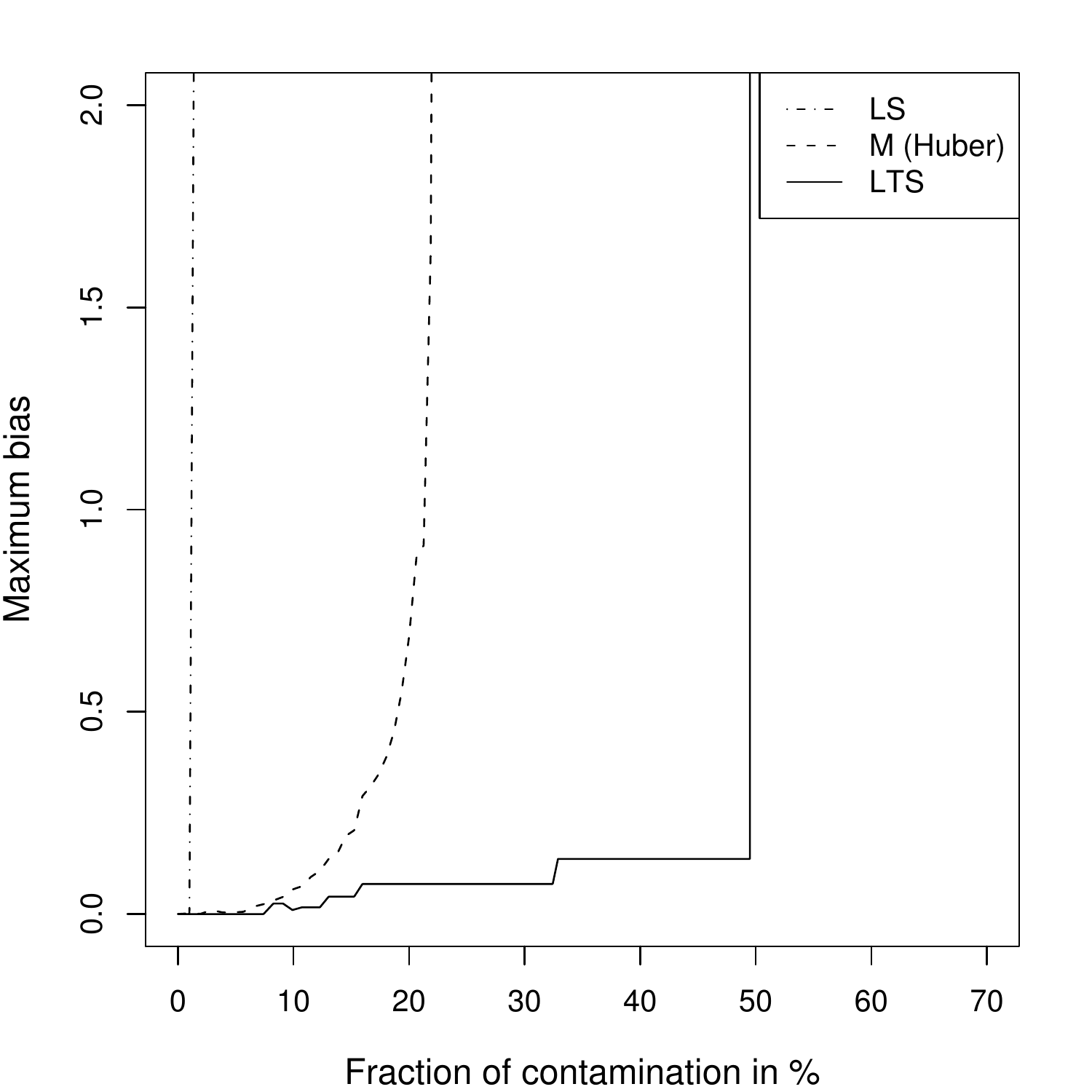}
\caption{\label{maxbias} The regression problem from Figure \ref{Swamp} after removal of the outliers detected there.}      
\end{center}
\end{figure}
In Figure 
\ref{maxbias} we see that the bias of the least squares estimator tends to infinity if a single observation is replaced by bad outliers. The other robust estimators exhibit a moderate bias up till a certain point where they also break down. To conclude we note that if the outliers' position in $x$ would have been put further away from the data cloud too (leverage points), then we would observe a faster breakdown for the robust M regression method displayed here, as this method is known to be only resistant to outliers in $y$. 

\subsubsection{Breakdown point}
\label{sec:bdpreg}

From the bias curves one observes that for each estimator, there exists a point where the bias tends to infinity with $\varepsilon$. This point is referred to as the {\em breakdown point}. Loosely, the breakdown point indicates which percentage of the data may be replaced with outliers before the estimator yields aberrant results. Based on the maxbias curve, for finite samples the breakdown point is given by: 
\begin{equation}\label{bdp}
\varepsilon^*_n(T,Z)=\min\left\{\frac{m}{n}; \mathrm{maxbias}(m,T,Z)=\infty\right\}.
\end{equation}
For $n\rightarrow\infty$ one obtains the {\em asymptotic breakdown point}, denoted $\varepsilon^*$. For least squares regression it holds that $\varepsilon^*=0$. The maximal possible value of the asymptotic breakdown point equals 1. However, estimators satisfying equivariance conditions\footnote{Equivariance conditions are deemed reasonable for most estimating problems, e.g. location, covariance and regression; for the definition of affine equivariance see Section \ref{sec:affeq} and for the definition of scale equivariance see Section \ref{sec:rmrs}.} have a maximal asymptotic breakdown point of 0.5, which means that the typical points must out-number the outliers in order to produce meaningful results. One of the goals in designing robust estimators is obtaining a high breakdown point. Howbeit, bounded influence and high breakdown should not result in a drastic decrease in efficiency. 

\subsubsection{Statistical efficiency}

An important property to any statistical estimator is the variance. It is well known that many parametric estimators have optimality properties at their underlying model. For instance, the maximum likelihood estimator for the linear regression model with normally distributed error terms is the least squares estimator. The least squares estimator is also the minimum variance unbiased estimator for the linear model (the Gau\ss -Markov theorem). This implies that predictions made by any other regression estimator for data which follow the linear model with normally distributed error terms, will have a higher uncertainty than the least squares predictions. So also robust estimators for regression are prone to an increase in variance compared to least squares. This statement may be generalised to other settings than regression; it can be stated that robust estimators always have a higher variance than classical parametric estimators if evaluated at the underlying model of the parametric estimator. They are said to be less {\em efficient} than parametric estimators.
So as to design robust estimators it is important not only to investigate the robustness properties but also the efficiency properties. One could even conjecture that for chemometric  robust estimators, efficiency is a more important property than robustness in the breakdown sense. Data sets for multivariate calibration hardly ever contain 50\% of outliers. Moreover, a few outliers very far away from the data cloud would readily be detected by inspecting the data. What is occurring more frequently is a data set which slightly deviates from normality (the most frequently assumed underlying distribution) without any gross outliers being present. For such data a robust estimator will outperform the classical estimator because the latter is only optimal at the exact normal model, given it is efficient, such that the effect of increase in variance of the robust estimator does not compensate for the classical estimator's loss in precision due to deviation from normality. 
 
\section{Robust regression}\label{sec:robreg}

Regression assumes a key position in chemometrics. Apart from being applied as
a method in its own right, it is also a part of more complex estimators such
as partial least squares or three-way methods. One way to develop robust
alternatives to these methods is by replacing the classical regressions by
robust ones. In that case the properties of the entire method derive from the
type of robust regression used. In this section we present an overview of some
of the most useful robust estimators for regression.

The data of a regression situation are the $n\times p$ matrix $\mathbf{X}$ of
predictor variables with elements $x_{ij}$ and the $n-$vector $\mathbf{y}$
with elements $y_{i}$. For regression with intercept we assume that
the first column of the data matrix is a column of ones. Call  $\mathbf{x}%
_{i}=\left(  x_{i1}, \ldots ,x_{ip}\right)  ^T$ the  column vector
containing the elements of the $i$-th row of $\mathbf{X}$. The linear
regression model is then given by
\begin{equation}
y_{i}=\mathbf{x}_{i}^T\boma{\beta}+e_{i}%
,\ i=1,...,n,\label{regmodel1}%
\end{equation}
where the unknown regression parameter $\boma{\beta}$ is a $p$-vector  and
$e_{i}$ denotes the error terms, which are assumed to be i.i.d. random variables.

For a given estimator ${\hat{\boma{\beta}}}$ call $r_{i}=r_{i}\left(
{\hat{\boma{\beta}}}\right)  =y_{i}-\mathbf{x}_{i}^T{\hat{\boma{\beta}
}}$ the $i$-th residual. Most regression estimators are based on a
minimisation of the size of the residuals. The classical least squares (LS)
estimator is defined as
\begin{equation}
\hat{\boma{\beta}}_{LS}=\arg\min_{\boma{\beta}}\sum_{i=1}^{n}r_{i}\left(
\boma{\beta}\right)  ^{2}.\label{defLS}%
\end{equation}
An ancient alternative to LS is the $L_{1}$ estimator defined as%
\begin{equation}
\hat{\boma{\beta}}=\arg\min_{\boma{\beta}}\sum_{i=1}^{n}\left\vert
r_{i}\left(  \boma{\beta}\right)  \right\vert .\label{defL1}%
\end{equation}

\subsection{M-estimators}

The LS estimator is not robust, in the sense that atypical observations may
uncontrollably affect the outcome. The reason is that a large residual would
dominate the sum (\ref{defLS}). One way out of this difficulty is to define a
more general family of estimators. Note that (\ref{defLS}) and (\ref{defL1})
can be written as:
\begin{equation}
\hat{\mathbf{\boma{\beta}}}=\arg\min_{\boma{\beta}}\sum_{i=1}^{n}\rho(r_{i}\left(
\boma{\beta}\right)  ),\label{defM}%
\end{equation}
where $\rho(r)=r^{2}$ for LS and $\rho\left(  r\right)  =\left\vert
r\right\vert $ for $L_{1}.$ By taking other $\rho$ functions different
estimators are obtained. In fact, equation (\ref{defM}) is the definition of a
whole class of estimators commonly referred to as \emph{M-estimators}.

Note that if $\hat{\boma{\beta}}$ is the LS estimate, then transforming $\mathbf{y}$ to $t\mathbf{y}$ with $t\in\mathfrak{R}$ transforms $\hat{\boma{\beta}}$ to $t\hat{\boma{\beta}}$. This property, called {\em regression equivariance}, is not shared by the estimator \eqref{defM}, except when $\rho(z)=|z|^a$ for some $a$. To make estimator \eqref{defM} scale equivariant, we define in general a regression M-estimator by%
\begin{equation}
\hat{\boma{\beta}}=\arg\min_{\boma{\beta}}\sum_{i=1}^{n}\rho\left(
\frac{r_{i}\left(  \boma{\beta}\right)  }{\hat{\sigma}}\right)  ,
\label{defMequivar}%
\end{equation}
where $\hat{\sigma}$ is a robust scale estimator of the residuals, that can be
estimated either previously or simultaneously with the regression parameters.

The function $\rho$ must be chosen adequately. Recall that we want an estimate
to be (a)\ robust in the sense of being insensitive to outliers, and (b)
efficient in the sense of being similar to LS when there are no outliers. For
(a) to hold, $\rho\left(  r\right)  $ must increase more slowly than $r^{2}$
for large $r$, and for (b), $\rho\left(  r\right)  $ must be approximately
quadratic for small $r$.

Differentiating (\ref{defMequivar}) with respect to $\boma{\beta}$ we get
that the estimate fulfils the system of  \emph{M-estimating equations}%
\begin{equation}
\sum_{i=1}^{n}\psi\left(  \frac{r_{i}\left(  \boma{\beta}\right)  }%
{\hat{\sigma}}\right)  \mathbf{x}_{i}=0\label{M-equation}%
\end{equation}
where $\psi=\rho^{\prime}.$ For LS, $\psi\left(  r\right)  =r$, and
(\ref{M-equation}) are the well-known normal equations. We may then in general
interpret (\ref{M-equation}) as a robustified version of the normal equations,
where the residuals are curbed. For $L_{1}$ we have $\psi\left(  r\right)
=\mathrm{sign}\left(  r\right)  .$ In general, solutions of (\ref{M-equation})
are \emph{local} minima of (\ref{defMequivar}), which may or may not coincide
with the global minimum.

Put $W\left(  r\right)  =\psi\left(  r\right)  /r.$ Then (\ref{M-equation})
may be rewritten as%
\begin{equation}
\sum_{i=1}^{n}w_{i}\left(  y_{i}-\mathbf{x}_{i}^T\boma{\beta}\right)
\mathbf{x}_{i}=0\label{WtNormEq}%
\end{equation}
with $w_{i}=W\left(  r_{i}\left(  \boma{\beta}\right)  /\hat{\sigma}\right)
.$ Then (\ref{WtNormEq}) is a weighted version of the normal equations, and
hence the estimator can be seen as weighted LS, with the weights depending on
the data . For LS, $W$ is constant. For an estimator to be robust,
observations with large residuals should receive a small weight, which implies
that $W\left(  r\right)  $ has to decrease to zero fast enough for large $r.$

If $\psi$ is an increasing function, the estimate is called \emph{monotonic.}
A family of monotonic estimators which contains LS and $L_{1}$ as extreme
cases is the \emph{Huber} family, with $\rho^{\prime}$ given by%
\begin{equation}
\psi_{\mathrm{H},k}\left(  r\right)  =\left\{
\begin{array}
[c]{ccc}%
r & \mathrm{for} & \left\vert r\right\vert \leq k\\
k~\mathrm{sign}\left(  r\right)   & \mathrm{otherwise} &
\end{array}
\right.  \label{defHubPsi}%
\end{equation}
The extreme cases $k\rightarrow\infty$ and $k\rightarrow0$ correspond to LS
and $L_{1},$ respectively.

Monotonic estimates have the computational advantage that (\ref{M-equation})
gives the \emph{global} minima of (\ref{defMequivar}). But they may lack
robustness if $\mathbf{X}$ contains atypical rows (the so-called
\emph{leverage points}). The intuitive reason is that if some $\mathbf{x}_{i}$
is \textquotedblleft large\textquotedblright, then the $i$-th term will
dominate the sum in (\ref{WtNormEq}), which would be unfortunate if $\left(
\mathbf{x}_{i},y_{i}\right)  $ is atypical (a \textquotedblleft bad leverage
point\textquotedblright). For this reason it is better to use M-estimators
given by (\ref{defMequivar}) with a \emph{bounded }$\rho.$ An example is the
\emph{bisquare} family, with%
\begin{equation}
\rho_{\mathrm{B},k}\left(  r\right)  =\left\{
\begin{array}
[c]{ccc}%
\left(  \frac{r}{k}\right)  ^{2}\left(  3-3\left(  \frac{r}{k}\right)
^{2}+\left(  \frac{r}{k}\right)  ^{4}\right)   & \mathrm{for} & \left\vert
r\right\vert \leq k\\
1 & \mathrm{else} &
\end{array}
\right.  .\label{defBisRho}%
\end{equation}
Bounded $\rho$s present computational difficulties which will be discussed in the next Sections.

When $k\rightarrow\infty$ in (\ref{defHubPsi}) or (\ref{defBisRho}), the
corresponding estimate tends to LS and hence becomes more efficient and at the
same time less robust. Thus $k$ is a tuning parameter the choice of which is a
compromise between efficiency and robustness. The usual practice is to choose
$k$ to attain a given efficiency, such as 0.90.

\subsection{Computing M-estimators}

Equation (\ref{WtNormEq}) suggests an iterative procedure to obtain local
minima. Assume we have an initial value $\hat{\boma{\beta}}_{0}.$ Call
$\hat{\boma{\beta}}_{m}$ the approximation at iteration $m.$ Then given
$\hat{\boma{\beta}}_{m},$ compute the residuals $r_{i}=r_{i}\left(
\hat{\boma{\beta}}_{m}\right)  $ and then the weights $w_{i}=W\left(
r_{i}/\hat{\sigma}\right)  ,$ and solve (\ref{WtNormEq}) to obtain
$\hat{\boma{\beta}}_{m+1}$. The procedure is called \emph{iterative
reweighted least squares }(IRWLS), and converges if $W\left(  z\right)  $ is a
decreasing function of $\left\vert z\right\vert $ 
(Maronna et al., 2006)\cite{MaronnaMY06}.

If $\psi$ is monotonic, the choice of $\hat{\boma{\beta}}_{0}$ influences
the number of iterations, but not the final outcome. But if $\rho$ is bounded,
then $\psi$ tends to zero at infinity, which implies that there may be many
local minima, and therefore the choice of $\hat{\boma{\beta}}_{0}$ is
crucial. Using a non-robust initial estimator like LS may yield
\textquotedblleft bad\textquotedblright\ local minima.

A good initial estimator is also necessary to obtain the scale $\hat{\sigma}.$
If there are no leverage points one could use $L_{1}$ as an initial
$\hat{\boma{\beta}}_{0},$ and compute $\hat{\sigma}$ as a robust scale of
the residuals $r_{i}\left(  \hat{\sigma}\right)  $ (e.g. the MAD). But
otherwise we need other choices. The initial estimator should not need a
previous residual scale. Before initial estimators can be considered, we present some
further concepts.

\subsection{Robust measures of residual size}\label{sec:rmrs}

Given $\mathbf{r=}\left(  r_{1},...,r_{n}\right)  $ we shall define a scale
$\sigma\left(  \mathbf{r}\right)  $ such that $\sigma\left(  t\mathbf{r}%
\right)  =|t|\sigma\left(  \mathbf{r}\right)  $ for $t\in\mathfrak{R}$ (called {\em scale equivariance}). We shall consider
two types of scales.

\subsubsection{Scales based on ordered values}

Call $\left\vert r\right\vert _{\left(  i\right)  }$ the ordered absolute
values of the $r_{i}$s: $\left\vert r\right\vert _{\left(  1\right)  }%
\leq...\leq\left\vert r\right\vert _{\left(  n\right)  }.$ The simplest scale
is a \emph{quantile} of $\mathbf{r:}$
\begin{equation}
\sigma\left(  \mathbf{r}\right)  =\left\vert r\right\vert _{\left(  h\right)
}\label{quantscale}%
\end{equation}
for some $h\in\left\{  1,..,n\right\}  .$ For $h=n/2$ we have the median.
Other choices will be considered below.

A smoother alternative is to consider a scale more similar to the standard
deviation, namely the \emph{trimmed squares scale}%
\begin{equation}
\sigma\left(  \mathbf{r}\right)  =\left(  \frac{1}{n}\sum_{i=1}^{h}\left\vert
r\right\vert _{\left(  i\right)  }^{2}\right)  ^{1/2},\label{trimscale}%
\end{equation}
which for $h=n$ gives the familiar root mean squared error (RMSE).

\subsubsection{Scale M-estimators}\label{sec:smest}

Henceforth a $\rho$-\emph{function} will denote a function $\rho$ such that
$\rho(x)\ $is a nondecreasing function of $\left\vert x\right\vert ,$
$\rho(0)=0,$ and $\rho(x)$ is (strictly) increasing for $x>0$ such that
$\rho(x)<\rho(\infty).$ if $\rho$ is bounded, it is also assumed that
$\rho(\infty)=1.$

An M-estimator of scale (an \emph{M-scale} for short) is defined as the
solution $\sigma$ of an equation of the form%
\begin{equation}
\frac{1}{n}\sum_{i=1}^{n}\rho\left(  \frac{r_{i}}{\sigma}\right)
=\delta\label{defMscale}%
\end{equation}
where $\rho$ is a $\rho$-function and $\delta\in\left(  0,\rho\left(
\infty\right)  \right)  .$ The choice $\rho\left(  z\right)  =z^{2}$ and
$\delta=1$ yields the RMSE. The choice $\rho\left(  z\right)  =\mathrm{I}%
\left(  \left\vert z\right\vert >1\right)  $ and $\delta=0.5$ yields
$\sigma=\mathrm{med}\left(  \left\vert r\right\vert \right),$ where ``med" denotes the median. We shall be
interested in estimates with \emph{bounded} $\rho.$

Equation (\ref{defMscale}) is nonlinear, but it is easy to solve iteratively.
Put
\begin{equation}
W_{\sigma}\left(  z\right)  =\frac{\rho\left(  z\right)  }{z^{2}%
}.\label{defWsig}%
\end{equation}
Then (\ref{defMscale}) can be rewritten as%
\[
\sigma^{2}=\frac{1}{n\delta}\sum_{i=1}^{n}w_{i}r_{i}^{2}%
\]
with $w_{i}=W_{\sigma}\left(  r_{i}/\sigma\right)  ,$ which displays
$\hat{\sigma}$ as a weighted RMSE. Given some starting value $\sigma_{0},$ an
iterative procedure can be implemented as was done for regression M-estimators.

\subsubsection{Calibrating scales for consistency}

If $z\sim\mathrm{N}\left(  0,\sigma^{2}\right)  ,$ then the standard deviation
of $z$ is $\sigma$ by definition. The median of $\left\vert z\right\vert $ is
instead 0.675$\sigma.$ Hence if we have a sample $z_{1},...,z_{n},$ then
$\hat{\sigma}=\mathrm{med}\left(  \left\vert z_{1}\right\vert ,...,\left\vert
z_{n}\right\vert \right)  $ will tend to 0.675$\sigma$ for large $n,$ and
therefore $\hat{\sigma}/0.675$ would be an approximately unbiased estimate of
$\sigma$ for normal data. 

In general, given a scale estimate $\hat{\sigma}$ it is convenient to
\textquotedblleft normalise\textquotedblright\ it by dividing it through a
constant $c$ so that $\hat{\sigma}/c$ estimates the standard deviation at the
normal model. The M-scale (\ref{defMscale}) with $\rho=\rho_{\mathrm{B},1}$ has $c$=1.65.

\subsection{Regression estimators based on a robust residual scale}

Given $\mathbf{\beta,}$ let $\mathbf{r}\left(  \boma{\beta}\right)  =\left(
r_{1}\left(  \boma{\beta}\right)  ,...,r_{n}\left(  \boma{\beta}\right)
\right)  .$ We shall consider an estimator of the form%
\begin{equation}
\mathbf{\hat{\beta}=}\arg\min_{\boma{\beta}}\hat{\sigma}\left(
\mathbf{r}\left(  \boma{\beta}\right)  \right)  \label{MinScalReg}%
\end{equation}
where $\hat{\sigma}$ is a robust scale.

\subsubsection{The LMS and LTS estimators}\label{sec:LTS}

If $\hat{\sigma}$ is given by (\ref{quantscale}), we have the \emph{least
quantile estimator.} The case $h=n/2$ is the \emph{least median of squares}
(LMS) estimate. Actually, to attain maximum breakdown point one must take%
\begin{equation}
h=\left[  \frac{n+p+1}{2}\right]  \label{hBest}%
\end{equation}
where $[t]$ is the integer part of $t$. Estimators of this class are very
robust in the sense of having a low bias, but their asymptotic efficiency is zero.

If $\hat{\sigma}$ is given by (\ref{trimscale}), we have the \emph{least
trimmed squares }(LTS) estimator.\emph{ }Again, the optimal $h$ is given by
(\ref{hBest}). Their asymptotic efficiency is about 7\%.

\subsubsection{Regression S estimators}\label{sec:S}

Regression estimators with $\hat{\sigma}$ given by (\ref{defMscale}) are
called \emph{S-estimators. }It can be shown that they satisfy the M-estimating
equations (\ref{M-equation}) with $\psi=\rho^{\prime};$ and it follows that,
given an initial approximation, they can be computed by means of the IRWLS
algorithm. The boundedness of $\rho$ is necessary for the robustness of the
estimate.  But if $\rho$ is bounded, then $\psi$ is not monotonic, which
implies that the equations yield only \emph{local} minima of $\hat{\sigma
}\left(  \mathbf{r}\left(  \boma{\beta}\right)  \right)  ,$ and hence a
reliable starting approximation is needed. An approach to obtain an initial
estimator is given in the next Section.

The efficiency of the S-estimator with $\rho$ the bisquare function is about
29\%. In general, it can be shown that the efficiency of S-estimators cannot
exceed 33\%. Although better than LMS and LTS, S-estimators do not allow the
user to choose a desired high efficiency. This goal is attained by the
estimates to be described in Section \ref{secMM}.

\subsection{The subsampling algorithm}

The approach to find an approximate solution to (\ref{MinScalReg}) is to
compute a "large" finite set of candidate solutions$,$ and replace the
minimisation over $\mathbf{\beta\in}R^{p}$ by minimising $\widehat{\sigma
}(\mathbf{r(\beta}))$ over that finite set. To compute the candidate solutions
we take subsamples of size $p$
\[
\left\{  \left(  \mathbf{x}_{i},y_{i}\right)  :i\in J\right\}  ,\ \ J\subset
\left\{  1,...,n\right\}  ,\ \#\left(  J\right)  =p.
\]
For each $J$ find $\boma{\beta}_{J}$ that satisfies the exact fit\textbf{\ }%
$\mathbf{x}_{i}^T\boma{\beta}_{J}=y_{i}$ for $i\in J.$ Then the
problem of minimising $\widehat{\sigma}(\mathbf{r(\beta))}$ for $\mathbf{\beta
\in}R^{p}$ is replaced by the finite problem of minimising $\widehat{\sigma
}(\mathbf{r(\beta}_{J}\mathbf{))}$ over $J.$ Since choosing all $\binom{n}{p}$
subsamples would be prohibitive unless both $n$ and $p$ are rather small, we
choose $N$ of them at random: $\left\{  J_{k}:k=1,..,N\right\}  $ and the
initial estimate $\widehat{\boma{\beta}}_{J_{k^{\ast}}}$ is defined by
\begin{equation}
k^{\ast}=\arg\min\left\{  \widehat{\sigma}\left(  \mathbf{r}\left(
\boma{\beta}_{J_{k}}\right)  \right)  :k=1,...,N\right\}  .\label{sigbetJ}%
\end{equation}

Suppose the sample contains a proportion $\varepsilon$ of outliers. The
probability of an outlier-free subsample is $\alpha=(1-\varepsilon)^{p},$ and
the probability of at least one outlier-free\ subsample is $1-(1-\alpha)^{N}$.
If we want this probability to be larger than $1-\gamma,$ we must have
\[
\ln\gamma\geq N\ln\left(  1-\alpha\right)  \approx-N\alpha
\]
and hence
\begin{equation}
N\geq\frac{\left\vert \ln\gamma\right\vert }{\left\vert \ln\left(  1-\left(
1-\varepsilon\right)  ^{p}\right)  \right\vert }\approx\frac{\left\vert
\ln\gamma\right\vert }{\left(  1-\varepsilon\right)  ^{p}}\label{Npeps}%
\end{equation}
for $p$\ not too small. Therefore $N$ must grow exponentially with $p$ if
robustness is to be ensured.

\subsection{Regression MM-estimators\label{secMM}}

We are now ready to define a family of estimators attaining both robustness and controllable
efficiency. We shall deal with (\ref{defMequivar}) where $\rho\left(
z\right)  $ is a  \emph{bounded} $\rho$-function. We assume that there is a
previous estimator $\hat{\boma{\beta}}_{0}$ which is robust but possibly
inefficient (e.g. an S-estimator)$.$ Compute $\hat{\boma{\beta}}_{0}$ and
the corresponding residuals $r_{i}.$ Compute $\hat{\sigma}$ as an M-scale
(\ref{defMscale}) with $\rho=\rho_{\mathrm{B},1}.$ Let $\tilde{\sigma}%
=\hat{\sigma}/c_{0}$ with $c_{0}=1.65$. Now let $\rho=$ $\rho_{\mathrm{B},k}$
with $k$ chosen to have a given efficiency $\gamma$. We recommend
$\gamma=0.85$ which implies $k=$ 3.44. Then compute $\hat{\boma{\beta}}$ as
a local solution of (\ref{defMequivar}) using the IRWLS starting from
$\hat{\boma{\beta}}_{0}.$ The resulting estimator has the BP of
$\hat{\boma{\beta}}_{0}$ and the asymptotic efficiency $\gamma.$

It is shown by Maronna et al. (2006)\cite{MaronnaMY06} that if $\hat{\boma{\beta}}_{0}$ is the
bisquare S-estimator, then the resulting MM-estimator with efficiency 0.85 has
a contamination bias not much larger than that of $\hat{\boma{\beta}}_{0}.$

\subsection{Robust location and covariance}\label{sec:loccov}

Multivariate location and covariance play a central role in multivariate
statistics because many multivariate methods directly build on these
estimates. For example, principal component analysis is carried
out on the centred data, and the standard method uses a decomposition
of the covariance matrix to find the principal components.
Outliers or deviations from a model distribution can lead to 
very different results, and thus it is necessary to robustly estimate
multivariate location and covariance. Many methods have been proposed
for this purpose. Before discussing various approaches, we will first
think about desired 
properties of robust location and covariance estimators. Aside from
robustness issues, a central property is affine equivariance which
will be discussed below.

\subsubsection{Affine equivariance}\label{sec:affeq}

It is desirable that location and covariance estimates
respond in a mathematically convenient form to
certain transformations of the data. For example, if a constant is
added to each data point, the location estimate of the modified data
should be equal to the location estimate of the original data plus
this constant, but the covariance estimate should remain unchanged. 
Similarly, if each data point is multiplied by a constant,
the new location estimate should be equal to the old one multiplied by
the same constant, and the new variances should be the constant squared
times the old variances. More general, one can define a 
transformation that is using a nonsingular $p\times p$ matrix 
$\mathbf{A}$ and a vector $\mathbf{b}$ of length $p$ to transform the
$p$-dimensional observations $\mathbf{x}_1,\ldots ,\mathbf{x}_n$
by $\mathbf{Ax}_j + \mathbf{b}$. This transformation performs any desired
nonsingular linear transformation of the original data. Thus, if $\mathbf{t}$
denotes a location estimator, it is requested that
\begin{equation}\label{affineloc}
\mathbf{t}(\mathbf{Ax}_1+\mathbf{b},\ldots ,\mathbf{Ax}_n+\mathbf{b})=
\mathbf{A}\cdot \mathbf{t}(\mathbf{x}_1,\ldots ,\mathbf{x}_n)
+\mathbf{b} ,
\end{equation}
and for a covariance estimator $\mathbf{C}$ we require
\begin{equation}\label{affinecov}
\mathbf{C}(\mathbf{Ax}_1+\mathbf{b},\ldots ,\mathbf{Ax}_n+\mathbf{b})=
\mathbf{A}\cdot \mathbf{C}(\mathbf{x}_1,\ldots ,\mathbf{x}_n)
\cdot \mathbf{A}^T.
\end{equation}
Location and covariance estimators that fulfil (\ref{affineloc}) and
(\ref{affinecov}) are called {\it affine equivariant} estimators.
These estimators transform properly under changes of the origin,
the scale, or under rotations. 

\begin{figure}
\begin{tabular}{cc}
\resizebox{0.5\textwidth}{!}{\includegraphics{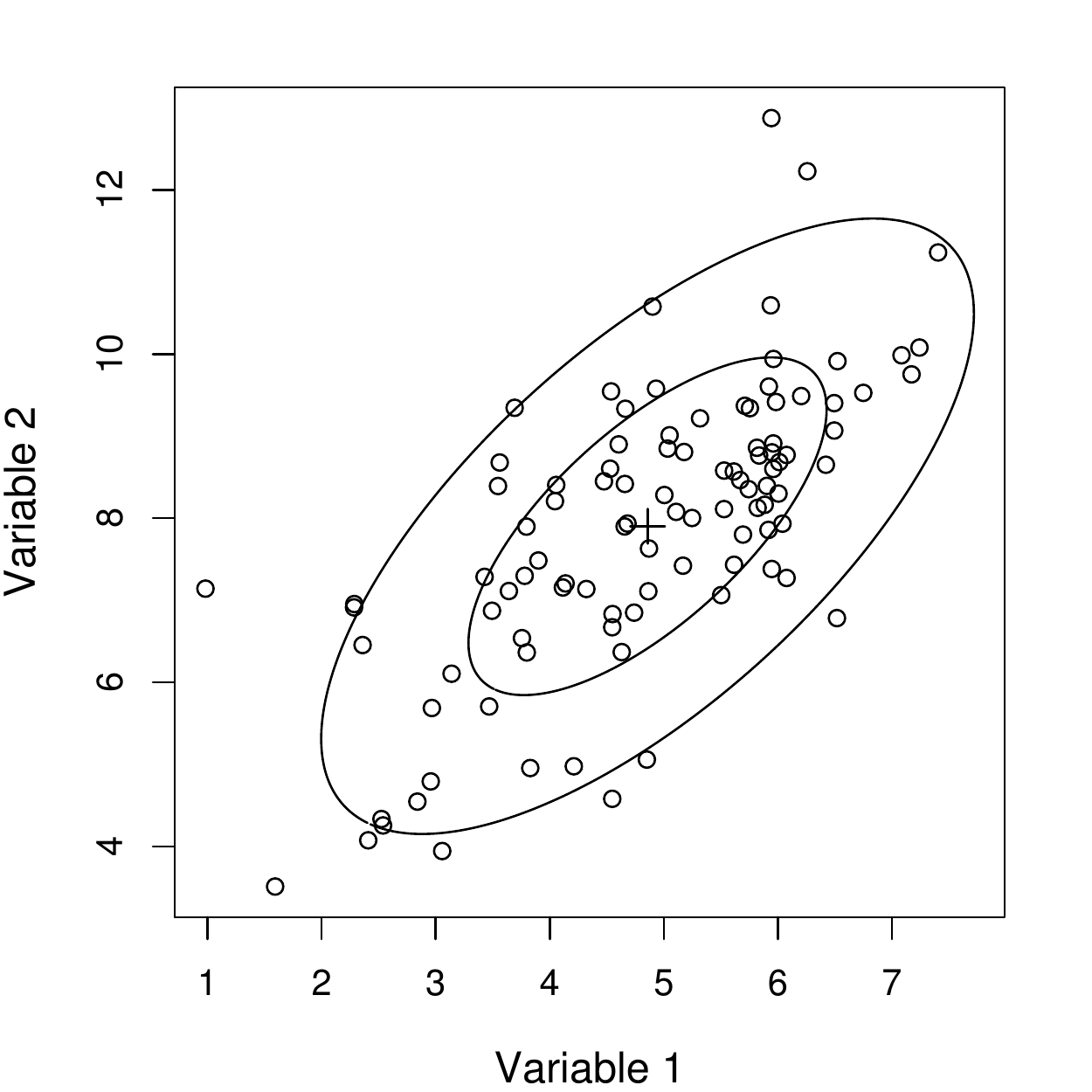}} &
\resizebox{0.5\textwidth}{!}{\includegraphics{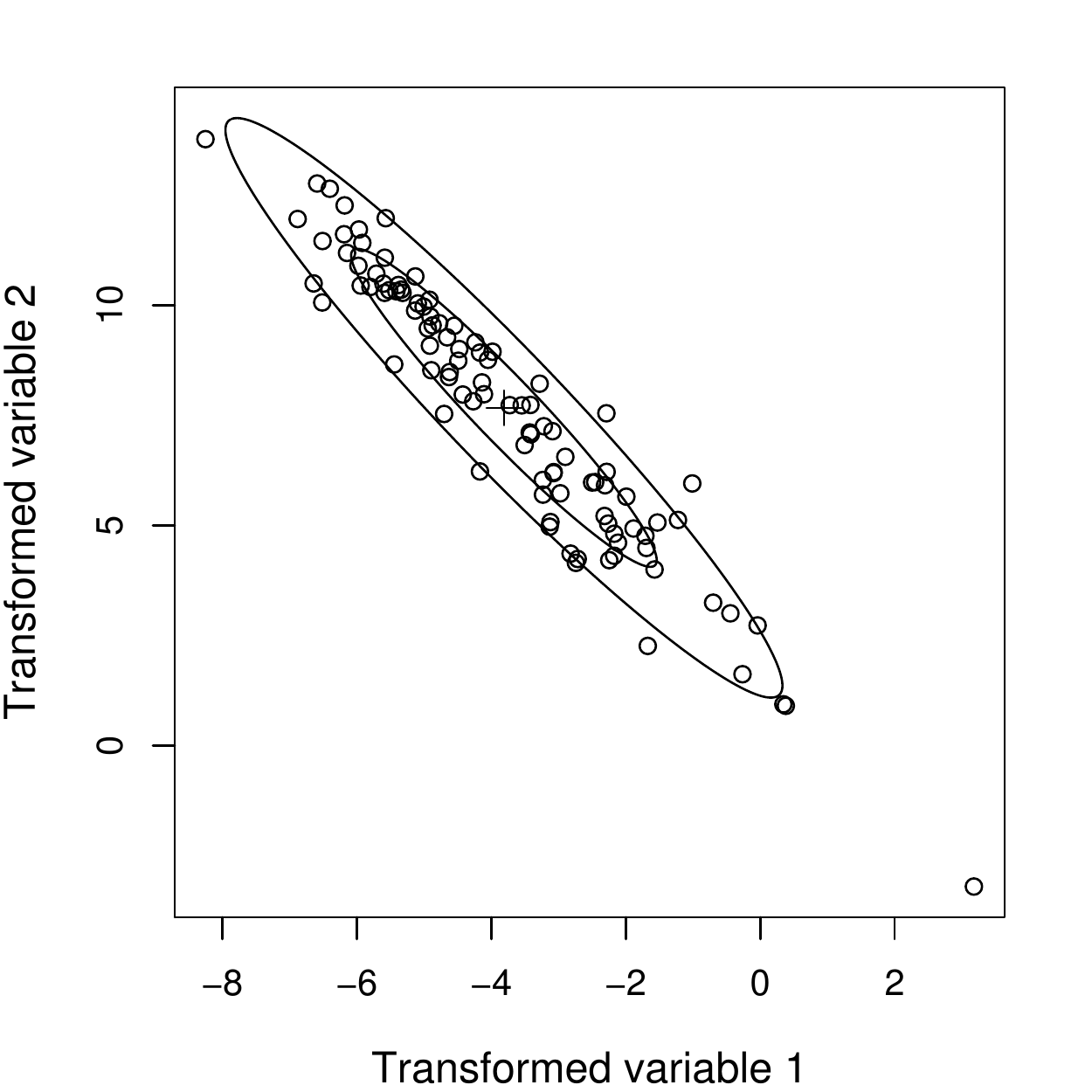}}\\
(a) & (b)
\end{tabular}
\caption{\label{affinefig} Bivariate data with estimated location ($+$) and
covariance matrix (visualised by tolerance ellipses); plotted are (a) the 
original data and (b) the transformed data, together with the transformed
estimates.}
\end{figure}

Figure \ref{affinefig}a shows a bivariate data set where the location
($+$) was estimated by the arithmetic mean and the covariance by
the sample covariance matrix. The latter is visualised by so-called
{\it tolerance ellipses}: In case of normally distributed data the
tolerance ellipses would contain a certain percentage of data points
around the centre and according to the covariance structure.
Here we show the 50\% and 90\% tolerance ellipses.
Figure \ref{affinefig}b pictures the data after applying the transformation
$$
\mathbf{Ax}_j+\mathbf{b}=
\left(
\begin{array}{rr}
-2 & 3 \\
1 & -1 \\
\end{array}
\right)
\mathbf{x}_j + \mathbf{b}
$$
to each data point. The location and covariance estimates were not
recomputed for the transformed data but were transformed according
to the equations (\ref{affineloc}) and (\ref{affinecov}).
It is obvious from the figure that the transformed estimates are
the same as if they would have been derived directly from the
transformed data. Note that the transformation matrix $\mathbf{A}$
is close to singularity because the spread of the data becomes 
very small in one direction. The property of affine equivariance
is only valid for nonsingular transformation matrices.

Affine equivariance is not only important for estimation of location
and covariance but also for multivariate methods like discriminant
analysis or canonical correlation analysis. The results of these
methods will remain unchanged under linear transformations. This
is different for principal component analysis which is 
{\it orthogonal equivariant} but not affine equivariant. The results will 
only be properly transformed under orthogonal transformation matrices
$\mathbf{A}$.

\subsubsection{Asymptotic breakdown point}

The breakdown point (for simplicity, we will omit ``asymptotic in this section) was already discussed in section \ref{sec:bdpreg} in 
the context of regression, and it is also used as an important characterisation
of robustness for location and covariance estimators. Clearly, the breakdown
points of the classical estimators, the arithmetic mean and the sample
covariance matrix, are both 0 because even a single observation placed
at an arbitrary position in space can completely spoil these estimators.

The simplest choice for a very robust location estimator 
would be the median, computed for each variable.
Since the median for univariate data has a breakdown point of 0.5,
the coordinate-wise median would also have this
breakdown point. However, this estimator is not affine equivariant.
For the data used in Figure \ref{affinefig}a the median for the variables
is ($4.97,8.14$) and for the transformed data in Figure \ref{affinefig}b
it is ($-4.12,7.98$). The transformation of the coordinate-wise median
of the original data results in ($-3.79,7.76$). This difference would
in general become larger if less data points were available.

The univariate median is that point which minimises the sum of the
distances to all data points.
A natural extension of this concept to higher dimensions is called
{\it spatial median} or $L_1$-median. It is defined as that point 
in the multivariate
space which minimises the sum of the Euclidean distances
to all data points. The spatial median has good statistical properties:
it has a breakdown point of 0.5.
However, this multivariate location estimator is only orthogonal
equivariant but not affine equivariant. Note that in the context of
principal component analysis or partial least squares
this would be sufficient because these
methods are only orthogonal equivariant.

\subsubsection{The MCD estimator}

An estimator of multivariate location and covariance which is affine
equivariant and has high breakdown point is the {\it Minimum Covariance
Determinant} (MCD) estimator. The idea behind this estimator is in fact
related to the LTS estimator from Section \ref{sec:LTS}. Here,
one is searching for those $h$ data points
for which the determinant of the (classical) covariance matrix is
minimal. The location estimator $\mathbf{t}$ is the mean of these
$h$ observations, and the covariance estimator $\mathbf{C}$ is 
given by the covariance matrix with the smallest determinant, but
multiplied by a constant to obtain consistency for normal distribution.
The parameter $h$ determines the robustness but also the efficiency of
the resulting estimator. The highest possible breakdown point can be
achieved if $h\approx n/2$ is taken, but this choice leads to a low
efficiency. On the other hand, for higher values of $h$ the efficiency
increases but the breakdown point decreases. Therefore, a compromise
between efficiency and robustness is considered in practice.

The computation of the MCD estimator is not trivial. While for a low
number of samples in low dimension in principle all subsets of
$h$ data points can be considered in order to find the subset
with smallest determinant of its covariance matrix, this is no
longer possible for large $n$ or in higher dimension. For this
situation, a fast algorithm has been proposed which finds an 
approximation of the solution \cite{RousVDr}. 

It is important to note that the MCD estimator can only be applied
to data sets where the number of observations is larger than the
number of variables, which is a serious limitation for many applications
in chemometrics. The reason is that if $p>n$ then also $p>h$, and
the covariance matrix of any $h$ data points will always be singular,
leading to a determinant of 0. Thus, each subset of $h$ data points
would lead to the smallest possible determinant, resulting in a
non-unique solution. In fact, a non-trivial solution can only be obtained if
$h$ is smaller than the rank of the data.

\begin{figure}
\begin{tabular}{cc}
\resizebox{0.5\textwidth}{!}{\includegraphics{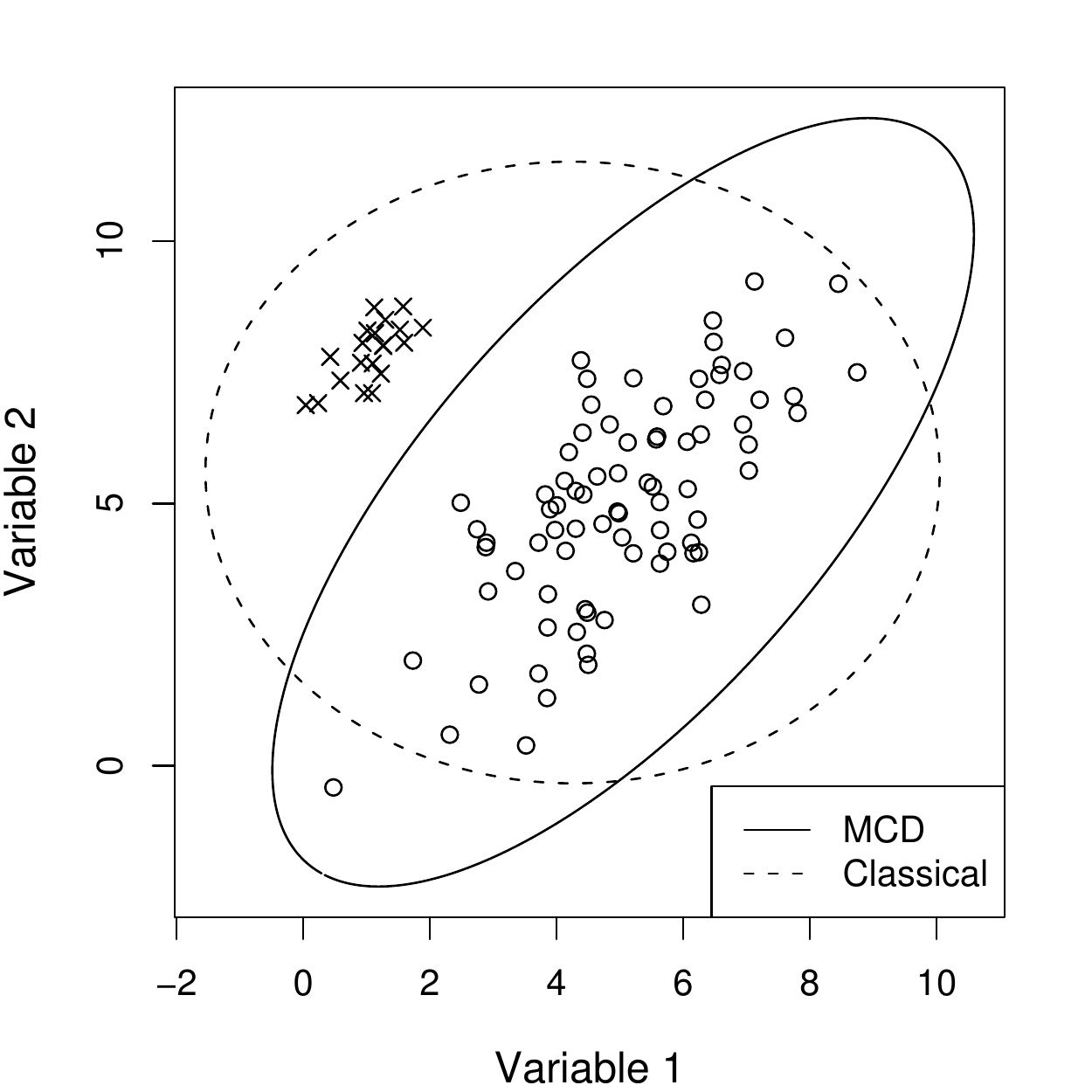}} &
\resizebox{0.5\textwidth}{!}{\includegraphics{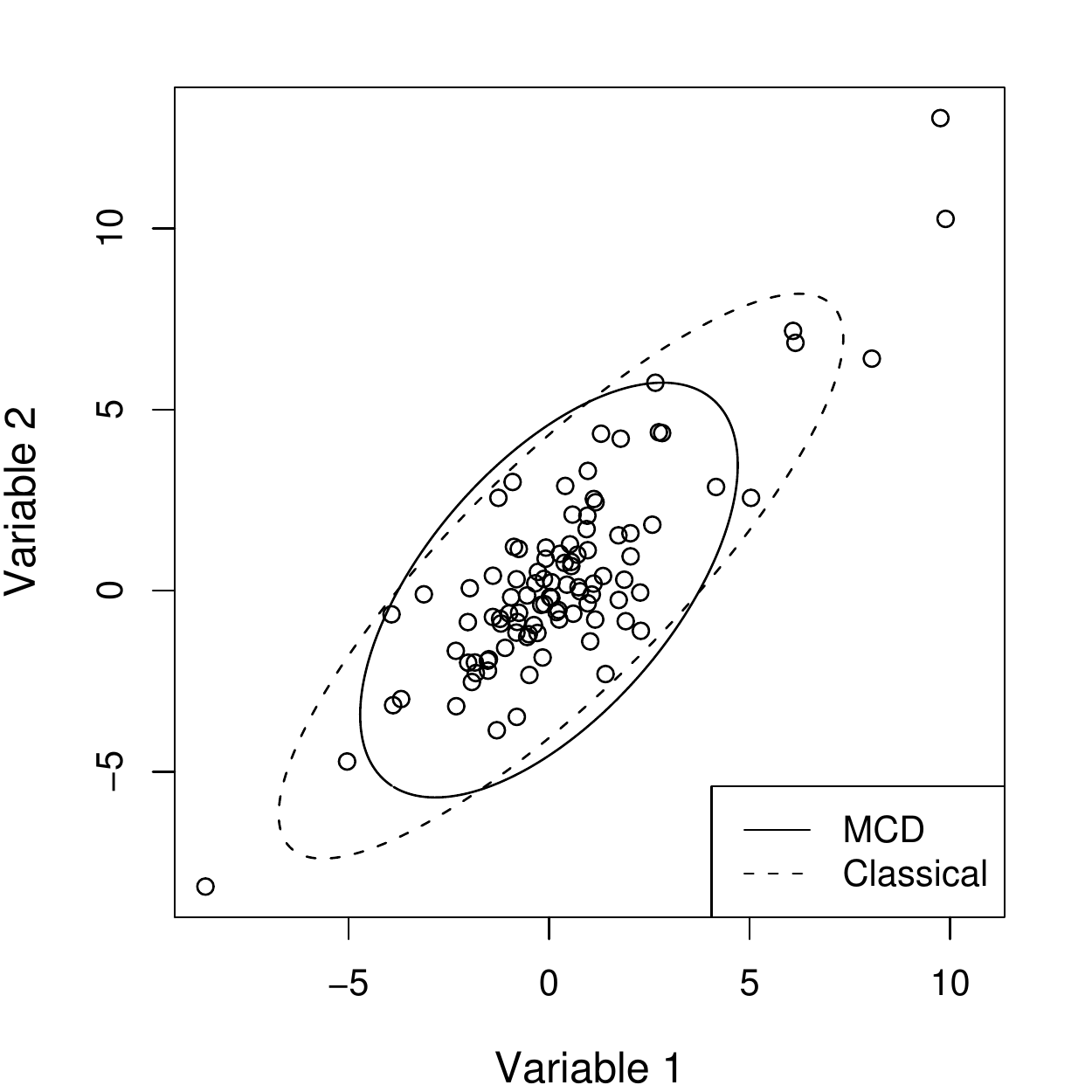}}\\
(a) & (b)
\end{tabular}
\caption{\label{mcdfig} Tolerance ellipses (97.5\%) based on the MCD estimator
and on the classical sample mean and sample covariance matrix for
(a) bivariate normally distributed data with outliers and 
(b) bivariate $T_2$ distributed data.}
\end{figure}

Figure \ref{mcdfig} shows a comparison between the MCD estimator and 
classical location and covariance estimation for two simulated data
sets. The covariance estimates are visualised by 97.5\% tolerance
ellipses, the location estimates are the centres of the ellipses.
In Figure \ref{mcdfig}a a bivariate normally distributed data set is used,
where 20\% of the data points are generated with a different mean and
covariance. Note that these deviating data points cannot be identified
as outliers by inspecting the projections on the coordinates. Not only
the location estimate is influenced by the deviating points, but especially
the covariance structure. The data coming from the outlier distribution
are inflating the tolerance ellipse based on the classical
estimators while that based on the MCD is much more compact and
reflects the structure of the majority of data. 

The second example shown in Figure \ref{mcdfig}b is simulated from a 
bivariate $T$ distribution with 2 degrees of freedom with a certain
covariance structure. Also here the inflation of the classical ellipse
due to some very distant points is visible. 

We can also compute the correlation coefficient using the classical and
robust covariance estimates. In the first example the classical correlation
is $0.00$ while the MCD gives a correlation of $0.70$, which is also
the result of the classical correlation for the data without outliers.
For the second example we obtain a value of $0.84$ for the classical correlation 
and $0.60$ for the robust correlation.

\subsubsection{Multivariate S estimators}

Similar as in the regression context (see Section \ref{sec:S}), it is possible to define S estimators 
in the context of robust location and covariance estimation \cite{Davies,Lopuhaa}.
The idea is to make the Mahalanobis distances small. The Mahalanobis or multivariate
distances are defined as
$$
d(\mathbf{x}_i,\mathbf{t},\mathbf{C})=
(\mathbf{x}_i-\mathbf{t})^T \mathbf{C}^{-1} (\mathbf{x}_i-\mathbf{t}) 
\quad \mbox{ for } i=1,\ldots ,n
$$
for a location estimator $\mathbf{t}$ and a covariance estimator $\mathbf{C}$. Note that $d$ is actually a squared distance.
Thus, in contrast to the squared Euclidean distance
$$
d(\mathbf{x}_i,\mathbf{t})=
(\mathbf{x}_i-\mathbf{t})^T (\mathbf{x}_i-\mathbf{t})
\quad \mbox{ for } i=1,\ldots ,n
$$
the Mahalanobis distance also accounts 
for the covariance structure of the data.
Small Mahalanobis distances can be achieved by using a scale estimator $\sigma$
and minimising
$\sigma(d(\mathbf{x}_1,\mathbf{t},\mathbf{C}),\ldots ,d(\mathbf{x}_n,\mathbf{t},\mathbf{C}))$
under the restriction that the determinant of $\mathbf{C}$ is 1.
Davies \cite{Davies} suggested to take for the scale estimator $s$ an 
M estimator of scale \cite{Huber,HRRS} which has been defined in Equation \eqref{defMscale}.

S estimators are affine equivariant, for differentiable $\rho$ they are asymptotically
normal, and for well-chosen $\rho$ and $\delta$ they achieve maximum breakdown point.

\subsubsection{Multivariate MM estimators}

Like in robust regression (see Section \ref{secMM}), a drawback of S estimators is that their asymptotic
efficiency might be rather low. MM estimators for multivariate location and
covariance combine both high breakdown point and high efficiency
\cite{Lopuhaa92}.
The resulting estimators are affine equivariant and have bounded
influence function.
The solution for the estimators can be found by an iterative algorithm.

\subsubsection{The Stahel-Donoho estimator}

The name of this estimator for multivariate location and covariance origins
from independent findings of Stahel \cite{Stahel81} and Donoho \cite{Donoho82}.
The idea is based on downweight outlying observations in the classical estimation
of multivariate location and covariance. Outlying observations are observations
which are deviating from the multivariate data structure with respect to the majority 
of data points. Note that multivariate outliers are not necessarily univariate
outliers, since they can be ``hidden'' in the multivariate space. An example are 
the outliers in Figure \ref{mcdfig}a that could not be identified as univariate
outliers by inspecting the values on the original coordinates.

Finding multivariate outliers is in some sense strongly related to multivariate
location and covariance estimation. Once reliable estimates have been derived,
the Mahalanobis distances can be computed, and observations with large values of the Mahalanobis
distance can be considered as potential multivariate outliers. All methods discussed
so far possessing high breakdown point are potentially suitable for multivariate
outlier detection.

The Stahel-Donoho estimator first identifies multivariate outliers
in a very simple way: Each observation is projected to the one-dimensional
space and a measure of outlyingness is computed. Of course, there are
infinitely many possible projection directions from multivariate to one
dimension, and thus an infinite number of measures of outlyingness for each
observation is obtained. Thus, the goal is to identify the supremum over all possible
projection directions $\mathbf a \in \mathfrak{R}^p$ with $\|\mathbf{a}\|=1$
of the measure of outlyingness, 
\begin{equation}\label{eq:defOutl}
\mbox{out}(\mathbf{x}_i,\mathbf{X})=\sup_{\mathbf{a}}
\frac{|\mathbf{x}_i^T\mathbf{a}-m(\mathbf{Xa})|}
{s(\mathbf{Xa})}
\end{equation}
for observation $\mathbf{x}_i$ ($i=1,\ldots ,n$) of the data set $\mathbf{X}$. 
Here, $m$ and $s$ and robust univariate location and scatter estimators, respectively,
e.g. the median and the MAD. Using an appropriate weight function $w$, each observations
receives a weight $w_i=w(\mbox{out}(\mathbf{x}_i,\mathbf{X}))$, depending on its
outlyingness. The location estimator is then defined as
$$
\mathbf{t}=\frac{1}{\sum_{i=1}^n w_i}\sum_{i=1}^n w_i\mathbf{x}_i
$$
and the covariance estimator as
$$
\mathbf{C}=\frac{1}{\sum_{i=1}^n w_i}\sum_{i=1}^n w_i(\mathbf{x}_i-\mathbf{t})
(\mathbf{x}_i-\mathbf{t})^T .
$$
If high breakdown point estimators are used for $m$ and $s$, and if an appropriate
weight function $w$ is chosen, the Stahel-Donoho estimator can achieve the maximum
breakdown point. 

The disadvantage of this estimator is its high computational cost. Although
approximate algorithms have been developed, it will be difficult to deal with 
high-dimensional data sets that are typical in chemometric applications.
On the other hand, unlike the previously discussed robust estimators for 
multivariate location and covariance, the Stahel-Donoho estimator can
handle data sets with more variables than observations, which makes
it attractive for chemometrics.

\subsubsection{Using spatial signs}\label{sec:spatialsign}

Also spatial sign covariances can handle data with more variables than observations.
They are very fast to compute, have a bounded influence function,
can deal with a moderate fraction of outliers in the data, are not relying
on the assumption of multivariate normality, but are not affine 
equivariant. 
The spatial sign $\mathbf{S}(\mathbf{x})$ of a multivariate observation 
$\mathbf{x}$ with respect to the centre $\mathbf{t}$ of the data is defined as 
$$
\mathbf{S}(\mathbf{x})=\frac{\mathbf{x}-\mathbf{t}}
{\|\mathbf{x}-\mathbf{t}\|}
$$ 
(see \cite{Visuri00}). This is a unit vector pointing in the direction 
$\mathbf{x}-\mathbf{t}$.
For the location estimator $\mathbf{t}$ one can take the spatial median
which is the solution of minimising 
$\sum_{i=1}^n \|\mathbf{x}_i-\mathbf{t}\|$. The spatial median has maximum
breakdown point and is orthogonal equivariant. Another choice for 
$\mathbf{t}$ could be the co-ordinate-wise median.

At the basis of spatial signs the so-called {\it spatial sign covariance
matrix} can be constructed:
\begin{itemize}
\item
compute the sample covariance matrix from the spatial signs
$\mathbf{S}(\mathbf{x}_1), \ldots ,\mathbf{S}(\mathbf{x}_n)$,
and find the corresponding eigenvectors $\mathbf{u}_j$, for
$j=1,\ldots ,p$, and arrange them as columns in
the matrix $\mathbf{U}$,
\item
project the observations on the $j$-th eigenvector (scores) and estimate robustly
the spread (eigenvalues) by using e.g. the MAD,
$$
\lambda_j=\mbox{MAD}(\mathbf{x}_1^T\mathbf{u}_j,\ldots ,
\mathbf{x}_n^T\mathbf{u}_j)^2
$$
for $j=1,\ldots ,p$. Arrange them in the diagonal of a squared matrix,
i.e. $\mathbf{\Lambda}=\mbox{diag}(\lambda_1,\ldots ,\lambda_p)$,
\item
The covariance matrix estimate is
$$
\mathbf{C}=\mathbf{U\Lambda U}^T .
$$
\end{itemize}

Note that transformation of all data by projecting onto a unit sphere, inherently changes the topology of the data. Therefore, it is easy to grasp that, while this transformation leads to a covariance matrix with a high asymptotic breakdown point, it can already be significantly biased at low levels of contamination. To counter this effect, recently the concept of the sign covariance matrix has been extended to a more general class of radial transformations. The main idea behind this is to have a generalized radial transformation that projects outliers onto the unit sphere, while the bulk of the data remains unchanged. Raymaekers and Rousseeuw\cite{Raymaekers18} have shown that that approach preserves the eigenstructure of the data, has bounded influence functions and has the same finite sample breakdown point as the spatial sign covariance matrix.

\subsection{Projection pursuit}

As we have reviewed in the previous sections the basic idea behind the construction of most robust multivariate estimators consists of attributing weights to each data point separately. Weights can be computed in many ways; they can be continuous (e.g. M estimators, S estimators) or binary (estimators based on trimming such as LTS regression). With the computation of weights, the classical estimator is always to some extent a part of the robust estimator process. For example, M regression estimators are computed from iteratively re-weighted data. Within each iterative re-weighting step the classical estimator is computed. For LTS regression a subset of $h$ cases is sought for. Once this subset is found, the classical estimator is computed on the reduced data set. 

An entirely different approach to robustifying multivariate methods consists of projection pursuit. Initially, projection pursuit was developed as a data reduction technique for high dimensional data \cite{FT}. However, in a short time span application of projection pursuit has spread throughout many areas of statistics, such that in 1985 a review article could already report projection pursuit based approaches to density estimation, regression, estimation of the covariance structure and principal component analysis \cite{Huber85}. Its relative popularity can be explained by the method's versatility: depending on which criterion one uses to evaluate the projections (the {\em projection index}), a projection pursuit algorithm can yield (approximate) estimates to very different approaches. In practice, if a single projection pursuit algorithm is established, it can be used almost directly to produce a manifold of estimators. 

The basic idea of projection pursuit consists of reducing a problem of an intrinsically multivariate nature to many univariate problems by dint of projection. If the data are $p$ variate then a projection pursuit algorithm encompasses the following steps: construct all possible $p$ vectors ({\em directions}) and evaluate for each of these the projection index. The direction which yields the optimal value for the projection index is the solution. For instance, principal components are components which capture a maximum of variance. So in theory they can be found by computing all possible $p$ vectors (denoted $\boma{a})$ and then computing the variance of $\mathbf{Xa}$. The vector $\mathbf{a} \in \mathfrak{R}^p$ yielding the maximal value for $\mathrm{var}(\mathbf{Xa})$ is the first principal component. By evaluating a robust measure of spread (e.g. a scale M estimator, see Section \ref{sec:smest}), a robust method for principal component analysis is readily obtained. 

Of course, in practice only a finite number of directions can be constructed. Hence, the projection pursuit approach to all multivariate estimation procedures always yields approximate solutions. The quality of the approximation depends on the number of directions evaluated. The obtained solution evidently also depends on the choice of the directions which are scanned. Several algorithms have been proposed in literature. One can choose to construct directions randomly. However, in doing so one disregards the data structure due to which it is hard to ascertain that based on a small set, the obtained solution will be a good approximation to the true solution. A second approach consists of taking the $n$ directions contained in the data set as directions to evaluate and if necessary to augment these directions by random linear combinations of the original directions. This algorithm has been adopted successfully for principal component analysis \cite{CR} as well as for continuum regression \cite{SFCV}. A third approach is a so-called {\em grid algorithm}. The grid algorithm restricts the search for the optimum to a plane. Its consists of the following steps: 
\begin{enumerate}
\item Compute for each variable $\mathbf{x}_i$ the projection index based on the $n$ data points. This yields $p$ values for the projection index. 
\item Sort the variables in descending order according to the value they yield for the projection index. 
\item Find the optimum direction in the plane spanned by the first two sorted variables. This yields the first approximation to the optimal direction: $\mathbf{a}^{(1)}=(\gamma_1^{(1)}\ \gamma_2^{(1)} \ \mathbf{0}_{p-2}^T)$.
\item For $j=3:p$, find the optimal directions in the plane spanned by the vectors $\mathbf{X}\mathbf{a}^{(j-1)}$ and the $j$th sorted variable $\mathbf{x}_{(j)}$. The next solution is then given by: 
\begin{equation} \mathbf{a}^{(j)}= \begin{pmatrix} \prod_{k=1}^j\gamma_1^{(k)} &\gamma_2^{(1)}\prod_{k=2}^j\gamma_1^{(k)}&\gamma_2^{(2)}\prod_{k=3}^j\gamma_1^{(k)}&\cdots  &\gamma_2^{(j)},&\mathbf{0}_{p-j}^T\end{pmatrix}.\end{equation}
\item When all variables have passed the previous phase, restart evaluating all entries $a_i$ of $\mathbf{a}$ by searching the optimal direction in the plane spanned by the vectors $\mathbf{X}\mathbf{a}^{(j-1)}$ and $\mathbf{x}_j$. Stop the iteration if $\parallel\mathbf{a}^{(q)}-\mathbf{a}^{(q-1)}\parallel$ is smaller than a certain tolerance limit (e.g. $10^{-5}$).        
\end{enumerate}
To find the optimal directions in the plane itself (required above in steps 3 through 6), the following algorithm is used: 
\begin{enumerate}
\item Consider a limited set of linear combinations of both variables $\gamma_1\mathbf{x}_1+\gamma_2\mathbf{x}_2$ and evaluate the projection index for each of these directions. The directions are chosen on a unit sphere (the side constraint $\gamma_1^2+\gamma_2^2=1$ should be satisfied) at regular intervals. For instance, if ten initial guesses are chosen these directions are at 0, 18, 36, ..., 162 degrees. 
\item Project the data onto the initial optimum.
\item Scan the same number of directions in a narrower interval, e.g. for ten directions: -45, -35, ..., 45 degrees. The angle in which the grid search is effectuated is made narrower until convergence is reached.  
\end{enumerate}
The grid algorithm has shown to be successful for principal component analysis \cite{PeterGrid}, where it is more precise than the algorithm based on choosing data points and linear combinations. An implementation of it for continuum regression has also been reported \cite{FSCV}. 

Although much research has been carried out on the algorithmic aspect of projection pursuit, a draw-back of the method is still its computational cost. With any of the algorithms above, still a large number of directions need to be constructed for the projection pursuit approximation to be reliable. As computer power is continuously increasing, at the moment the computational cost for ``normal" data sets encountered in chemometrics (e.g. size $100\times2000$) is not excessively high to construct the estimator once. Howbeit, for many applications of projection pursuit, a single computation of the estimator is not sufficient. For instance, if the estimator needs to be cross-validated, in each cross validation loops the estimator needs to be evaluated. Depending on the application, including a projection pursuit based estimator into a cross-validation routine may still require long computation times.     

\section{Robust alternatives to principal component analysis}\label{sec:PCA}

Recall that principal component analysis (PCA) proceeds by finding directions
in space which maximise or minimise the dispersion, measured by the variance.
The classical approach is based on the covariance matrix. The first principal
direction is the unit vector $\mathbf{b}_{1}$ such that $\mathrm{var}\left(
\mathbf{Xb}_{1}\right)  =\max.$ The directions $\mathbf{b}_{j}$ for $j>1$ are
the unit vectors such that $\mathrm{var}\left(  \mathbf{Xb}_{j}\right)  =\max$
under the restriction that $\mathbf{b}_{j}^T\mathbf{b}_{k}=0$ for
$k<j.$ Call $\lambda_{1}\geq\lambda_{2}\geq...\geq\lambda_{p}$ the eigenvalues
of $\mathbf{\Sigma}$ in descending order, and $\mathbf{e}_{1},...,\mathbf{e}%
_{p}$ the respective eigenvectors. Then it is shown that $\mathbf{b}%
_{j}=\mathbf{e}_{j}$ for $j=1,..,p.$ Since $\lambda_{j}=\mathrm{var}\left(
\mathbf{Xb}_{j}\right)  ,$ the number $q$ of components is usually chosen so
that the \textquotedblleft proportion of unexplained
variance\textquotedblright%
\[
u_{q}=\frac{\sum_{j=q+1}^{p}\lambda_{j}}{\sum_{j=1}^{p}\lambda_{j}}%
\]
is sufficiently small (say 10\%).

PCA may be viewed geometrically as searching for a $q$-dimensional linear
manifold that \textquotedblleft best\textquotedblright\ approximates the data.
The point of the manifold closest to $\mathbf{x}_{i}$ is its \textquotedblleft%
$q$-dimensional reconstruction{}\textquotedblright%

\begin{equation}
\mathbf{\hat{x}}_{i}=\mathbf{BB}^T(\mathbf{x}_{i}-\mathbf{\bar{x}}%
)+\mathbf{\bar{x}}\label{reconstruc}%
\end{equation}
where $\mathbf{\bar{x}}$ is the data average and $\mathbf{B}$ is the
orthogonal $p\times q-$matrix with columns $\mathbf{b}_{1},...,\mathbf{b}_{q}.$ 

Outliers in the data may uncontrollably alter the directions $\mathbf{b}_{j}$
and/or the eigenvectors and hence the choice of $q.$ There are many proposals
to overcome this difficulty. The simplest is to replace the covariance matrix
with a robust dispersion matrix. The eigenvectors of this robust dispersion
matrix will result in robust principal components. 
Croux and Haesbroeck (2000) \cite{CrouxH00} derived influence functions
and asymptotic variances for the robust estimators of eigenvalues and
eigenvectors.

Another approach is to maximise a robust dispersion measure instead of the variance.
This idea was proposed by Li and Chen (1985) \cite{LiChen85}. 
Croux and Ruiz-Gazen (2005) \cite{CR2}
derived theoretical properties of the estimators for the eigenvectors, eigenvalues
and the associated dispersion matrix. They introduced an algorithm for computation
which was improved by Croux et al. (2007) \cite{PeterGrid}.

Rather than describing the many procedures proposed in the literature, we give
two very simple methods.

The first was proposed by Locantore et al. (1999) \cite{Locantore} and is called \emph{spherical
principal components} (SPC). Let $\mathbf{\hat{\boma{\mu}}}$ be a robust location
vector. Let
\[
\mathbf{y}_{i}=\frac{\mathbf{x}_{i}-\mathbf{\hat{\boma{\mu}}}}{\left\Vert
\mathbf{x}_{i}-\mathbf{\hat{\boma{\mu}}}\right\Vert },
\]
(see Section \ref{sec:spatialsign}).
That is, the $\mathbf{y}_{i}$s are the $\mathbf{x}_{i}$s shifted to the unit
spherical surface centred at $\mathbf{\hat{\boma{\mu}}.}$ Compute the cross-products
matrix of the $\mathbf{y}_{i}$s
\[
\mathbf{C=}\sum_{i=1}^{n}\mathbf{y}_{i}\mathbf{y}_{i}^T%
\]
and its eigenvectors $\mathbf{b}_{j}$ ($j=1,...,p$). Let $\hat{\sigma}\left(
.\right)  $ be a robust dispersion measure like the MAD, and define
$\lambda_{j}=\hat{\sigma}\left(  \mathbf{Xb}_{j}\right)  ^{2}.$ Sort the
$\lambda_{j}$s in descending order (and the respective $\mathbf{b}_{j}$s
accordingly). Then proceed as in the classical case. It is shown that in the
case of an elliptic distribution, the $\mathbf{b}_{j}$s estimate the
eigenvectors of the covariance matrix (but not necessarily in the correct order).

A simple way to obtain $\hat{\boma{\mu}}$ is the coordinatewise median:%
\begin{equation}
\hat{\boma{\mu}}=\left(  \mu_{1},..,\mu_{p}\right)^T%
\ \mathrm{with\ }\mu_{j}=\mathrm{med}_{i}\left(  x_{ij}\right)
.\label{defCoorMed}%
\end{equation}
A better one is the space median, which is
\begin{equation}
\hat{\boma{\mu}}=\arg\min_{\boma{\mu}}\sum_{i=1}^{n}\left\Vert
\mathbf{x}_{i}-\boma{\mu}\right\Vert ,\label{defSpaMed}%
\end{equation}
see also Section \ref{sec:spatialsign}.

It is easy to compute $\mathbf{\hat{\mu}}$ iteratively. Start from some
initial $\mathbf{\mu}_{0}$ (e.g., the coordinatewise median). At iteration
$k,$ let $w_{i}=1/\left\Vert \mathbf{x}_{i}-\boma{\mu}_{k}\right\Vert $
($i=1,..,n)$ and compute $\mathbf{\mu}_{k+1}$ as the mean of the
$\mathbf{x}_{i}$s with weights $w_{i}.$The procedure converges quickly.

A more efficient robust PCA estimator (Maronna, 2005)\cite{Maronna05} is as follows. Given
$q,$ let $\mathbf{B}_{0}$ be a $p\times q$-matrix of principal directions and
$\boma{\mu}_{0}$ a robust location vector. For instance $\mathbf{B}%
_{0}=[\mathbf{b}_{1},...,\mathbf{b}_{q}]$ where the $\mathbf{b}_{j}$s are the
first principal directions given by the SPC method, and $\boma{\mu}_{0}$ is
the vector $\hat{\boma{\mu}}$ in (\ref{defCoorMed}) or (\ref{defSpaMed}). At
iteration $k+1,$ compute the reconstructions $\mathbf{\hat{x}}_{i}%
=\mathbf{B}_{k}\mathbf{B}_{k}^T(\mathbf{x}_{i}-\boma{\mu}%
_{k})+\boma{\mu}_{k},$ and the distances $r_{i}=\left\Vert \mathbf{x}%
_{i}-\mathbf{\hat{x}}_{i}\right\Vert ^{2}.$ Let $\hat{\sigma}$ be an M-scale
(\ref{defMscale}) of $\left(  r_{1},...,r_{n}\right)  .$ Let $w_{i}=W_{\sigma
}\left(  r_{i}/\hat{\sigma}\right)  $ with $W_{\sigma}$ defined in
(\ref{defWsig}). Let $\boma{\mu}_{k+1}$ and $\mathbf{\Sigma}_{k+1}$ be the
mean and covariance matrix of the $\mathbf{x}_{i}$s with weights $w_{i},$ and
let $\mathbf{B}_{k+1}$ be the first $q$ eigenvectors of $\mathbf{\Sigma}%
_{k+1}.$ And so on. Simulations have shown that this procedure is robust and is more efficient than SPC
for normal data. Recent theoretical results back up the simulations by Maronna\cite{Maronna05}. It has been shown that this procedure is qualitatively robust in the sense that it has a bounded influence function in those subspaces where outliers are to be expected. Moreover, it is possible to tune the subspace selection parameter such that a high breakdown point and a high asymptotic relative efficiency can be reached simultaneously\cite{croux2017mpca}.    

\section{Robust alternatives to partial least squares}\label{sec:PLS}

\subsection{A brief introduction to PLS}\label{sec:intropls}

Partial least squares (PLS) is one of the most successful tools in chemometrics. Historically it started as a method to estimate structural relations between several blocks of variables \cite{Wold}, in which sense it is applied in the fields of marketing and econometrics \cite{Esposito}. However, when a structural PLS model is set up between two groups of variables, it can also be used for prediction and is thus a regression technique. In chemometrics, virtually all applications of PLS fall in the latter category. Hence, in the current section we will limit ourselves to partial least squares for models between two groups of variables, $\mathbf{x}$ and $\mathbf{y}$. It will be assumed that these variables are of dimensions $p$ and $q$, respectively and are assumed to be centred. The data are a set of $n$ samples measured at these variables. 

Data from chemometrics are often of a high dimensional nature. Usually $p$ is big and may exceed $n$ (this applies to most spectrophotometrical applications); moreover the $p$ variables in $\mathbf{x}$ may be multicollinear. For such data it is well known that the least squares estimator fails. A viable approach to overcome these problems is first to estimate a new set of uncorrelated latent variables $\mathbf{t}$ and $\mathbf{u}$ from the original variables, between which standard regression can be carried out. Depending on how the latent variables are defined, different regression techniques are obtained. 

In virtually all regression methods based on the estimation of latent variables, the latter are defined as linear combinations of the original variables: for all $i \in \{1,\min(n,p)\}$ it holds that $\mathbf{t}_i=\mathbf{X}\mathbf{v}_i$ and $\mathbf{u}_i=\mathbf{Y}\mathbf{w}_i$. In the special case of PLS, the latent variables are defined by the following criterion: 

\begin{subequations} \label{eq:two}
\begin{equation}\label{eq:twoa}
(\mathbf{v}_h,\mathbf{w}_h)=\argmax_{\mathbf{a}_h,\mathbf{b}_h} \left(\mathrm{cov}\left[\mathbf{X}\mathbf{a}_h,\mathbf{Yb}_h\right]\right)
\end{equation}
under the constraints that
\begin{equation}\label{eq:twob}
\parallel\mathbf{a}_h\parallel=1  \ \ \mbox{and} \ \ \parallel\mathbf{b}_h\parallel=1 \ \
\end{equation}
and 
\begin{equation}\label{eq:twoc}
\mathbf{a}_h^T\mathbf{X}^T\mathbf{X}\mathbf{a}_i=0 \ \ \mbox{for } 1 \leq h < i.
\end{equation}
\end{subequations}     

It is clear from the criterion that PLS can be seen as being a compromise between principal component analysis and regression. The criterion can be solved by dint of the Lagrange multiplier method, leading (among other results) to the conclusion that the X weighting vectors $\mathbf{v}_i$ are successive eigenvectors of the matrix $\mathbf{X}^T\mathbf{y}\mathbf{y}^T\mathbf{X}$. For the other entities engaged in the criterion (e.g. $\mathbf{w}$), analogous eigenvector relations can be determined.  

Let $\mathbf{U}$ and $\mathbf{T}$ denote the matrices which collect the above vectors $\mathbf{u}_i$ and
$\mathbf{t}_i$ in their columns.
PLS regression resides in the idea of carrying out a regression of $\mathbf{U}$ on $\mathbf{T}$ (which does not lead to problems since the number of latent variables $k$ is always $k<n$ and they are by definition uncorrelated). Although the regression is set up between the latent variables, it is possible to re-write the formul\ae\ in such a way that a direct relation between $\mathbf{X}$ and $\mathbf{Y}$ is obtained: 
\begin{equation}\label{eq:regpls}
\hat{\mathbf{Y}}=\mathbf{X}\mathbf{V}\left(\mathbf{V}^T\mathbf{X}^T\mathbf{XV}\right)^{-1}\mathbf{V}^T\mathbf{X}^T\mathbf{Y}.% 
\end{equation}
Here, $\mathbf{V}$ is the matrix with the weighting vectors $\mathbf{v}_i$ in its columns.
In order to do prediction, this equation is very practicable. From it, it can be seen as well that it is possible to define a matrix of PLS regression coefficients: 
\begin{equation}\label{eq:bpls}
\hat{\mathbf{B}}=\mathbf{V}\left(\mathbf{V}^T\mathbf{X}^T\mathbf{XV}\right)^{-1}\mathbf{V}^T\mathbf{X}^T\mathbf{Y}. 
\end{equation}

PLS uses the classical covariance in its definition (see criterion \eqref{eq:twoa}). The classical covariance is a nonrobust estimator; as all PLS estimates derive from this classical covariance it can be expected that the whole PLS procedure is nonrobust. Indeed, if one takes into consideration the PLS influence function (first derived by Serneels et al. \cite{SCV}), it can be seen that the PLS influence functions are unbounded and thus that PLS is nonrobust. 

Prior to proceeding to robust PLS, we note that PLS is a specific estimator fitting into a more general framework called {\em continuum regression}. The continuum embraces the whole range of regression methods between ordinary least squares (OLS) and principal component regression (PCR), PLS being half-way. However, as OLS, PCR and PLS are the only three methods from the continuum regression framework for which the maximization can be solved analytically, PLS comes in many cases out as the best compromise between modelling predictor variance (PCR), modelling relation to the predictand (OLS) and computational simplicity.  

\subsection{Robustifying PLS}

Several approaches are possible to obtain a robust alternative to PLS. By analogy to PCA, it is possible to do projection pursuit or to use a robust estimator for covariance. By analogy to regression, it is also possible to make partial versions (i.e. latent variables based versions) of robust regression estimators such as least absolute deviation (LAD) or robust M regression. In what follows, the key principles of the existing robust PLS methods, will be outlined. 

\subsubsection{Projection pursuit}

The power of projection pursuit (PP) is that it reduces an essentially multivariate problem to many ones of a bivariate nature: it suffices to replace the classical covariance in criterion \eqref{eq:twoa} by a robust estimator for covariance. By such an order of proceeding one obtains robust estimates for the weighting vectors. The scores, however, again contain the outliers (due to multiplication with $\mathbf{X}$). Thus, in order to obtain robust regression coefficients, a robust regression estimator needs to be used to perform the regression between the latent variables. 

A robust PLS regression estimator has hitherto only been published for univariate PLS regression (i.e. for the case where $q=1$), as a part of the robust continuum regression (RCR) framework \cite{SFCV}. Robust PLS is obtained there by setting the continuum parameter $\delta$ to 0.5. The projection index proposed by the authors is the trimmed covariance. In the final step, a robust M regression is performed to obtain the regression coefficients. 

Robust PLS as a part of the RCR framework can deal with high dimensional data and is robust both with respect to vertical outliers and leverage points. Its theoretical robustness properties have hitherto not been investigated. By analogy to projection pursuit PCA \cite{CR2} one can expect that the estimated weighting vectors will inherit the robustness properties of the projection index used. In this case this implies that the breakdown point can be expected to equal the percentage of trimming used. One can also expect the influence function to be bounded but nonsmooth, as is typical for estimators based on trimming. For the regression coefficients, however, this can be expected not to carry through as they also depend on the final robust M regression step. Further theoretical developments will shed more light on its properties. 

A final disadvantage is that RCR may be computationally slow, if it is needed to insert it in a cross-validation procedure. In that case it may be advisable to use a low number of PP directions during cross validation, and to compute a more precise estimate for calibration. Both RCR and the cross-validation procedure are publicly available as a part of the TOMCAT toolbox \cite{TOMCAT}.

\subsubsection{Robust covariance estimation}
It can be shown that all PLS estimators derive from two basic population entities: the shape of $\mathbf{X}$ and the covariance between $\mathbf{X}$ and $\mathbf{y}$. In fact, if one considers the augmented data $\mathbf{Z}=(\mathbf{X}, \mathbf{y})$, then these properties are summarised in the shape of $\mathbf{Z}$. It is well known that the covariance matrix of $\mathbf{Z}$ takes on the partitioned form
\begin{equation}\label{eq:covpart}
\boldsymbol{\Sigma}_{\mathbf{Z}}=\begin{pmatrix} \boldsymbol{\Sigma}_{\mathbf{X}}%
& \boldsymbol{\Sigma_{\mathbf{Xy}}}\\ \boldsymbol{\Sigma_{\mathbf{Xy}}}^T%
& \sigma_{\mathbf{y}}^2 \end{pmatrix}.   
\end{equation}
Hence, as all PLS estimators derive from $\boldsymbol{\Sigma}_{\mathbf{X}}$ and $\boldsymbol{\Sigma_{\mathbf{Xy}}}$, it suffices to use robust covariance estimates for these (or, more practicably, for $\boldsymbol{\Sigma}_{\mathbf{Z}}$) in order to obtain a robust PLS procedure. 

The approach of plugging in a robust covariance estimator has been explored. Both the affine equivariant robust and semi nonparametric covariance matrix estimators have been examined in the context of PLS.
  
\paragraph{Robust PLS based on the Stahel-Donoho estimator}

Gil and Romera \cite{GR} propose to plug in the Stahel-Donoho or Minimum Volume Ellipsoid estimators for covariance (see Section \ref{sec:loccov}), but they prefer the Stahel-Donoho estimator based on previous results (theoretical and simulation) by Maronna and Yohai \cite{MaYo}.

The robustness properties of the whole PLS procedure are not known. However, it has been shown that just like for PLS itself, all PLS influence functions derive from the influence functions of both covariance estimates involved. For the robust procedure this is also true such that the influence function will be behaving similar to the influence function of the robust covariance estimator plugged in. In this case one may expect the influence function to be analogous to the influence function of the Stahel-Donoho estimator. Statistical efficiency and breakdown have not yet been investigated for this method. 

A major drawback to the method is that it is only fit for data for which $n>p$, hence precluding almost any application to spectrophotometry. A full algorithm is not publicly available. 

\paragraph{Robust PLS based on the sign covariance matrix}

Another approach is to use a sign covariance matrix \cite{Visuri00}, see Section \ref{sec:spatialsign}. The sign covariance matrix is a semi nonparametric covariance matrix estimate which is an attempt to generalise the bivariate correlation estimators such as the Spearman and Kendall correlation to a multivariate estimator. The sign covariance matrix is based on the concept of the spatial sign, and is the simplest of six proposals of sign covariance matrices made by Visuri et al. \cite{Visuri00}. Therein, eventually a covariance matrix based on the Oja median is preferred over the spatial sign, because the former is affine equivariant. Howbeit, as the PLS method is not affine equivariant altogether, this property is not a prerequisite for the construction of a robust PLS method. Hence, it is possible to use the spatial sign covariance matrix instead, without loss of good properties. In fact, using the spatial sign covariance matrix has two advantages over the method based on Oja medians: 
\begin{enumerate}
\item the spatial sign covariance matrix has a very simple mathematical definition, which implies that the mathematical treatment, but above all, the computational algorithm, becomes very simple; 
\item in contrast to the method based on Oja medians, the spatial sign covariance matrix has a bounded influence function, such that the resulting PLS procedure will be robust. 
\end{enumerate}

Recollect that the spatial sign covariance matrix consists of a transformation of the data to their spatial signs, followed by a computation of the (classical) covariance matrix. As all PLS estimators derive from the covariance estimator, the same carries through to the whole PLS procedure, i.e. a robust PLS based on the spatial sign covariance matrix is equivalent to the following steps: 
\begin{enumerate}
\item Transform each observation in the data to its spatial sign, i.e. replace each row $\underline{x}_i$ from $\boldsymbol{X}$ by%
\begin{equation}\label{eq:three}
\sgn(\underline{x}_i)=\begin{cases}\underline{x}_i/\parallel\underline{x}_i\parallel & \text{if $\underline{x}_i\neq\underline{0}$},\\
\underline{0} & \text{if $\underline{x}_i=\underline{0}$},\end{cases}
\end{equation} 
where underlined characters denote row vectors;

\item carry out PLS on the transformed data. 
\end{enumerate}

Thanks to this property robust PLS based on a spatial sign covariance matrix becomes extremely efficient in the computational sense: in terms of computational efficiency, it is the fastest existing robust alternative to PLS. Because the method consists of a transformation of the data prior to normal PLS, the projection to the spatial sign can be seen as a form of data preprocessing. Hence, the method also carries the name of {\em spatial sign preprocessing} (\cite{SDV}).

The influence function of the spatial sign transformation has been determined \cite{SDV}. In fact, it is analogous to the influence function of PLS\cite{SCV}: a sequential set of influence functions each of which derive from the influence function of the spatial sign covariance matrix. The influence functions are bounded and smooth. 

The breakdown and efficiency properties have not been theoretically investigated, but simulation results are available. These indicate that, while the spatial sign transformation imparts a high empirical breakdown point to the resulting PLS procedure, its empirical bias curve shows significantly more bias at low levels of contamination compared to PRM or RSIMPLS. Given these results, when computations are tractable, one might prefer a method whose robustness properties can be tuned. Spatial sign pre-processing as originally published\cite{SDV} merely consists of transforming all data onto the unit sphere, such that its properties cannot be tuned. However, as highlighted at the end of Section \ref{sec:spatialsign}, recently more general radial transformations have been investigated that do depend on tuning parameters\cite{Raymaekers18}. Incorporating these generalized spatial sign transformations into the spatial sign preprocessing framework, may remedy the above drawbacks and can be an interesting direction for further research on this topic.         

\subsubsection{Robust PLS by robust PCA}

It has been heeded in Section \ref{sec:intropls} that the computation of the PLS estimators comes down to eigenvector and eigenvalue computations. Since principal component analysis corresponds to an eigenanalysis of the covariance matrix; PLS can thus be seen as PCA applied to covariance matrices of a special type such as $\mathbf{X}^T\mathbf{YY}^T\mathbf{X}$. 

A possible means to create a robust version of PLS resides thus in using a robust PCA method for these purposes. This approach has been followed by Hubert and  Vanden Branden (2003), who have proposed the RSIMPLS method \cite{Hubert}. A full mathematical treatment would be too exhaustive for this summary, but the method can vaguely be described as using the ROBPCA method \cite{Hubert2} to compute the PLS components and adapting the algorithm such that further steps are consistent with the SIMPLS algorithm \cite{deJ}. 

The method is entirely robust, both with respect to vertical outliers and leverage points. The algorithm is fairly fast in terms of computation such that high dimensional data are tractable with RSIMPLS. An implementation is publicly available as shareware as a part of the LIBRA toolbox (a MATLAB Library for Robust Analysis) \cite{LIBRA}. 

The robustness properties of the method are not fully known. The influence function of the ancillary ROBPCA method is known and is shown to be bounded but nonsmooth (reflecting the two estimation stages in the ROBPCA algorithm) \cite{Debruyne}. The influence function of a very closely related method for robust PLS has been established by the same authors \cite{Debruyne}, showing analogous behaviour. They do not describe the true influence functions of RSIMPLS for reasons of mathematical tractability but no surprises seem to be expected: one can assume the RSIMPLS influence function to be as well bounded but nonsmooth. The method has a tuneable parameter which presumably determines the breakdown point of the estimator. Simulations (see next section) corroborate these assumptions.  

\subsubsection{Robust PLS as a partial version of robust regression}

The name partial least squares suggests that PLS be a partial version of the least squares regression estimator. One way to interpret this statement is by seeing the word {\em partial} in the sense of being {\em restricted to the space spanned by the latent variables} $\mathbf{t}_i$. Indeed, PLS regression consists of an estimation stage of the latent variables followed by a regression between these latent variables (or of $\mathbf{y}$ on $\mathbf{T}$ if univariate). 

A straightforward approach to robustify PLS thus consists of constructing a partial version of a robust regression estimator. Two such estimators have been proposed: the partial least absolute deviation (PLAD) estimator \cite{Dodge} and the partial robust M (PRM) regression estimator \cite{SCFV}. The latter is more interesting since it leads to a computationally simple method. It is well known that computation of M estimators for regression can be completed with the use of an iteratively reweighted least squares algorithm \cite{DLR}. Hence, to compute the partial robust M regression estimator, it suffices to perform an iteratively re-weighted partial least squares algorithm, analogous to the earlier IRPLS method \cite{CA}, but by using robust starting values and by using weights which depend on distances both in the residual and score spaces \cite{SCFV}. This makes the method robust with respect to both vertical outliers and leverage points.  

The theoretical robustness properties of the PRM regression method are not yet known. The method has a tuneable parameter and in fact, could easily be generalised to different weighting schemes than the one proposed by the authors. Simulations on the original proposal indicate that the method combines a good efficiency to a high robustness. In terms of computation the method is fairly fast and can handle high dimensional data sets, but is still outperformed by robust PLS based on the sign covariance matrix (spatial sign preprocessed PLS) \cite{SDV}. 

A drawback to the method is that models are not nested, i.e. a three-component PRM model on a given dataset is not identical to the first three components of a 4 component model on the same data.  

Recently, an alternative weighting scheme has been proposed \cite{AA} based on a disparity metric \cite{Markatou}. The results therein by and large corroborate the robustness properties as described in \cite{SCFV} and \cite{SDV}.

Summarizing, it can be stated that partial robust regression estimators, typically calculated in iuterative reweighting schemes, yield a good tradeoff between robustness properties and statistical efficiency, while coming at a moderate computational cost. They have proven to perform well in applications, as shown in \cite{Liebmann}. 

\subsection{An application}

\subsubsection{Application of robust techniques}

In contrast to theoretical statistics, chemometrics is a field of research driven by applications. Robust statistics and thus robust chemometric methods can still be considered to form a niche inside the field of chemometrics, basically because practical situations in which application of robust techniques are appropriate, are much less common then applications where they are not needed. Hence, at present the commercial software which provides tools typically built for chemometric applications, does not include robust estimation methods. Nevertheless, though suitable applications of robust methods are more rare those of classical chemometric methods, there exist different experimental situations in which a gain can be made by applying robust tools instead of the classical ones. Situations in which robust methods are worthwhile can be of varying natures, but the most common occurrences fall into three categories: 
\begin{enumerate}
\item A group of outliers is present in the data. This likely, but not necessarily, causes the masking effect (see Section \ref{Sec:Mask}). Due to the masking effect, usage of classical chemometric tools for outlier detection will not detect any outliers. Ensuing application of classical calibration or classification methods will not be successful as these techniques will be influenced by the undetected group of outliers. In this respect the benefits of using a robust method are twofold: (i) The robust method can be used to efficiently detect the outliers prior to classical calibration or (ii) The robust method can be used to perform classification or calibration as such, since it is robust to the presence of outliers. Depending on the robustness properties of the robust estimator as well as on the time available for the analysis, one can opt for either of both strategies (the second option inevitably causes a wider uncertainty due to loss in efficiency). But we note that deletion of suspect points may also cause a loss in efficiency.
\item The process of data generation can be controlled well, but there is a wide source of natural variation in the data which cannot be excluded from the analysis. In such a case the distribution is likely to have wider tails than the normal and maybe to generate some outliers. Good examples for this type of data are data concerning e.g. biological organisms, where Nature's variation may be larger than in laboratory controlled experiments. If the data are more likely to follow e.g. a $t$ distribution instead of a normal distribution, application of robust estimators may already yield better results than applications of the classical estimator as the latter is only optimal at the normal model, but performance of the classical estimators tend to deteriorate fast as then distribution differs from the normal.    
\end{enumerate}

In what follows we will consider an example which falls into both categories: the benefit of robust calibration is shown to a data set containing a group of outliers for the calibration of one predictand, whereas it can be assumed to be close to normal for calibration to another predictand. 

\subsubsection{Prediction of the concentration of metal oxides in arch\ae ological glass vessels}\label{sec:appprm} 

The data set under consideration consists of 180 EPXMA spectra of arch\ae ological glass vessels. The experimental conditions under which the spectra were measured, are beyond the scope of the current example, but have been described in detail in \cite{Janssens}. It is noteworthy that the data contains samples which appertain to four different classes of arch\ae ological glass: sodic glass, potassic glass, calcic glass and potasso-calcic glass. The vast majority of samples belongs to the sodic group as only 10, 10 and 15 samples belong to the other three classes, respectively. Of the raw data set, the last 25 samples were identified to be outliers as they had been measured with a different detector efficiency. Hence, they are outlying in the space of the spectra, not the concentrations. In the statistical sense these spectra are thus leverage points. It is important to note that all outliers are samples of sodic glass. The detection of these outliers was done by a tedious procedure: each spectrum was evaluated separately. After removal of the outliers, classical partial least squares calibration was performed and was shown to yield good agreement between predicted and true concentrations for a test set \cite{Lemberge}. 

In the current section we will show how the use of robust statistics could have sped up this procedure, making the manual spectrum evaluation step superfluous, without causing a sizeable loss in prediction capacity. With these purposes the partial robust M regression estimator was applied, as it is both efficient and can cope with leverage outliers in the data. 

In order to perform calibration, the data were split up into two sets. A set for calibration was constructed as in Lemberge \emph{et al.} \cite{Lemberge}, where it had been decided that the training set should contain 20 sodic samples and 5 samples of each of the other types of glass. In addition to these 35 spectra, six spectra were added which belonged to the group of sodic samples measured at a different detector efficiency (bad leverage points). The remaining samples with correct detector efficiency were left for validation purposes. 

In the original analysis, univariate PLS calibration was performed for all of the main constituents of the glass. These are sodium oxide, silicium dioxide, potassium oxide, calcium oxide, manganese oxide and iron (III) oxide. Here we will limit the description by showing the results for the prediction of sodium oxide and iron (III) oxide. The reason for this is the following: the lighter the element\footnote{In EPXMA one observes the characteristic peaks element-wise; it is thus equivalent to write ``model for sodium" and ``model for sodium oxide".} for which to calibrate, the bigger the influence of the leverage points. Hence, showing the results for sodium and iron covers the trends which can be observed from this set of models, as sodium is the lightest and iron the heaviest element to be modelled. Why the model for sodium should be affected more by the outliers than the model for iron, can be explained by physics: a decrease in the detector efficiency function is caused by a contamination layer on the detector's surface. The number of X-ray photons which reach the detector, is inversely proportional to the thickness of the contamination layer. However, highly energetic photons will not be absorbed by the contamination layer. The characteristic energies for Na K$_\alpha$ and Fe K$_\alpha$ photons are 1.02 KeV and 6.4 KeV, respectively. Hence, one may expect the peaks corresponding to iron photons to be affected far less by the lower detector efficiency than the sodium peak.

We come to describing the results for this data set. A PLS and a PRM model were constructed for the calibration of the sodium and iron concentrations. In the former model eight latent variables were used, whereas in the latter seven were used, as in \cite{Lemberge}. The concentrations of Na$_2$O and Fe$_2$O$_3$ were estimated for the validation set. The root mean squared errors of prediction are given in Table \ref{tab:glass}.  We also state the respective RMSEP's obtained by Lemberge \emph{et al.} posterior to removal of the leverage points from all of the data from the training  set. The latter are reported in Table \ref{tab:glass} under the heading ``cleaned'' data, in contrast to the ``original'' data.

\begin{table}
\begin{center}
\begin{tabular}{|c|cc|cc|}
\hline
 & \multicolumn{2}{c}{Na$_2$O} \vline & \multicolumn{2}{c|}{Fe$_2$O$_3$} \\
\cline{2- 5}
 & Original & Cleaned & Original & Cleaned\\
\hline
PLS & 2.66  & 1.26 & 0.14 & 0.12\\
PRM & 1.50  & -- & 0.10 & -- \\
\hline
\end{tabular}
\end{center}
\caption{\label{tab:glass} Root mean squared errors of prediction for the EPXMA data set using the PLS and PRM-estimator, once using the original training sample and once using a clean version of the training sample, as in \cite{Lemberge}.}
\end{table}

It is observed (see Table \ref{tab:glass}) that indeed for sodium oxide, the classical PLS model is vastly affected by the leverage points, as the root mean squared error of prediction is almost double compared to the cleaned data set. Using a robust method (PRM), the effect of the leverage points can to a big extent be countered, although in comparison to the cleaned classical model there is still a non-negligible increase in RMSEP. 

For iron (III) oxide, the results also show the expected trends: the ``outliers" are not as outlying as in the model for sodium oxide (due to the higher energy of the iron K$_\alpha$ characteristic photons) and thus PLS performs only slightly inferior on the data set containing outliers than on the cleaned data set, whereas PRM surprisingly performs best of all.   

What we can conclude from these results is that at least a lot of time could have been gained by usage of a robust method. Depending on the requested accuracy, PRM could have been used directly for calibration or for detection of the outliers, which in both cases would have eliminated the tedious spectrum evaluation step in which the outliers were detected manually. 

\subsubsection{Summary}

In Table \ref{tab:sumpls} all discussed methods are listed, and for each method its most important pro and con are given. 

\begin{table}
\begin{center}
\caption{\label{tab:sumpls} Overview of the different robust PLS methods discussed in this section with their most important benefit and drawback}
\begin{tabular}{|c|ccc|}
\hline%
Method & Pro & Contra & Reference  \\%
\hline%
PP-PLS & Part RCR framework & High computational cost & \cite{SFCV}\\%
Gil-Romera & Uses Stahel-Donoho & Not for $p>n$ & \cite{GR}\\%
Spatial sign & Fastest method & Not tuneable & \cite{SDV}\\%
PRM & High efficiency & Not nested & \cite{SCFV}\\%
RSIMPLS & Tuneable robustness & Nonsmooth IF & \cite{Hubert2}\\% 
\hline
\end{tabular}
\end{center}
\end{table}

\section{Robust approaches to discriminant analysis}
\label{sec:da}

\subsection{Discriminant analysis}

In discriminant analysis we observe observations coming from several
groups or populations. The group membership is known for the
observed observations, they form the {\it training sample}.
At the basis of this training sample it is desired to construct
discriminant rules which allow to classify new observations
with unknown group membership to one of the populations.

Suppose that $p$ characteristics or variables have been measured,
and that $n$ observations are available from $g$ different populations
$\pi_1,\ldots ,\pi_g$. Since the group memberships are known for
the $n$ training data, we can split them into the $g$ groups,
resulting in subsamples of size $n_1,\ldots ,n_g$ with 
$\sum_{j=1}^g n_j=n$. The observations will thus be denoted by
$\mathbf{x}_{ij}$ with index $j=1,\ldots ,g$ representing the groups
and index $i=1,\ldots ,n_j$ numbering the samples within a group.

We will first consider the {\it Bayesian discriminant rule} to
classify new observations, and later the {\it Fisher discriminant rule}.
Let us assume that the observations from group $j=1,\ldots ,g$ have been sampled
from a population $\pi_j$ with an underlying density function $\mathbf{f}_j$,
and denote $p_j$ as the prior probability of this group with 
$\sum_{j=1}^g p_j =1$. If the subsample
size $n_j$ reflects the prior probability of a group,
$n_j/n$ can be used to estimate $p_j$. However, if the data have not been
sampled completely at random from the mixture, this estimate can be
quite unrealistic.
For the density $\mathbf{f}_j$ we usually assume a $p$-dimensional normal
density with mean $\boma{\mu}_j$ and covariance matrix $\mathbf{\Sigma}_j$.
The Bayesian discriminant rule assigns an observation $\mathbf{x}$ to that
population $\pi_k$ for which the expression $\mbox{ln}(p_j\mathbf{f}_j(\mathbf{x}))$
is maximal over all groups $j=1,\ldots ,g$. 
With the assumption of normal distribution this rule translates to
maximising the quadratic discriminant scores $d_j^Q(\mathbf{x})$, defined as
\begin{equation}
\label{QDAscores}
d_j^Q(\mathbf{x})=-\frac{1}{2}\mbox{det}(\mathbf{\Sigma}_j)-
\frac{1}{2}(\mathbf{x}-\boma{\mu}_j)^T\mathbf{\Sigma}_j^{-1}
(\mathbf{x}-\boma{\mu}_j) +\mbox{ln}(p_j),
\end{equation}
i.e. an observation $\mathbf{x}$ is assigned to that group $k$ for which
the quadratic discriminant score is maximal.

The method discussed so far is called {\it quadratic discriminant analysis} (QDA)
because the discriminant scores used for the assignment of an observation
are quadratic in $\mathbf{x}$.
Often, however, it can be assumed that the group covariances are equal,
i.e. $\mathbf{\Sigma}_1=\ldots =\mathbf{\Sigma}_g=\mathbf{\Sigma}$.
In this case the discriminant rule simplifies to using the linear
discriminant scores
\begin{equation}
\label{LDAscores}
d_j^L(\mathbf{x})=\boma{\mu}_j^T\mathbf{\Sigma}^{-1}\mathbf{x}-
\frac{1}{2}\boma{\mu}_j^T\mathbf{\Sigma}^{-1}
\boma{\mu}_j +\mbox{ln}(p_j)
\end{equation}
in an analogous way for assigning an observation $\mathbf{x}$
to a group. The scores are now linear in $\mathbf{x}$ and thus
the corresponding method is called {\it linear discriminant analysis} (LDA).

It is crucial how the parameters are estimated for the discriminant
scores (\ref{QDAscores}) or (\ref{LDAscores}). Classically, the population
means $\boma{\mu}_j$ are estimated by the arithmetic means
$\bar{\mathbf{x}}_j=\frac{1}{n_j}\sum_{i=1}^{n_j}\mathbf{x}_{ij}$, and
the population covariances $\mathbf{\Sigma}_j$ by the sample covariances
$\mathbf{S}_j=\frac{1}{n_j-1}\sum_{i=1}^{n_j}(\mathbf{x}_{ij}-\bar{\mathbf{x}}_j)
(\mathbf{x}_{ij}-\bar{\mathbf{x}}_j)^T$. As noted above, the prior probabilities
can be estimated by the relative frequencies of data points in a particular group.
If equal covariances can be assumed (LDA) it is necessary to estimate the
joint covariance matrix $\mathbf{\Sigma}$. This can be done by a pooled
estimate of the group covariance matrices,
$$
\mathbf{S}_{pooled}=\frac{1}{n_1+\ldots +n_g-g}\left( (n_1-1)\mathbf{S}_1 +
\ldots + (n_g-1)\mathbf{S}_g\right).
$$
Thus, the resulting LDA rule is to assign an observation $\mathbf{x}$
to population $\pi_k$ if the estimated LDA score
\begin{equation}
\label{LDAclass}
\hat{d}_k^L(\mathbf{x})=\bar{\mathbf{x}}_j^T\mathbf{S}_{pooled}^{-1}\mathbf{x}-
\frac{1}{2}\bar{\mathbf{x}}_j^T\mathbf{S}_{pooled}^{-1}
\bar{\mathbf{x}}_j +\mbox{ln}\left(\frac{n_j}{n}\right)
\end{equation}
is the largest of all LDA scores 
$\hat{d}_1^L(\mathbf{x}),\ldots ,\hat{d}_g^L(\mathbf{x})$.

\subsection{Robust LDA}

The LDA rule (\ref{LDAclass}) with the classical estimates of location and
covariance is vulnerable to outliers, and it can lead to a much higher
misclassification error in the sense that the spoiled LDA rule will
result in many wrongly classified observations. 

The discriminant rules (\ref{QDAscores}) and (\ref{LDAscores}) can easily
be robustified by plugging in robust estimates for the group means
and covariances. For obtaining a robust LDA rule the joint covariance matrix
has to be estimated. There are several proposals for this purpose:
one can derive a pooled estimate like in the classical case by averaging
the robust group covariances. Another possibility is based on centring the 
observations
of each group by their robust estimate of location, and then to derive a
robust estimate of covariance out of all centred observations \cite{HeF00}.
In an iterative procedure the group centres can again be updated by the
robust location estimate of all observations.
Finally, one could obtain a pooled covariance matrix by using weights
for the arithmetic group mean and a joint sample covariance matrix which
are derived from robust covariance estimation of all data points jointly
\cite{HawkinsM97}.

The remaining question is which robust location and covariance matrix should 
be used for plugging in into the discriminant rules. MCD estimators were
used by \cite{HubertV04}, and S estimators by \cite{HeF00}, the latter
showing slightly better performance for the misclassification probabilities.

To illustrate the effects of classical and robust parameter estimation
for LDA and QDA
we generate two groups of data with 100 observations each. The data in each
group are sampled from bivariate normal distributions with certain means
and covariances (unequal). However, only 90 data points of the first group
are following the group distribution, but the remaining 10 points are generated
with the parameters of the second group. The idea behind this data generation
is a scenario where part of the available training data have been assigned incorrectly:
The 10\% ``outliers'' in the first group belong to the second group, but
have been assigned by an incorrect decision to the first group.
Note that this wrong assignment of training data can easily occur in
practice, and that a proportion of 10\% outliers can usually be considered as
rather low.

\begin{figure}
\begin{tabular}{cc}
\resizebox{0.5\textwidth}{!}{\includegraphics{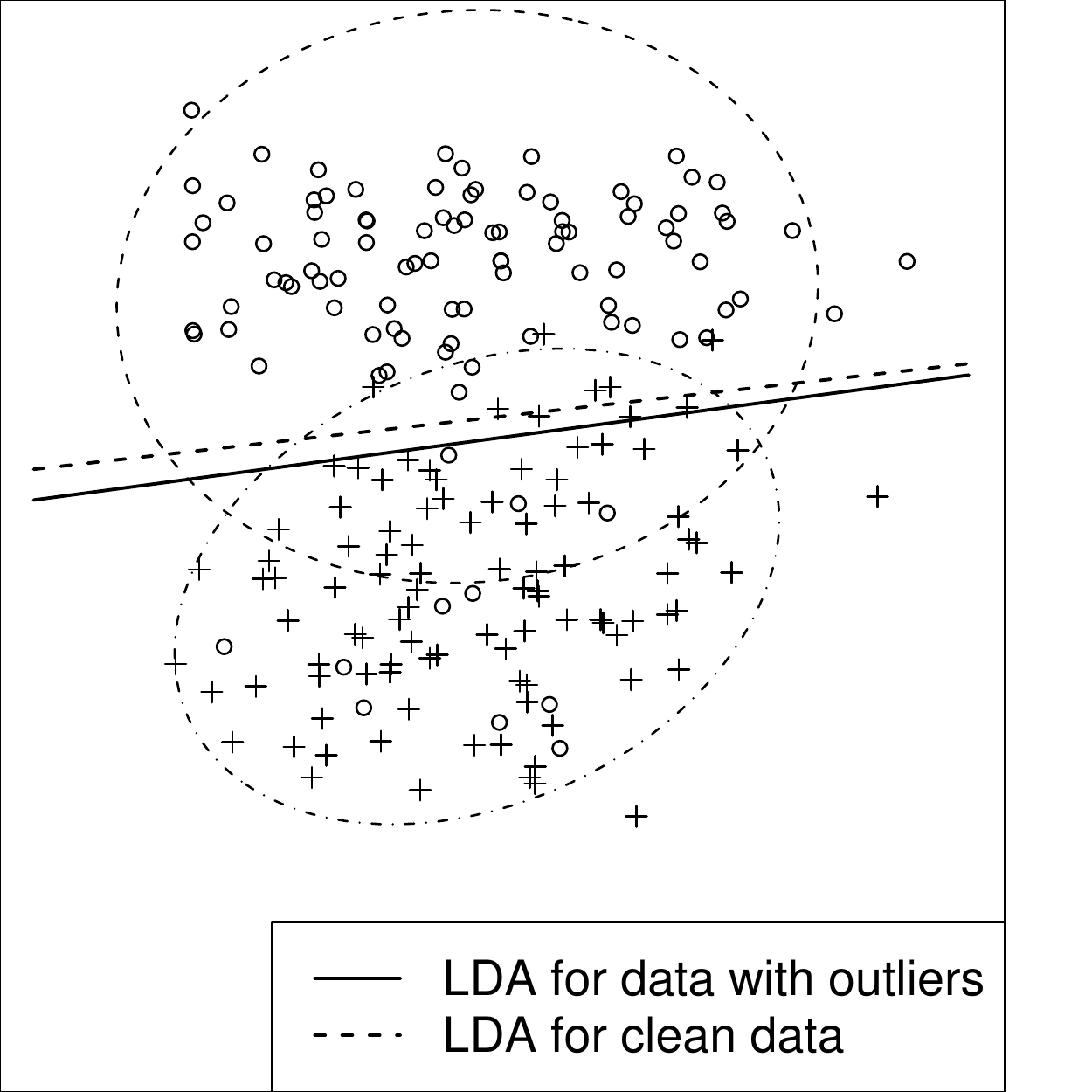}} &
\resizebox{0.5\textwidth}{!}{\includegraphics{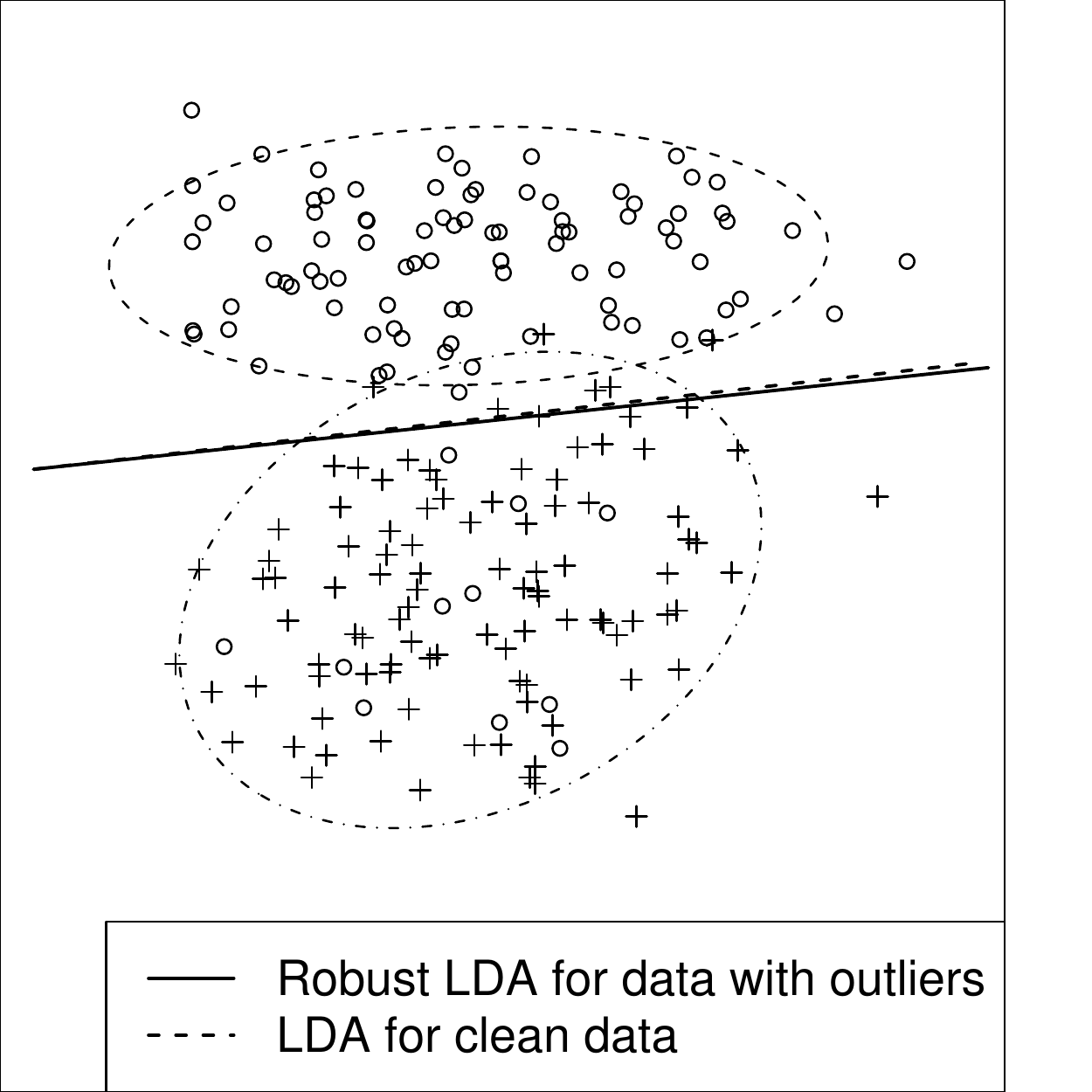}}\\
(a) & (b)
\end{tabular}
\caption{\label{figLDA} Two groups of data simulated from bivariate normal
distributions with different means and covariances. 10\% of the data points
from the first group ($\circ$) were sampled from the distribution of the
second group ($+$). LDA separation lines are shown when using classical
parameter estimates (a) and robust MCD estimates (b).}
\end{figure}

Figure \ref{figLDA} shows the two groups of data, the first group with $\circ$
and the second with $+$ as symbols. The ellipses shown in the figures are
90\% tolerance ellipses using the classical estimates (Figure \ref{figLDA}a)
and the MCD estimates (Figure \ref{figLDA}b). Clearly, the classical
tolerance ellipse for the first group is inflated by the outliers.
The two straight lines in each picture show the separation lines for the
data without outliers (dashed) and for the complete data set (solid).
The separation lines are obtained by computing the LDA scores (\ref{LDAscores})
for each point $\mathbf{x}$ in the two dimensional space, and by assigning
each point to the corresponding group. Points on the separation line would have 
equal group assignment. In Figure \ref{figLDA}a the parameters for
(\ref{LDAscores}) were estimated in the classical way by using 
formula (\ref{LDAclass}) for the estimated LDA scores. 
The difference in the resulting separation lines with and without outliers
is not big, it leads to 3 additional misclassifications of the second group.
The solid separation line
in Figure \ref{figLDA}b uses MCD estimates for the parameters while the
dashed line is obtained by classical estimates on the clean data. The latter line
almost coincides with the line obtained by robust estimation, and thus both
solutions result in the same number of misclassifications (which is unavoidable
because the groups are overlapping and since we applied LDA to data with different
covariance structures). 

\begin{figure}
\begin{tabular}{cc}
\resizebox{0.5\textwidth}{!}{\includegraphics{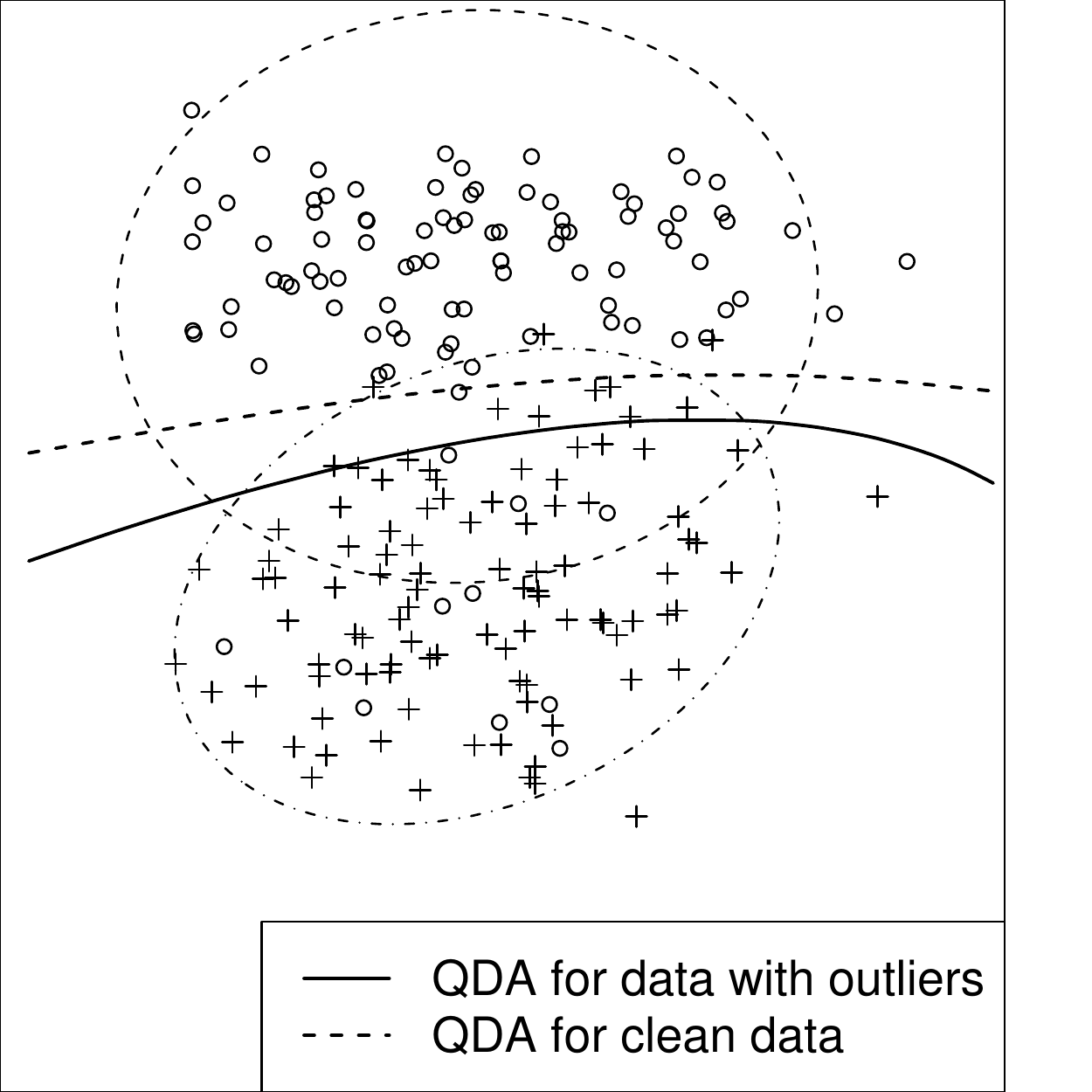}} &
\resizebox{0.5\textwidth}{!}{\includegraphics{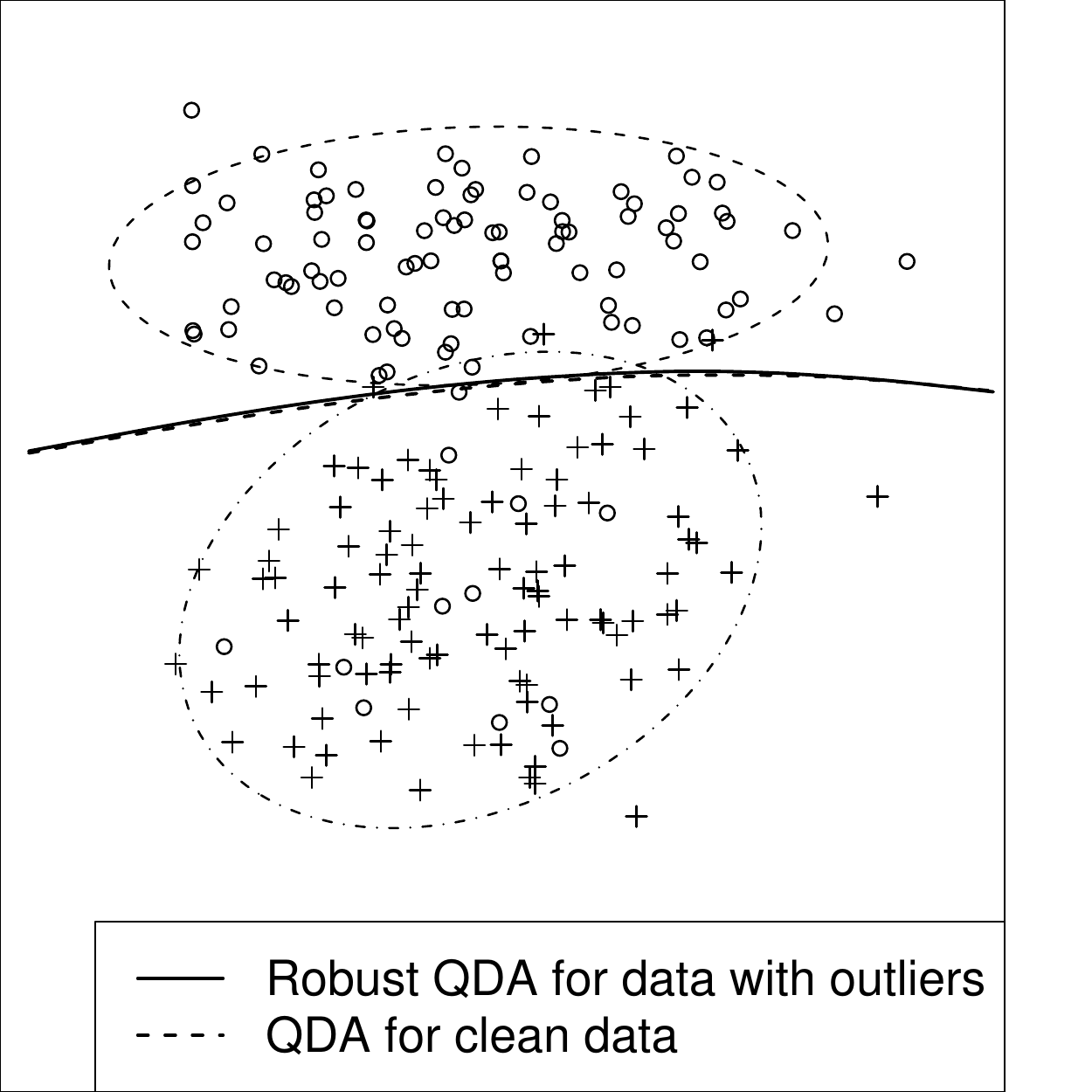}}\\
(a) & (b)
\end{tabular}
\caption{\label{figQDA} Two groups of data simulated from bivariate normal
distributions with different means and covariances. 10\% of the data points
from the first group ($\circ$) were sampled from the distribution of the
second group ($+$). QDA separation curves are shown when using classical
parameter estimates (a) and robust MCD estimates (b).}
\end{figure}

The resulting separation curves of classical and robust QDA estimation for
the same simulated data are shown in Figure \ref{figQDA}.
Similar to LDA, the QDA separation curves were obtained by estimating the
QDA scores (\ref{QDAscores}) with classical and robust MCD estimates, respectively,
for any data point in two dimensions, and assigning each data point to the
corresponding group. Points on the separation curve indicate equal group
membership. Like for LDA, the classical estimation of the QDA scores leads
to different solutions if all data were used or if only the non-outlying data
were analysed. Here, 8 additional observations were misclassified as a result
of using classical estimation for data with incorrect group assignment.
The robust estimation gives the same result as the classical estimation
on the clean data since the separation curves in Figure \ref{figQDA}b 
practically coincide.

Once more this example underlines the advantages of robust parameter estimation
over classical estimation. On purpose we did not construct data with extreme
outliers, but simply data with wrong group assignment. Thus, the difference
in the results is not dramatical, but still essential. Here, we only considered
misclassifications of the training data, but more important will be misclassifications
of future observations. Especially for observations which are close to the 
decision boundary the estimates to be used for the discriminant rule will
be important.

\subsection{Robust Fisher LDA}

Discriminant analysis was originally introduced by Fisher \cite{Fisher38}
for the separation of two groups. The extension of this method to several
groups was done by Rao \cite{Rao48}. The method leads to linear discriminant
functions, and it turns out that the problem can be formulated as an 
eigenvector/eigenvalue problem with a relatively simple solution.
Nevertheless, the method is frequently used in practical applications
and turns out to be powerful also in comparison to more advanced methods.

We use the same notation as in the previous sections. Additionally,
the overall weighted mean for all populations is denoted by
$\bar{\boma{\mu}}=\sum_{j=1}^g p_j\boma{\mu}_j$. 
Then the covariance matrix $\mathbf{B}$ describing the variation
{\it between the groups} is defined as
$$
\mathbf{B}=\sum_{j=1}^g p_j (\boma{\mu}_j-\bar{\boma{\mu}})
(\boma{\mu}_j-\bar{\boma{\mu}})^T ,
$$
and the {\it within groups covariance matrix} $\mathbf{W}$ is
defined as
$$
\mathbf{W}=\sum_{j=1}^g p_j \mathbf{\Sigma}_j .
$$
$\mathbf{W}$ can be seen as a pooled version of the group covariance
matrices.
Under the assumption that the group covariance matrices are all equal
it can be shown that the group centres can be best separated by
maximising the expression
\begin{equation}
\label{Fisherrule}
\frac{\mathbf{a}^T\mathbf{Ba}}{\mathbf{a}^T\mathbf{Wa}}
\quad \mbox{ for } \mathbf{a} \in \mathfrak{R}^p
\mbox{, } \mathbf{a\ne 0} .
\end{equation}
The solution of maximising (\ref{Fisherrule}) is given by the
eigenvectors $\mathbf{v}_1,\ldots ,\mathbf{v}_l$ of the matrix
$\mathbf{W}^{-1}\mathbf{B}$, scaled so that $\mathbf{v}_i^t
\mathbf{Wv}_i=1$ for $i=1,\ldots ,l$. The number $l$ of strictly positive
eigenvalues of the eigen-decomposition of $\mathbf{W}^{-1}\mathbf{B}$ can
be shown to be
$l\le \min(g-1,p)$.
By arranging the eigenvectors $\mathbf{v}_1,\ldots ,\mathbf{v}_l$ as columns
in the matrix $\mathbf{V}$ we can define the {\it Fisher discriminant scores}
for an observation $\mathbf{x}$ as
\begin{equation}
\label{Fisherscores}
d_j^F(\mathbf{x})=\Bigl((\mathbf{x}-\boma{\mu}_j)^T\mathbf{VV}^T
(\mathbf{x}-\boma{\mu}_j) -2 \mbox{ln} p_j \Bigr) ^{\frac{1}{2}}
\end{equation}
for $j=1, \ldots ,g$.
A new observation $\mathbf{x}$ to classify is assigned to population 
$\pi_k$ if its Fisher discriminant score $d_k^F(\mathbf{x})$
is the smallest among the scores for all groups
$d_1^F(\mathbf{x}),\ldots ,d_g^F(\mathbf{x})$.

Note that the Fisher discriminant scores (\ref{Fisherscores}) are
penalised by the term $-2\mbox{ln}p_j$ with the effect that an observation
is less likely to be assigned to groups with smaller prior probabilities.
Using this penalty term, the Fisher discriminant rule is optimal in the
sense of having a minimal total probability of misclassification for
populations which are normally distributed and have equal covariance 
matrices. This optimality is lost if fewer eigenvectors are taken for
computing the Fisher discriminant scores than the number of strictly
positive eigenvalues of $\mathbf{W}^{-1}\mathbf{B}$. On the other hand,
since $l\le \min(g-1,p)$, Fisher's method allows for a reduction of the
dimensionality which can be especially useful for graphical representations
of the observations in the new discriminant space. 
In fact, this is the main advantage of the Fisher method that it is able
to find a graphical representation of the data in a form that best captures
the differences among the groups.

For applying Fisher's rule to real data one has to estimate the group
means and covariances at the basis of the training data. Then the
eigenvectors of $\hat{\mathbf{W}}^{-1}\hat{\mathbf{B}}$ to the
strictly positive eigenvalues can be used to estimate the Fisher
discriminant scores $\hat{d}_j^F$ for each group $j$.
The classical way of estimating group means and covariances are sample
mean and sample covariance. In presence of deviations from the assumption
of multivariate normally distributed data of each group a robust approach
is recommended. As outlined in the previous section, several proposals
have been made in the literature for robustly estimating the parameters,
like MCD or S estimators.

% Kein Beispiel!

\subsection{Discriminant analysis if $p>n$}

Linear discriminant analysis is one of the most frequently used classification tools in multivariate statistics. Howbeit, direct application of a linear discriminant to sundry chemometric problems fails due to dimensionality prerequisites. There are essentially four ways in which one can proceed with a data set with dimensionality $p>n$ to which one wants to apply a linear discriminant classifier. These are subsequently highlighted in the ensuing paragraphs. 

\subsubsection{Singular value decomposition as a preprocessing step}

A simple way in which the dimensionality can be reduced without loss of information, is a singular value decomposition retaining all components. This comes down to decomposing the original data matrix $\boldsymbol{X}$ as
\begin{equation}\label{eq:SVDdatared}
\boma{X}^T=\boma{V} \boma{S} \boma{U}^T,
\end{equation}
where  $\boma{S}$ is a diagonal matrix whose diagonal elements are the $n$ singular values of $\boma{X}$, and $\boma{U}$  an $n \times n$ orthogonal matrix. Any further analysis can then be carried out on the reduced data matrix $\grave{\boma{X}}=\boma{U} \boma{S}$. Any estimated entity can be back-transformed to the space of the original variables. 

As this operation does not cause any loss of information from the original data, it can be seen as a preprocessing step which mainly serves to make the data set tractable and to speed up calculations. As such it is used as a standard preprocessing step in several robust multivariate estimation techniques such as RAPCA \cite{Sabine}, ROBPCA \cite{Hubert2}, Projection Pursuit PCA \cite{CR2} (for this method see also the discussion on SVD preprocessing in \cite{Ivana}), RSIMPLS \cite{Hubert} and PRM regression \cite{SCFV}. The publically available implementations of these methods as part of the LIBRA \cite{LIBRA} (includes RAPCA, ROBPCA and RSIMPLS) and TOMCAT toolboxes \cite{TOMCAT} (includes PP-PCA and PRM) do make usage of and SVD preprocessing step for the case $p>n$. 

This preprocessing step reduces the dimensionality of the data to size $\grave{\boldsymbol{X}} \in \mathfrak{R}^{n \times n}$. For several algorithms this dimensionality is sufficient and leads to an increase in computational efficiency (especially for algorithms based on projection pursuit or iterative reweighting, this is the case). However, as it merely comprises reduction of dimensions, robust estimation still has to follow. For many robust estimators (such as most robust estimators for the covariance matrix), more stringent requirements on the data dimensionality are necessary in order to have good robustness properties. Different strategies to cope with the high dimensionality for LDA based on such covariance matrices, have to be thought of. A simple strategy to follow is the approach of doing LDA in a PCA score space. 

\subsubsection{LDA in a PCA score space}

A second approach to perform LDA on undersampled data consists of the following two steps: 
\begin{enumerate}
\item do PCA on the original data; estimate the optimal number of PCs ($k$) and retain PCA scores and loadings of this dimensionality;
\item  perform LDA on the estimated scores.
\end{enumerate}

It should be noted that such an order of proceeding, as is the case for the material presented in the next two sections, is no longer equivariant to linear transformations.

This strategy is frequently applied on clean data sets; however, caution has to be taken when using this approach on data containing outliers. The reason is simple: PCA can be affected by the outliers such that a dimension reduction to a score space of $k$ components might lead to a loss of relevant information as the information retained can be to a far too great extent related to the outliers. 

This problem can readily be solved by using instead of classical PCA, one of the robust methods for PCA described in Section \ref{sec:PCA}. In this way one is sure that the relevant information concerning the non-contaminated data points, is contained in the robust score space. However, still classical LDA cannot be applied to this robust score space as the outliers will be projected into it further away from the centre than in a classical score space. This would cause the linear discriminant to be vastly affected by the outliers, and would probably lead to erroneous classification. The straightforward way to tackle this problem is by applying one of the robust methods for discriminant analysis described in the previous part of this Section in the robust score space. Summarising, a good order of proceeding consists of: 
\begin{enumerate}
\item do robust PCA on the original data; estimate the optimal number of PCs ($k$) and retain robust PCA scores and loadings of this dimensionality;
\item  perform robust LDA on the estimated scores.
\end{enumerate}

This order of proceeding is sufficient if only one data set is available. However, if one or more independent properties are measured on the same samples, such that also an $\boldsymbol{Y}$ data matrix exists, another option can consist of doing LDA in a PLS score space.

\subsubsection{LDA in a PLS score space}

When two sets of data matrices $\boldsymbol{X}$ and $\boldsymbol{Y}$ exist, $\boldsymbol{X}$ being undersampled and a linear discriminant classifier has to be applied, the dimension reduction can be more effective using a PLS score space. PLS components describe best this part of variation in the data which is relevant to the dependent variable $\boma{Y}$. It is thus a viable strategy to modify the approach described in the previous paragraphs to starting from a PLS score space. The same remarks hold, i.e. the PLS score space has to be estimated robustly by one of the methods described in Section \ref{sec:PLS}, whereafter a robust linear discriminant classifier has to be used. 

\subsubsection{D-PLS}

Finally, linear classification can also be adapted to the undersampled situation using a simpler strategy: doing regression for classification purposes by using a binary predictand. The binary predictand can be coded in different ways; no agreement exists on which is best. In practice mostly a [-1/1] or a [0/1] predictand is used, meaning that if a case belongs to the first class, the corresponding value in the predictand is set to 1, whereas in the other case, it is set to -1 or 0, respectively. In order to classify new samples, a rule of thumb has to be used. Several rules have appeared in the literature, but the most frequently applied seems to be choosing the arithmetic mean value between both as the decision boundary. In practice this implies that if a [-1/1] coding is chosen, future samples whose response is negative, are considered to belong to class 2, whereas samples whose predicted response is positive, are considered to class one. 

If least squares regression is used to do calibration and prediction, one obtains a method which is very closely related to linear discriminant analysis. If the data contain more variables than cases this is no longer possible, but is is of course straightforward to use one the regularised regression techniques which are appropriate for such a data structure, such as ridge regression, principal component regression or (probably the most popular one) partial least squares. If PLS is used for these purposes, it is commonly referred to as {\em discrimination partial least squares (D-PLS)} (see e.g. \cite{Naes}). D-PLS, and some generalisations, are frequently reported to perform well in applications to various types of data, for a few recent applications see e.g. \cite{SMVB,PGG}). The main topics of discussion concerning this method are how the decision boundary should be set, and whether univariate or multivariate PLS (PLS2) should be used. % Some recent publications, e.g. \cite{PGG}, favour the latter, although there is no universal agreement on the conclusions drawn there. 

If outliers are present in the data, the D-PLS method can be robustified by using a robust alternative to PLS regression, such as RSIMPLS or PRM with the same coded predictand. Note that if multivariate PLS is preferred, only RSIMPLS can be used as no multivariate equivalent of PRM exists. Although this approach to robust classification seems simple, up to our knowledge no papers have appeared describing its properties.

\section{Sparse robust methods}\label{sec:Sparse}
As data dimensions increase, it become increasingly challenging to analyze the data based on the entire set of variables considered. The latter can, in some instances, count in the hundreds of thousands, which makes it virtually impossible to be processed by a person. However, most frequently, when data dimensions are very high, a part of the data, sometimes even a large fraction, bears no information related to the entity of interest. In such cases, it is welcome to have estimators that intrinsically select that subset of variables that are relevant and this preferably in a model consistent way. The class of sparse estimation techniques does exactly that: yield estimates that are non-zero only for entries corresponding to relevant variables. The reader is referred to Filzmoser {\em et al.}\cite{FGT} for a good introduction to classical sparse chemometrics. This section will focus on how sparsity can be achieved in robust estimation techniques. 

\subsection{Sparse robust linear regression estimators}\label{sec:SpaRobReg}

Sparse estimates are generally obtained by adding a norm penalty into the maximization criterion they derive from. For example, consider the least squares regression vector. Recall \eqref{defLS} that the least squares regression vector is defined as the vector that minimizes squared residuals: 
\begin{equation}
\hat{\boma{\beta}}_{LS}=\arg\min_{\boma{\beta}}\sum_{i=1}^{n}r_{i}\left(
\boma{\beta}\right)  ^{2}.\nonumber%
\end{equation}
The definition of the vector of least squares regression coefficients can now be tweaked to yield sparse regression coefficients by adding an $L_1$ penalty to its norm: 
\begin{equation}
\hat{\boma{\beta}}_{LASSO}=\arg\min_{\boma{\beta}}\sum_{i=1}^{n}r_{i}\left(
\boma{\beta}\right)  ^{2} + \lambda \parallel \boma{\beta} \parallel_1.\label{defLASSO}%
\end{equation}
As $\lambda$ increases, the estimate of regression coefficients so obtained will be increasingly sparse. Sparsity can be introduced to other estimators similarly by $L_1$ penalization in the objective. The particular regression estimator defined in \eqref{defLASSO} is called the least absolute shrinkage and selection operator, or LASSO \cite{LASSO}. 

While the regression estimates obtained from \eqref{defLASSO} are generallly sparse, they suffer from the drawback that by definition, maximally $\min(n,p)$ coefficients can exactly equal zero. 
For very high dimensional flat data ($n << p$) this may be unsatisfactory. For instance, in genomics, data typically consist of far more genes than cases. As the number of genes analyzed often exceeds 100000, a sparse solution is required that facilitates interpretation by selecting a subset of supposedly impactful genes. Yet, given that the ratio $n/p$ can be small, analysts may want to investigate more than $n$ genes. A sparse estimator that has exact zero coefficients, but that can have more than $\min(n,p)$ non-zero coefficients, can be constructed by adding another shrinkage penalty into the objective: 
\begin{equation}
\hat{\boma{\beta}}_{ENet}=\arg\min_{\boma{\beta}}\sum_{i=1}^{n}r_{i}\left(
\boma{\beta}\right)  ^{2} + \lambda \parallel \boma{\beta} \parallel_1 + \mu \parallel \boma{\beta} \parallel_2.\label{defENet}%
\end{equation}
When the second shrinkage penalty is an $L_2$ penalty (also called {\em ridge} penalty in the context of regression) as in \eqref{defENet}, the resulting estimator is the Elastic Net\cite{ZouH}. 

Obtaining a sparse robust regression estimator can analogously be achieved by adding an $L_1$ norm penalty to the objective that defines the robust non-sparse regression estimator. In theory, it is possible to construct sparse versions of each of the approaches to 
robust regression described in Section \ref{sec:robreg}. However, as the field of sparse robust estimation techniques is still nascent, not all of these paths have been pursued. Alfons et al.\cite{Alfons} have provided a seminal publication in this space, constructing Sparse LTS, a sparse version of the least trimmed squares estimator, defined by adding an $L_1$ norm penalty term into \eqref{MinScalReg}. This idea has recently been generalized to also include an $L_2$ penalty term and as such, an LTS inspired version of the elastic net has been introduced\cite{Kurnaz18}.   

The optimal selection of the sparsity parameter $\lambda$ is typically unknown and can be determined through cross validation. The topic of model selection will be presented in more detail in Section \ref{sec:ModSel}. 

\subsection{Sparse robust PCA}
Sparse and robust alternatives to PCA have been proposed in the literature\cite{CFF,HRSV}. These methods achieve sparse dimensionality reduction and have shown to be practicable in explorative analysis for high dimensional data. A difficult aspect to tackle regarding these methods is how to optimize the sparsity parameter. Since no cost function can be defined based on a predictand, different choices of the sparsity parameter will lead to different sets of principal components to be retained. The literature cited here has proposed approaches to go about this issue in practical examples. 

\subsubsection{Example}

\begin{figure}[htbp]
\begin{center}
\includegraphics[width=.8\textwidth]{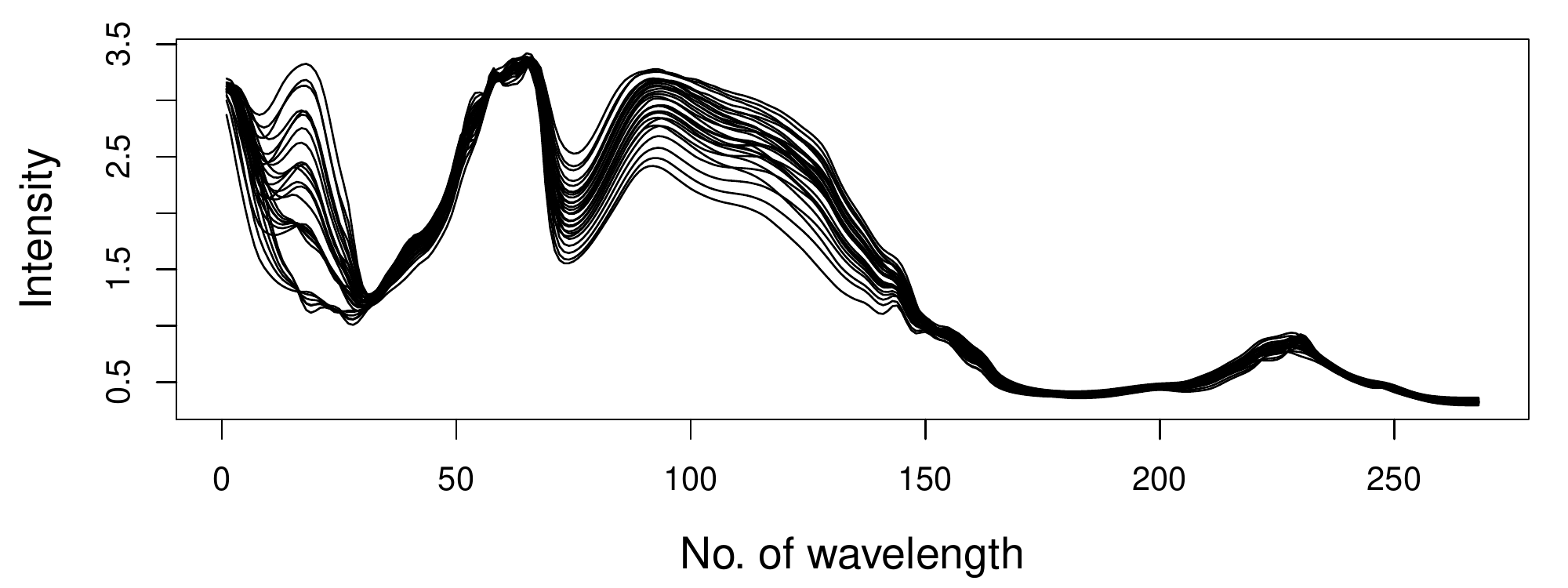}
\caption{\label{fig:spca1} NIR spectra of 28 pet yarns. Each line represents the spectral
information of one observation.}      
\end{center}
\end{figure}

A data set of 28 pet yarns, measured with near-infrared (NIR) spectroscopy\cite{SwiW99} 
is analyzed using PCA as well as sparse robust PCA\cite{CFF}.
There are 268 wavelengths available, and the original data are shown
in Figure~\ref{fig:spca1}. There is bigger variability in particular around wavelength
number 20, and again around wavelength number 110.

\begin{figure}[htbp]
\begin{center}
\includegraphics[width=.8\textwidth]{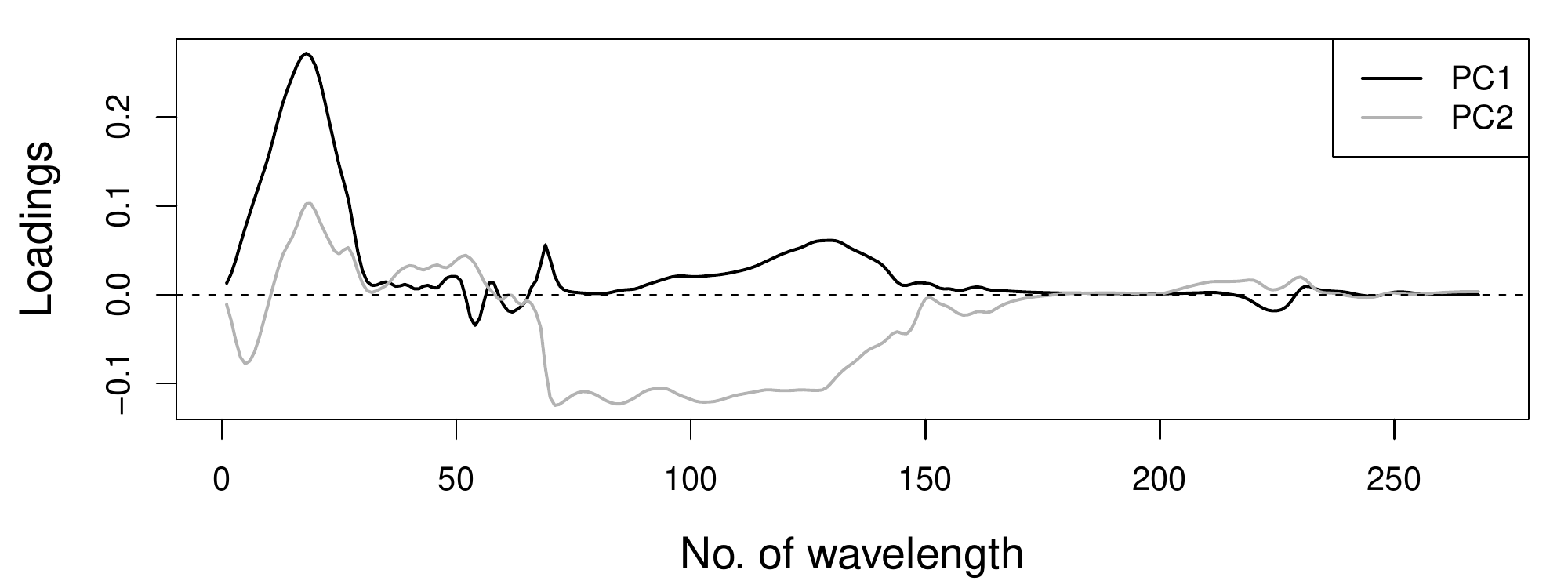}
\caption{\label{fig:spca2} Loadings of the first two classical PCs for the yarn data.} 
\end{center}
\end{figure}

The first two loading vectors from a classical PCA are shown in Figure~\ref{fig:spca2}.
The corresponding PCs account for almost 99\% of the total variability. The plot shows
that indeed both PC1 and PC2 get contributions from the aforementioned wavelength ranges.

\begin{figure}[htbp]
\begin{center}
\includegraphics[width=.8\textwidth]{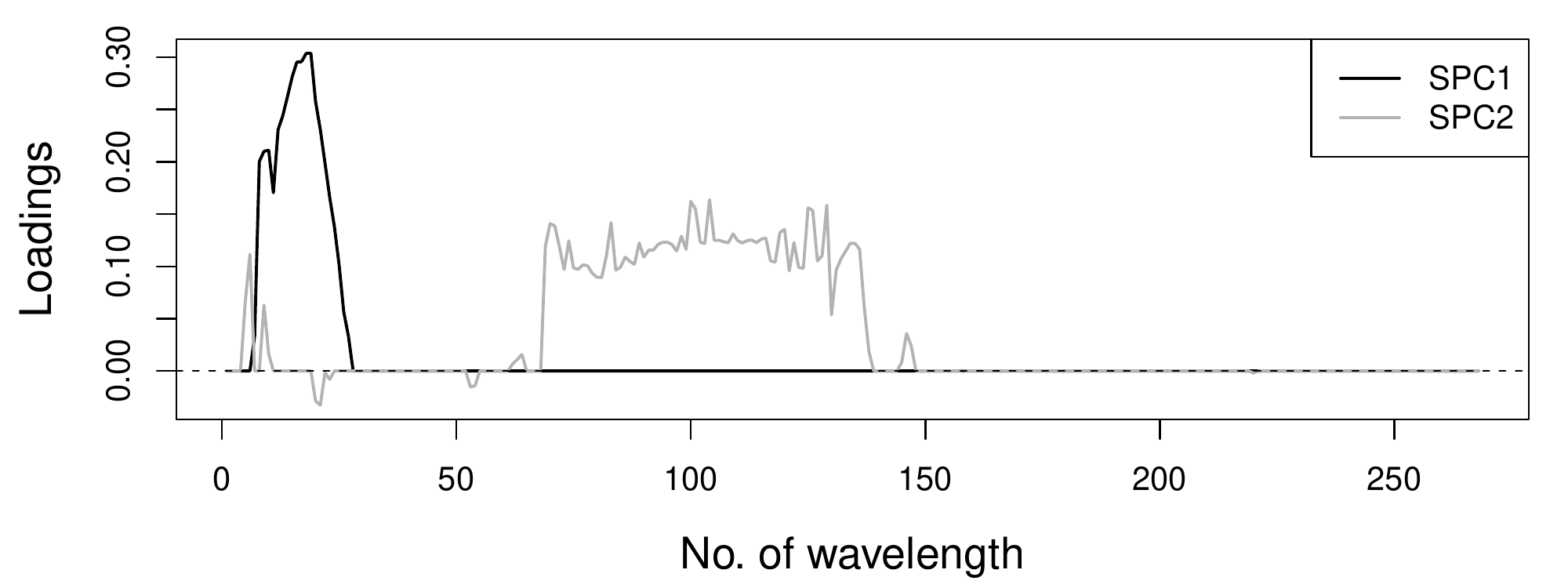}
\caption{\label{fig:spca3} Loadings of the first two sparse robust PCs for the yarn data.} 
\end{center}
\end{figure}

The picture looks different for sparse robust PCA, see Figure~\ref{fig:spca3}.
Again, the two mentioned wavelength ranges lead to an increase of the loadings,
but this time the first wavelength batch contributes almost only to the first component,
and the second batch exclusively to the second component. 
This is convenient in terms of an interpretation of the components, since they can
be uniquely assigned to specific wavelength ranges.
The (robust) variances of the 
corresponding scores are similar to those from the classical PCA, and thus also
the explained variance is similar. 

\subsection{Sparse robust PLS} 
As opposed to linear regression, that only yields estimates for regression coefficients and an intercept, partial least squares is a complex method that consists of a set of estimators: weighting vectors ($\boma{v}_h$), scores, loadings and regression coefficients. Sparse alternatives can be constructed for each of these separately, or for the entire set. 

The oldest proposal for a sparse partial least squares estimator\cite{Cao} consists of imposing a sparsity penalty on the vector of regression coefficients only. While that may be sufficient for regression tasks, one of the main reasons why PLS is often favored over other regularized regression methods in chemometric applications, is the much more sophisticated interpretability. For instance, analyzing loadings may yield insights as to how ranges in the spectra relate to entities of interest. It is obvious that such an interpretation can only be consistent with respect to dismissing uninformative variables, if loadings and regression coefficients are both sparse and that in a consistent way: the zero entries should correspond to the same subset of variables. 

Sparse PLS (SPLS) variants that yield such model consistent sparse estimates, have been introduced in the literature\cite{ChunKeles,Allen}. They derive from addition of an $L_1$ sparsity penalty to objective \eqref{eq:twoa}, albeit the problem is being reformulated slightly for mathematical elegance. A nice property of the sparse PLS as proposed by Chun and Kele\c{s}, is that for a univariate predictand, the objective can be solved analytically, which has led to the very fast sparse NIPALS (SNIPLS) algorithm\cite{HFSV}.  

An alternative to PLS that is both sparse and robust, can be designed by combining the ideas of robust PLS from Section \ref{sec:PLS}  with $L_1$ penalization. This path has been pursued regarding the combination of $L_1$ penalization of PLS as proposed by Chun and Kele\c{s}\cite{ChunKeles} with iterative reweighting. The resulting method has been coined sparse partial robust M regression\cite{HSFC} (SPRM). SPRM has later been extended into the space of classification as the SPRM-DA classifier\cite{HFSV}. 

As opposed to PCA, the sparsity parameter can readily be tuned in sparse PLS and sparse PRM. It involves a two-parameter cross-validation to a robustified prediction target criterion such as root mean-squared error of prediction (RMSEP). 
 
Because fewer variables contribute to the model, sparse estimators will yield more parsimonious models that come with a lower prediction variance. Another advantage sparse estimators offer, is enhanced interpretability. It is easier to analyze standardized regression coefficients from a sparse model, just because there are fewer individual coefficients to analyze. Note, however, that standardizing coefficients for a complex model like SPRM is non-trivial, and would require robustified resampling methods to estimate the variance of the regression coeffcients. Adequate robust resampling algorithms will be discussed in more detail in Section \ref{sec:robboot}. 

A significant advantage PLS based methods, such as SPRM, offer over regression estimators without dimension reduction step such as the LASSO, is that the latent components can be analyzed. Just like for PCA, SPRM scores and loadings can be visualized in a biplot. Again, biplots obtained from the sparse method (SPRM) will be a lot more straightforward to interpret than those obtained from the non-sparse equivalent (PRM). A concise example 
will illustrate this statement. 

\subsubsection{Example}

The target of the study that this example has been taken from, is to predict 
protein expression from gene expression for sixty human cancer cell lines. 
The data are provided online by the The National Cancer Institute 
(\url{http://discover.nci.nih.gov/cellminer/}). From the original data, 
the 40$^{\mathrm{th}}$ case was omitted due to missing values. Previous research \cite{lee2011sparse} had shown that it is viable to only retain 
the subset of variables from the gene expression data with the 
top 25\% highest variances, such that the modelling data ended up having 
dimensions $n=59$ and $p=5571$. The protein data consists of measurements of 162 expression levels. Since SPRM has only been designed for univariate response, the relationship for each protein expression was modeled separately and 162 models were obtained for each of the competitive methods.

\begin{figure}
  \centering
  \includegraphics[width=.45\linewidth]{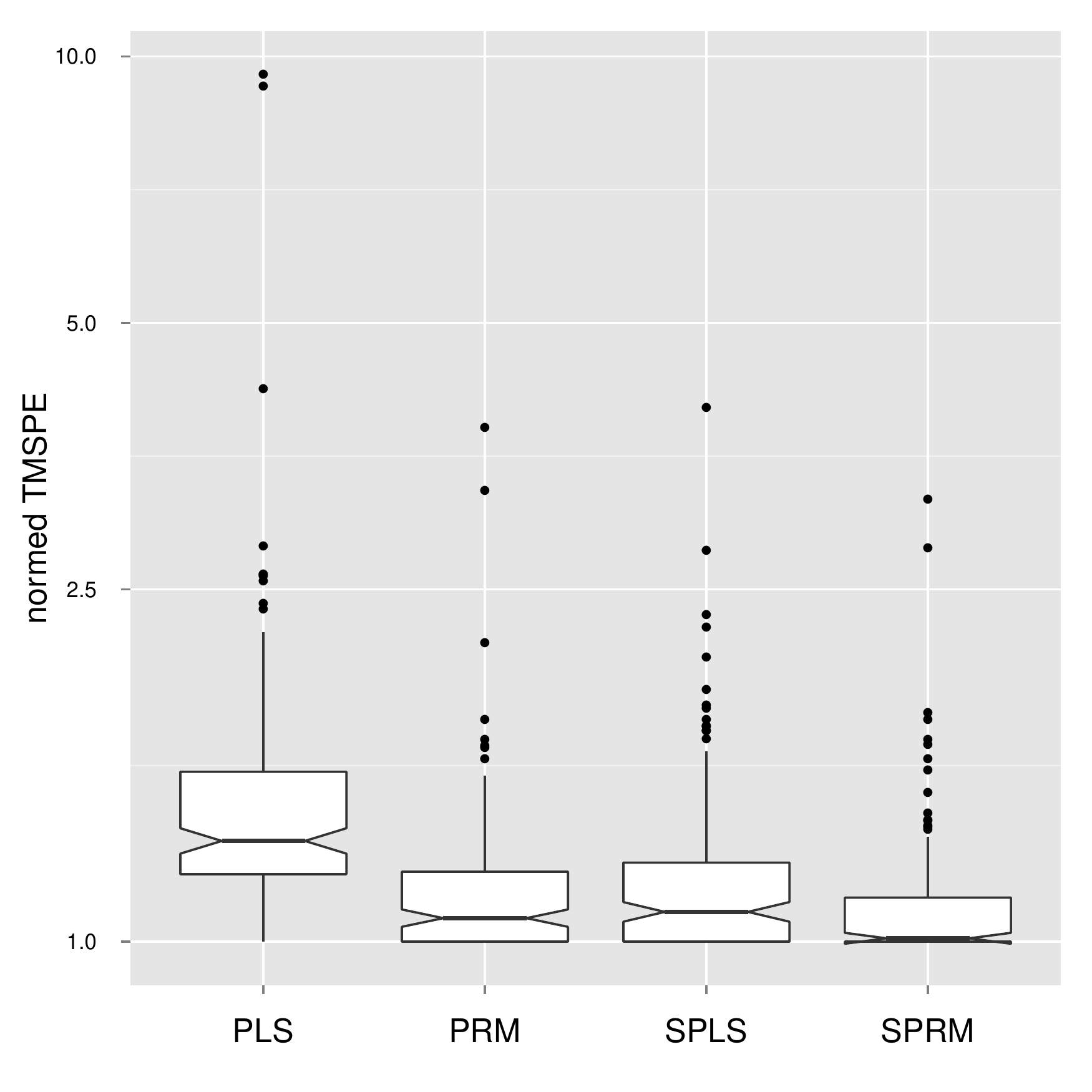}
\caption{Boxplots of normed TMSPE of 162 responses from the NCI data for PLS, PRM, SPLS and SPRM.\label{fig:normedTMSPE}}
\end{figure}

At first, four models will be compared for predictive performance across the 162 responses. These models are: PLS, PRM, Sparse PLS (SPLS) and SPRM.  For each of the methods, parameter selection was done using 10-fold cross validation (see Section \ref{sec:CV}). Predictive performance on non-outlying cases is quantified by the 15\% trimmed mean-squared prediction error (TMSPE). Then the TMSPEs were normed by dividing each model's TMSPE by the smallest of the four TMSPEs per response. This normed TMSPE is equal to 1 (for the best method) or larger comparable across the different responses (see Figure \ref{fig:normedTMSPE}). For a majority of the 162 models, the sparse and robust method (SPRM) outperforms the other techniques. The SPRM models have a median of the normed TMSPEs very close to 1 and therefore,  for 50\% of the models, SPRM is either the best or very close to the best model. Note that PLS does not perform well for these data.

\begin{figure}
\begin{subfigure}{.5\textwidth}
  \centering
  \includegraphics[width=.95\linewidth]{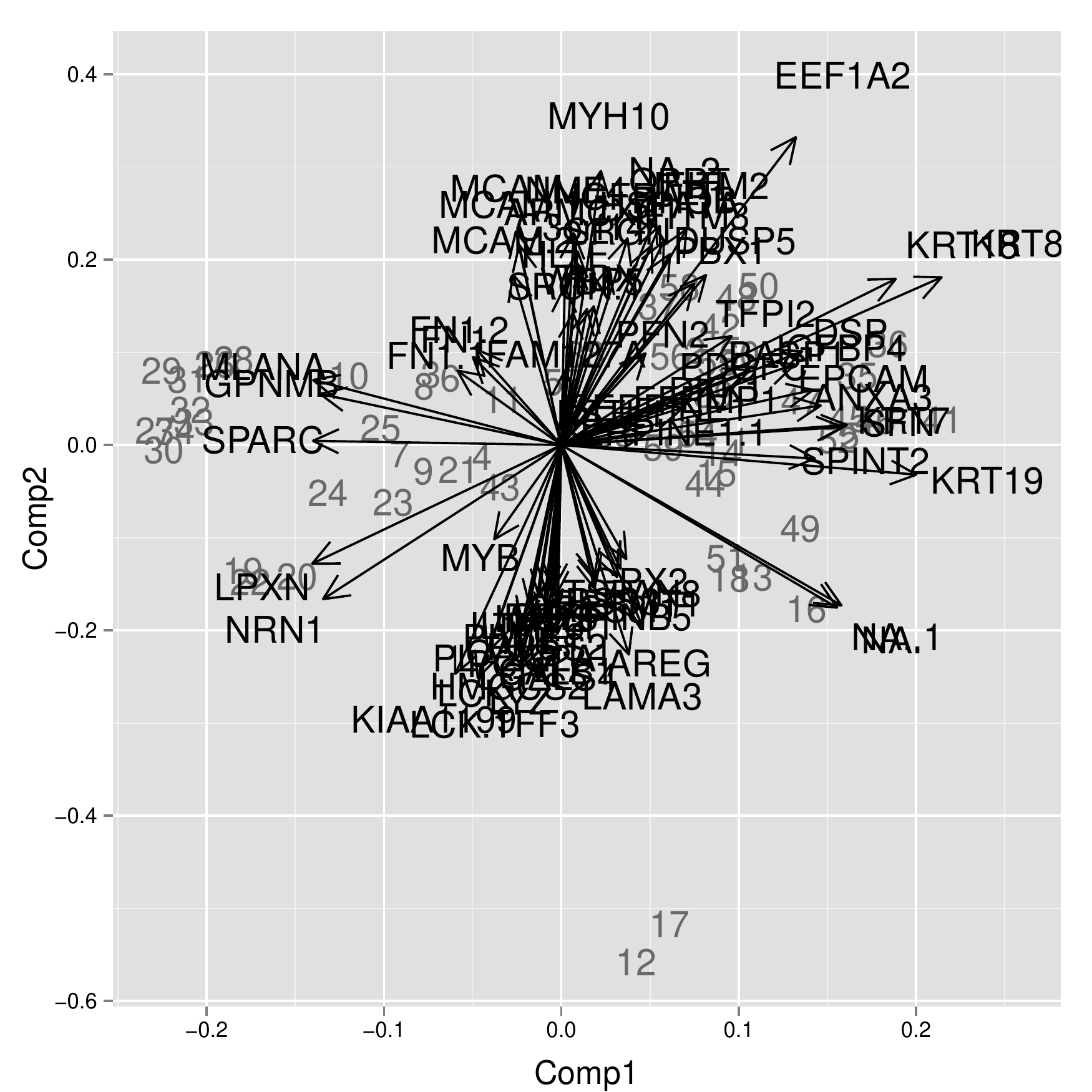}
  \caption{SPLS}
  \label{fig:nicSPRMb}
\end{subfigure}%
\begin{subfigure}{.5\textwidth}
  \centering
  \includegraphics[width=.95\linewidth]{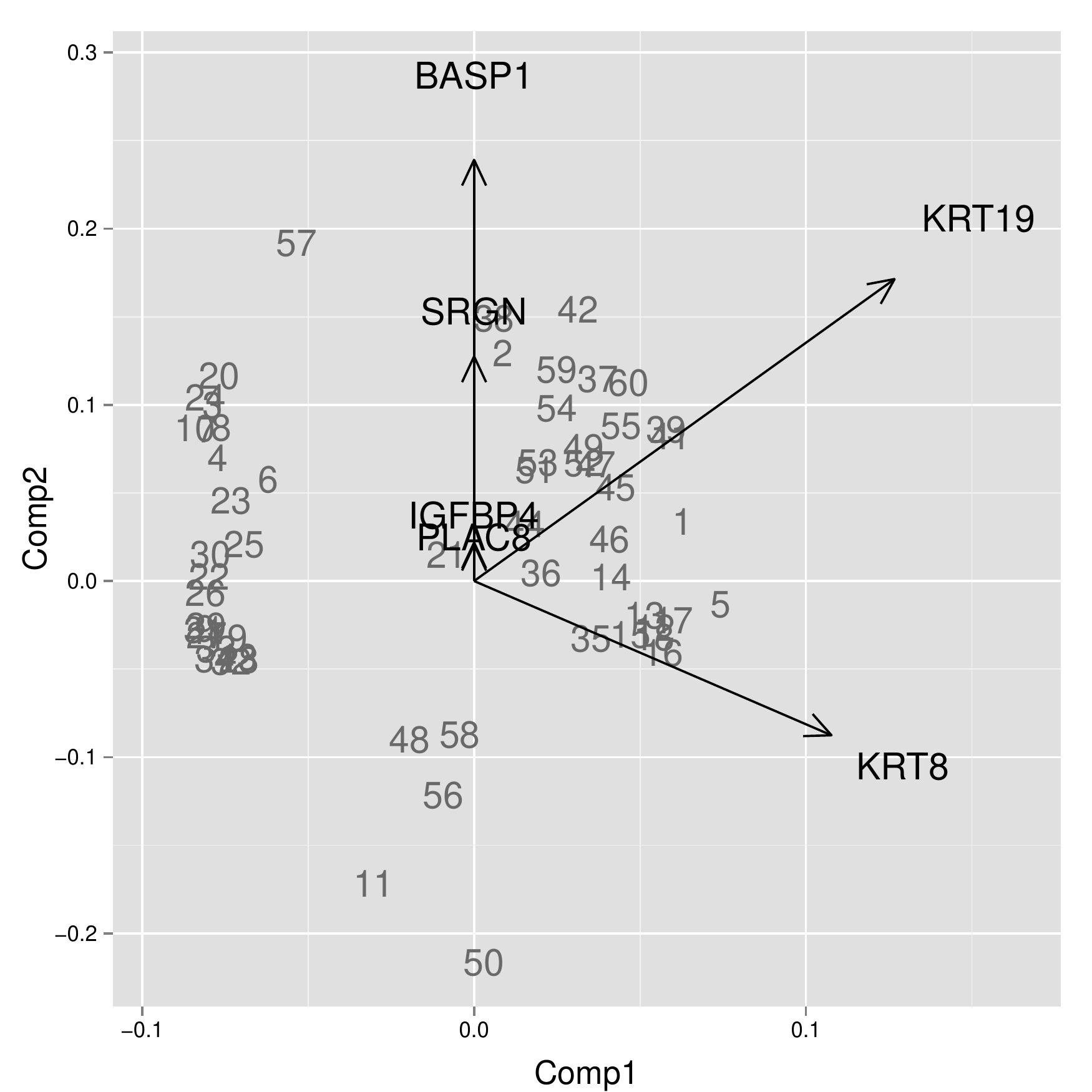}
  \caption{SPRM}
  \label{fig:nicSPRMa}
\end{subfigure}%
\caption{\label{fig:nicSPRM} The SPLS and SPRM biplots for the gene data example with protein expression of Keratin 18 as response.}
\end{figure}

Keratin 18 will now be focused upon as response, because it has the single highest variance and is often used as indicator for carcinoma \cite{oshima1996oncogenic}. For Keratin 18, SPRM yields significantly lower prediction errors than either PLS, SPLS and PRM. Notably, the SPRM only retains six out of 5571 variables. Figure \ref{fig:nicSPRM} shows the biplot of scores and directions for the first two latent components of the SPLS and the SPRM model. 

The sparse and robust method is clearly better interpretable than the non-robust variant (SPLS). Whereas SPLS biplots illustrate that the first two SPLS latent components are linear combinations of many input variables and still challenging to interpret, the first SPRM component is only composed of variables KRT8 and KRT19. The SPRM result corresponds to domain expertise: the expression of these genes is known to be closely related to the protein expression of Keratin 18 and they are used for the identification and classification of tumor cells \cite{schelfhout1989expression,oshima1996oncogenic}. KRT8 has previously been reported to play an important role in sparse and robust regression models of these data \cite{Alfons}. The biplot further unveils some clustering in the scores and provides insight into the multivariate structure of the data. The fact that the SPRM biplot leads to a much clearer interpretation than the SPLS biplot illustrates how outliers can distort even complex linear models. 

\subsection{Sparse robust discriminant analysis}

The discriminant analysis methods outlined in Section~\ref{sec:da} in general lead
to solutions that involve all variables. For example, the linear discriminant scores
from Equation~(\ref{LDAscores}) yield a linear combination of $\mathbf{x}$
in terms of a function $d_j^L(\mathbf{x})=\boldsymbol{\beta}^T\mathbf{x}+c$, with a 
constant $c$ depending upon the mean, the inverse covariance matrix, and the prior probability,
and with coefficients $\boldsymbol{\beta}=\mathbf{\Sigma}^{-1}\boma{\mu}_j$ defining
the contributions of the components in $\mathbf{x}$. In general, these contributions
will be different from zero, and thus the solution is not sparse.

Sparse robust discriminant methods have been introduced for high-dimensional data,
where the concept of sparseness is particularly relevant. One method which was mentioned
already in the previous section is SPRM-DA\cite{HFSV}. It can be used for the 
two-group case, is based on the idea of the
sparse PRM (SPRM) method, and uses the sparse NIPALS (SNIPLS) 
algorithm which yields sparsity.
The main steps of the whole procedure are as follows:
\begin{itemize}
\item Group-wise center and scale (robustly) the data; derive initial weights for
each observation based on Mahalanobis distances in a lower-dimensional PCA space
constructed for each group.
\item Robustly center (and possibly scale) the data, multiply them with the case
weights, and perform the SNIPLS algorithm to obtain the scores and sparse weighting
vectors.
\item Split the scores into the two data groups, compute robust Mahalanobis distances
for the observations of both groups, and derive weights based on these distances.
Another set of weights is obtained based on the first component of the scores,
which is most informative for the group separation; observations with a potentially
wrong group label receive a lower weight. Both types of weights are combined.
\item With these weights, the LDA rule is applied in the space of the 
scores, but on the weighted observations.
\end{itemize}
This results in a sparse classifier for a two-group problem, which is robust
against data outliers, but also robust against possible mislabeling.

\medskip
Another approach for a robust sparse discriminant method has been proposed
in Kurnaz et al.\cite{Kurnaz18}. In fact, this method also provides a robust version
for the Elastic Net problem \eqref{defENet} in the context of linear regression.
For the classification task, the Elastic Net is used within the logistic regression
model for the two-group case. The method minimizes a trimmed sum of the deviances 
to achieve robustness. Within the algorithm, robust weights are computed in a similar
way as for the robust Bianco-Yohai estimator for logistic regression\cite{Croux03}.
The complete algorithm is quite involved, but follows from the structure the 
sparse LTS algorithm\cite{Alfons}. In contrast to the Lasso estimator, the Elastic 
Net estimator is able to select blocks of correlated variables, which may be an
advantage for the interpretation of the results.

\subsubsection{Example}

In the framework of the COSIMA project\cite{cosima2015},
meteorite samples from the two meteorites Ochansk and Renazzo, stored in the
Natural History Museum Vienna, have been taken and analyzed by mass 
spectroscopy\cite{HFSV}. In total, 1540 variables are produced by the instrument,
for 110 spectra from the Ochansk meteorite samples, and 160 spectra from the 
Renazzo meteorite samples. A randomly selected training sample of 75 (Ochansk)
and 105 (Renazzo) observations is used to build an SPRM-DA model, which afterwards
is used to predict the group membership of the remaining test data.

\begin{figure}[htpb]
\begin{subfigure}{.5\textwidth}
  \centering
  \includegraphics[width=\linewidth]{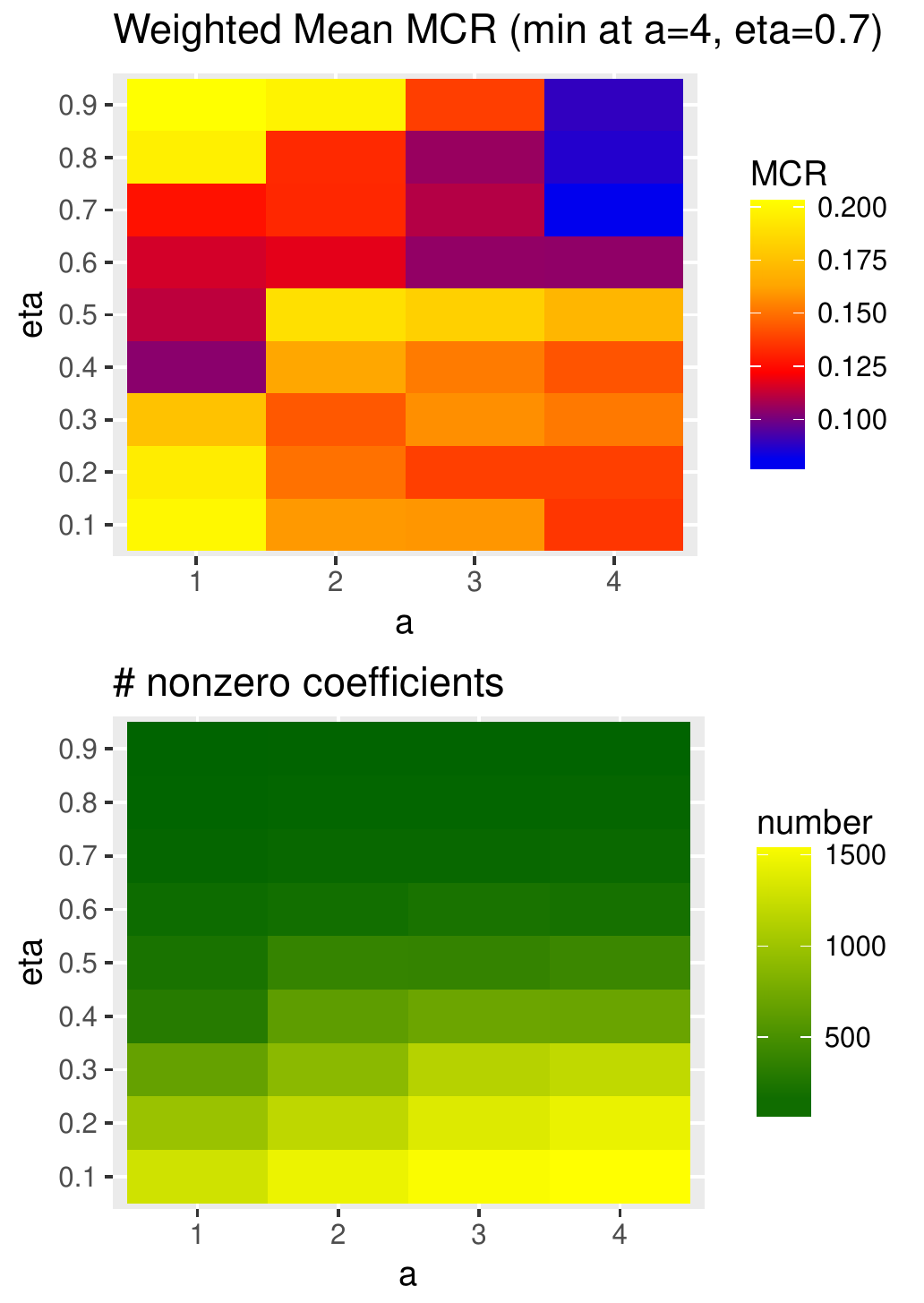}
  \caption{Selection of the tuning parameters}
  \label{fig:sprmdaa}
\end{subfigure}%
\begin{subfigure}{.5\textwidth}
  \centering
  \includegraphics[width=\linewidth]{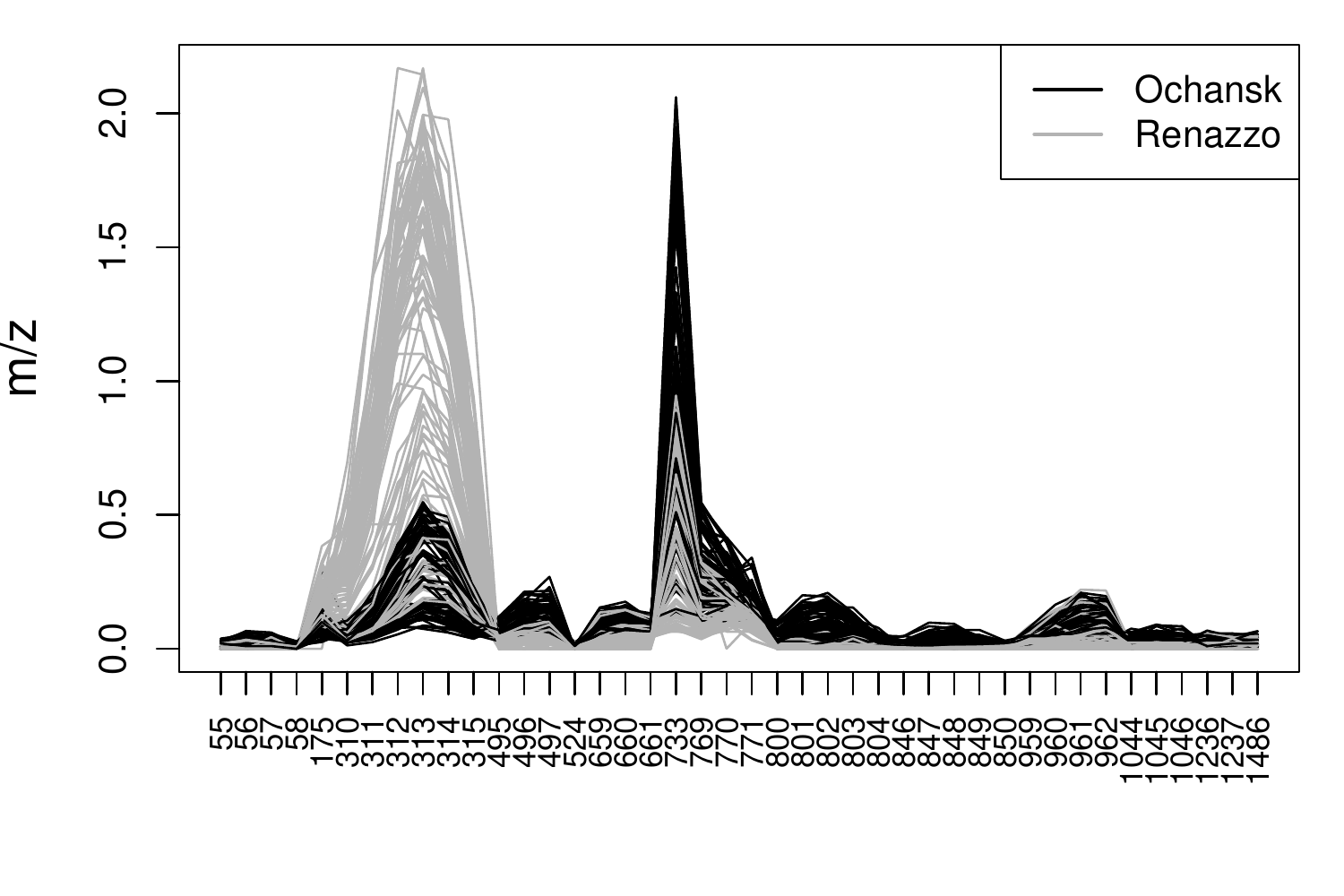}
  \caption{Spectra for the selected variables}
  \label{fig:sprmdab}
  
  \centering
  \includegraphics[width=.8\linewidth]{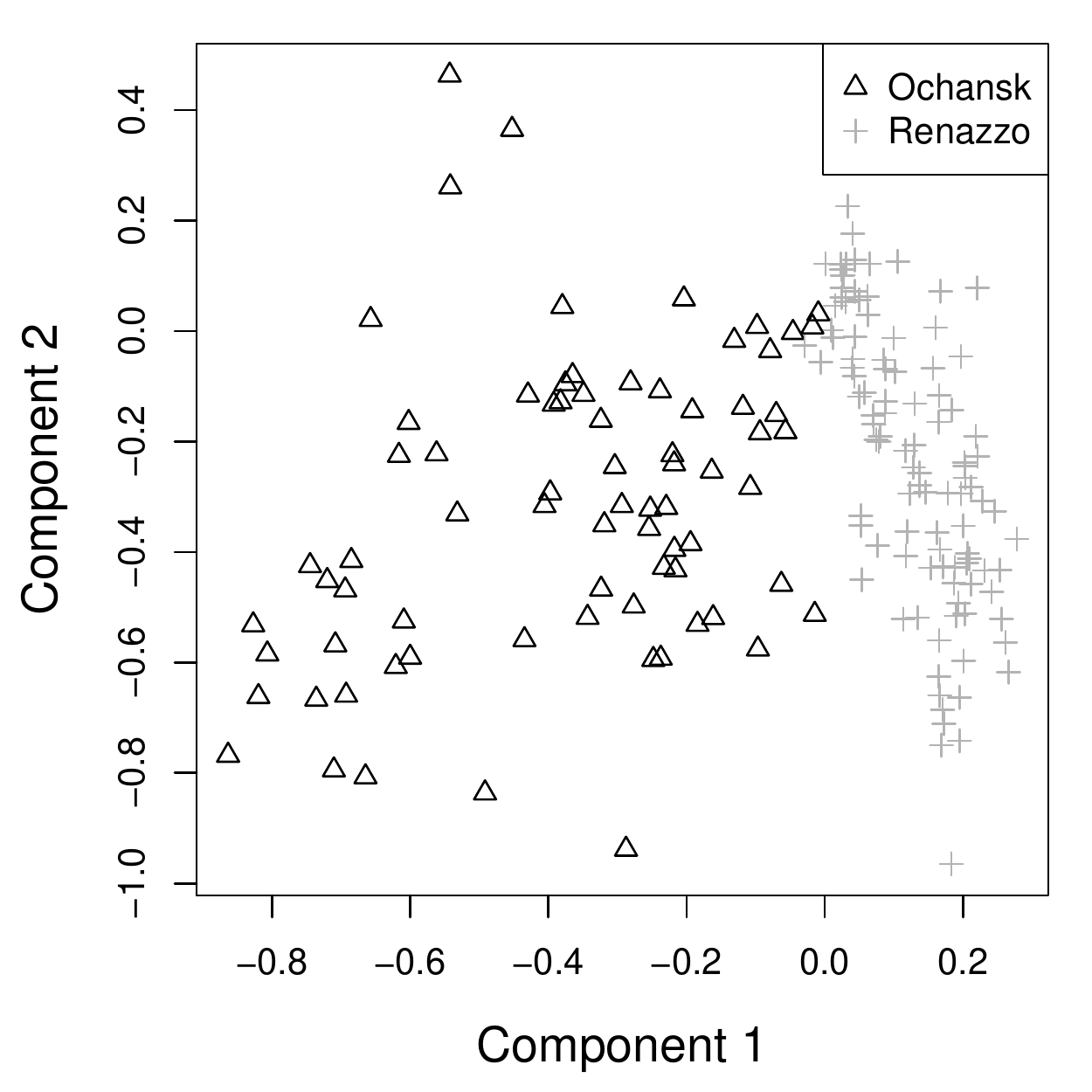}
  \caption{Projection on the scores}
  \label{fig:sprmdac}
\end{subfigure}%
\caption{\label{fig:sprmda} SPRM-DA applied to the mass spectra of the meteorite samples.}
\end{figure}

Figure~\ref{fig:sprmda}(a) shows some summaries for model fitting: For different
numbers of PRM components (horizontal axes) and different sparsity parameters ``eta''
(vertical axes), the upper plot summarizes the resulting robustified misclassification
rates (MCR), and the lower plot the resulting numbers of non-zero variables in the model.
The optimum choice minimizing the MCR is with $a=4$ components and ``eta''$=0.7$.
Only 42 coefficients are different from zero, and those point at specific masses of
the mass spectrum. The indexes of the corresponding variables are shown on the
horizontal axis of Figure\ref{fig:sprmda}(b), where all observations of both groups
are presented as line plots for the selected variables. It is visible that the groups
indeed clearly differ in these variables. SPLS-DA also returns scores, and a plot
of the first two score vectors is shown in Figure~\ref{fig:sprmda}(c), where the
two groups can clearly be distinguished. 
Finally, the application of the model to the test data results in only 3 (Renazzo) 
misclassified observations out of 90 samples.

\section{Validation}

\subsection{Precision and uncertainty}

\subsubsection{How to evaluate uncertainty for robust estimators?}
\label{sec:CV}

Robust methods of regression are often used for quantitative purposes: to actually {\em predict} one or several entities from data. However, if a robust method is used for prediction, then the following question will automatically be raised: ``Just how precise are these predictions?" Even if a robust method is used for outliers detection this question could make sense as well. If the score space is used to detect the outliers, it is maybe worthwhile to know how big the uncertainty of the scores for the individual samples is whilst drawing any inference from them. Note that accompanying scores with uncertainties is currently not common practice -- but can make sense. 

In any case, if uncertainties are required, the variance of the estimator has to be known. For classical estimators the variance is well known and is generally given by well tractable closed form expressions (e.g. for least squares regression it is known that the variance of the vector of regression coefficients equals 
$\mathrm{var}(\hat{\boldsymbol{\beta}})=\hat{\sigma}^2(\boldsymbol{X}^T\boldsymbol{X})^{-1}$, with the
estimated residual variance $\hat{\sigma}^2$). Already for some classical estimators an exact expression cannot be obtained. For instance, the partial least squares regression estimator is not linear in the predictand, due to which an exact equation cannot be derived. The most precise estimate of variance for the PLS regression coefficients can be constructed by retaining the first two terms of a first order Taylor series expansion from which the Jacobian matrix is computed. Note that in the case of PLS, a fast algorithm exists to estimate the variance of the vector of regression coefficients by this method \cite{SLV}. 

For robust estimators, the situation is in most cases even worse. Not only are the estimators frequently nonlinear in the predictand, but they are often also implicitly defined, such that it is very hard, or even impossible in some cases, to obtain exact or approximate analytical expressions for the estimators' variance. Nonetheless, one can still rely on numerical methods, the most popular of which is certainly the bootstrap. But also related methods like cross-validation or the jackknife (which is in fact an approximation to the bootstrap), are used in practice.  

\subsubsection{The bootstrap}\label{sec:bootalg}

The bootstrap is a numerical method which provides estimates of uncertainty for an estimator (denoted $T$) under consideration. Its main idea is the following. In statistics, data are virtually always assumed to consist of a sample of $n$ cases, which are $n$ observations of a $p$ variate distribution $G$. In theory, the precision of any estimates obtained from these $n$ samples could be checked easily by repeating this process many times, for instance by drawing 5000 samples, all of which contain $n$ cases, from the distribution. Then for each of these samples, the estimator under consideration can be evaluated, such that 5000 evaluations of this estimator $T(G_n^{(i)})$ are obtained (here, the notation $G_n^{(i)}$ means the distribution which puts weight $1/n$ at each case in sample $i$). By then computing the standard deviation of these 5000 evaluations, an estimate of scale for the estimator $T$ is obtained. 

Of course, in practice it is impossible to have 5000 samples of size $n$: this would imply that 5000$n$ times a measurement has to be done. But what can be done, is to mimic this process of drawing samples from the distribution, only using the one sample of $n$ cases which is available. The idea of the bootstrap thus consists of the following: resample a large number of times ($m$) from the available sample, evaluate for all resamples the estimator and eventually compute the scale of these $m$ estimates. Resampling is done in practice by sampling with replacement. 

A basic outline of a bootstrap algorithm is: 

\begin{enumerate}
\item compute the required estimate $T(G_n)$ from the original data;
\item select a random number $\check{n} \leqslant n$ of samples from the data matrix $\boma{X}$ (or from the augmented data matrix $(\boma{X}, \boma{y})$ if a predictand exists) and replace them with $\check{n}$ randomly chosen samples from the same data matrix;
\item repeat the previous process $m$ times, hence constructing $m$ bootstrap data matrices $\boma{X}^{(i)}$ (and if appropriate $\boma{y}^{(i)}$), to which correspond $m$ empirical distributions $G_n^{(i)}$;
\item compute $m$ boostrap estimates of $a_i=T(G_n^{(i)})$ from these bootstrap data matrices;
\item compute a measure of scale $S(\{a_i\})$ from these $m$ estimates; henceforth it is assumed that $S(\{a_i\})=\sqrt{\widehat{\mathrm{var}}(T(G))}$.  
\end{enumerate}

The basic algorithm can be modified in several ways, such as a correction for bias and an acceleration. For more details we refer the interested reader to a monograph on the bootstrap \cite{ET}, but we note that the basic algorithm presented here already performs well in most situations. 

Some questions need to be answered concerning the bootstrap method. At first, which scale estimator should be used? If asymptotic normality of the estimator is assumed, it is probably most appropriate to use the classical standard deviation. In other cases, if it is not sure which distribution can be expected, another
approach can be followed, by simply taking the $100\frac{1-\alpha}{2}$th and $100\frac{\alpha}{2}$th percentiles as confidence bands for the estimator in question, the explicit computation of a scale estimator thus being by-passed. 

Secondly, how many bootstrap resamples should be constructed? The quality of the bootstrap is evidently proportional to the number of bootstrap samples that are being constructed. It is considered good practice to use $m=2000$ bootstrap samples if one envisages the construction of confidence intervals \cite{ET}.

The bootstrap has been successfully applied to virtually all branches of statistics, and is gaining acceptancy in the applied fields as well. For typical chemometric tools the bootstrap has been evaluated to provide good estimates of uncertainty, for PLS see \cite{Denham}, for PARAFAC see \cite{Kiers} and for tri-PLS see \cite{SV}. 

\subsubsection{The robust bootstrap}\label{sec:robboot}

The bootstrap is a straightforward method for (approximately) obtaining uncertainties of an estimate. However, one of its basic assumptions is that all cases in the data are taken from a single underlying distribution $G$. When outliers are present in the data, these outlying cases may be assumed not to have been generated by that distribution, such that the basic principle of the bootstrap is not respected. Hence it follows that the bootstrap estimator for the uncertainty, may break down as well for contaminated data sets \cite{Matias2002}. 

The reason for breakdown of the bootstrap is the following: one replaces an arbitrary number of cases $\check{n}$ from the original $n$ cases by some randomly chosen set of $\check{n}$ cases from the original data. This new set of $\check{n}$ cases may accidentally be identical to the original set, such that in fact the bootstrap sample is the original data matrix $\boma{X}$. But it may also happen that, by coincidence, the randomly chosen set of $\check{n}$ cases only consists of outliers. Now assume that the estimator $T$ can resist a fraction $\nu=\tilde{n}/n$ of outliers, where $\tilde{n}$ denotes the number of outliers, and that $\frac{\check{n}}{n}>\nu$. This means that $T$ will break down for the bootstrap sample coincidentally containing  $\check{n}$ outliers, whereas it does not break down for the original data. This violates the basic principles of the bootstrap, as for doing the bootstrap one tries to mimic $m$ times the way in which the original set of samples has been drawn. As in the latter only $\ell$ outliers were present, so should this be the case for the bootstrap samples, otherwise the latter are drawn in a different way from the population. 

The effect of coincidental concentration of outliers in the bootstrap is illustrated for the data set described in Section \ref{sec:appprm}. Five hundred bootstrap data matrices were constructed with the basic algorithm described above, using the implementation from the MATLAB Statistics Toolbox (The MathWorks, Natick, MA, USA). As has been heeded in Section \ref{sec:appprm}, the PRM regression estimator can resist well to the leverage points in the data and thus does not break down. For each of these bootstrap data matrices, the vector of regression coefficients was estimated by PLS as well as by PRM, by virtue of which vectors the sodium oxide concentration was predicted. This means that for each of these predicted concentrations, a set of 500 bootstrap predicted concentrations was available, allowing us to have an idea of the distribution of these predicted concentrations. Although no exact theoretical results are available as regards the distribution of predicted responses by PLS nor by PRM, it seems viable to expect that the distribution be unimodal and symmetric. 
\begin{figure}
\begin{center}
\begin{tabular}{cc}
\resizebox{7.5cm}{!}{%
\includegraphics{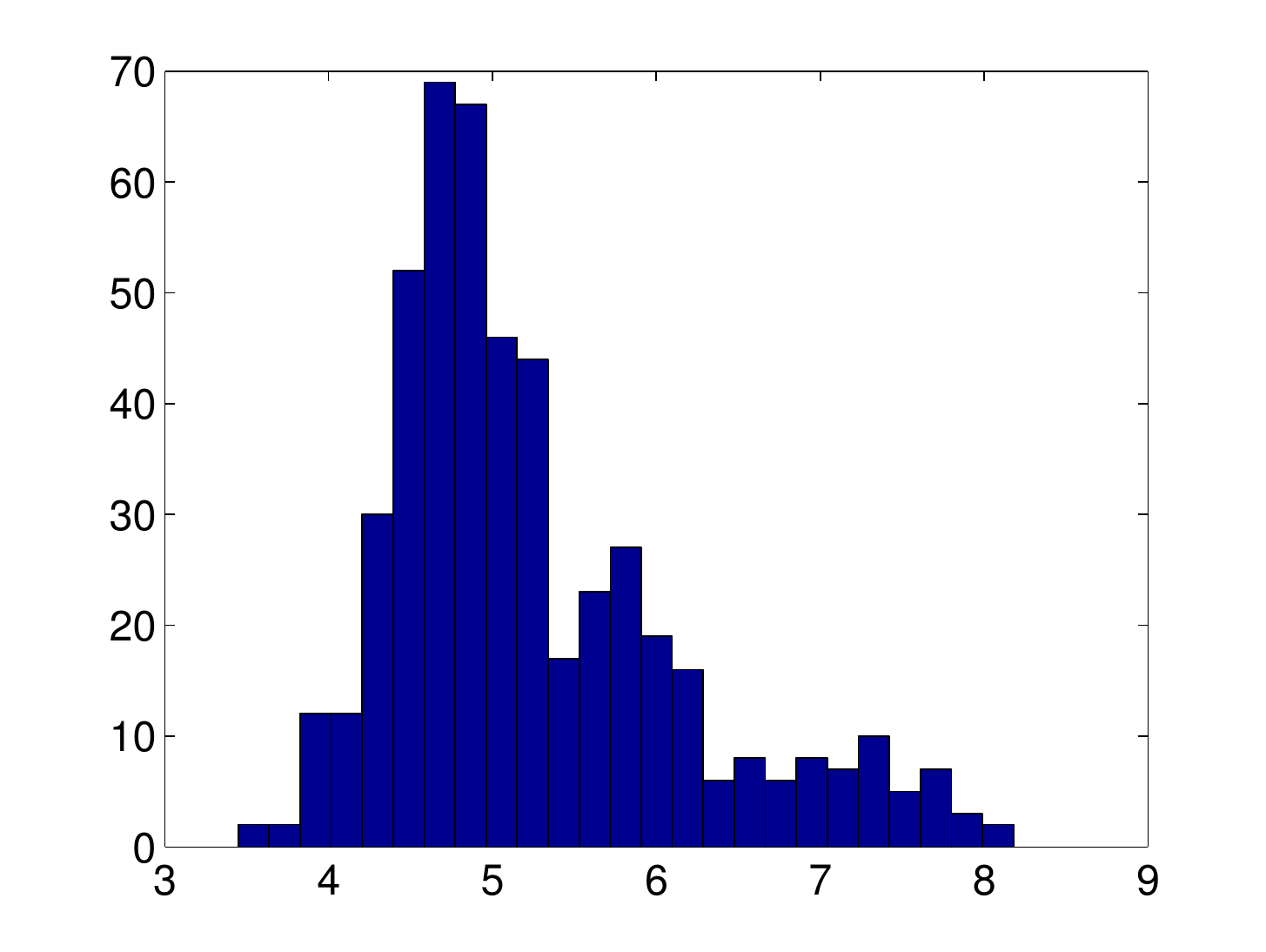}}&%
\resizebox{7.5cm}{!}{%
\includegraphics{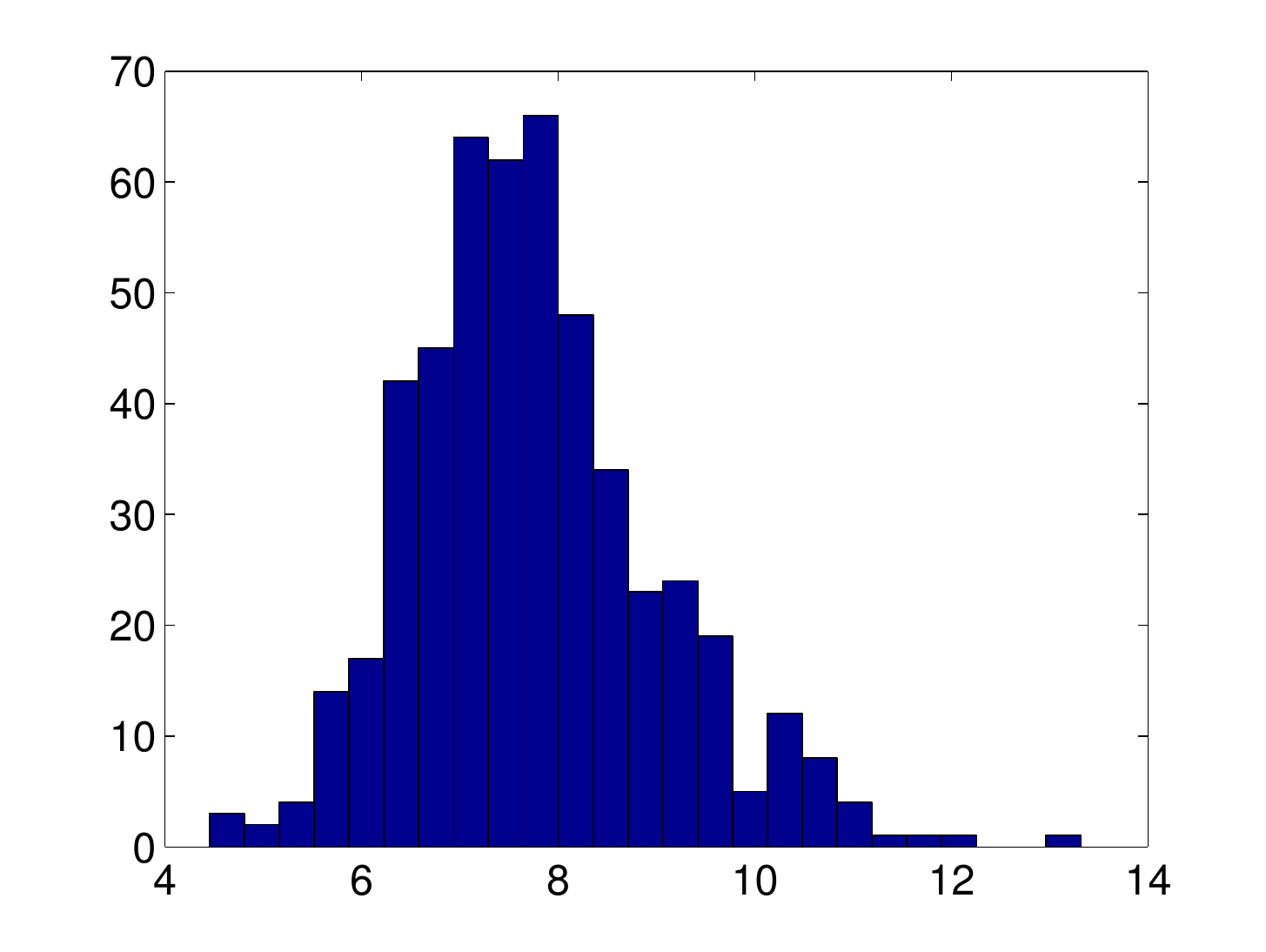}}\\
(a) & (b)\\
\resizebox{7.5cm}{!}{%
\includegraphics{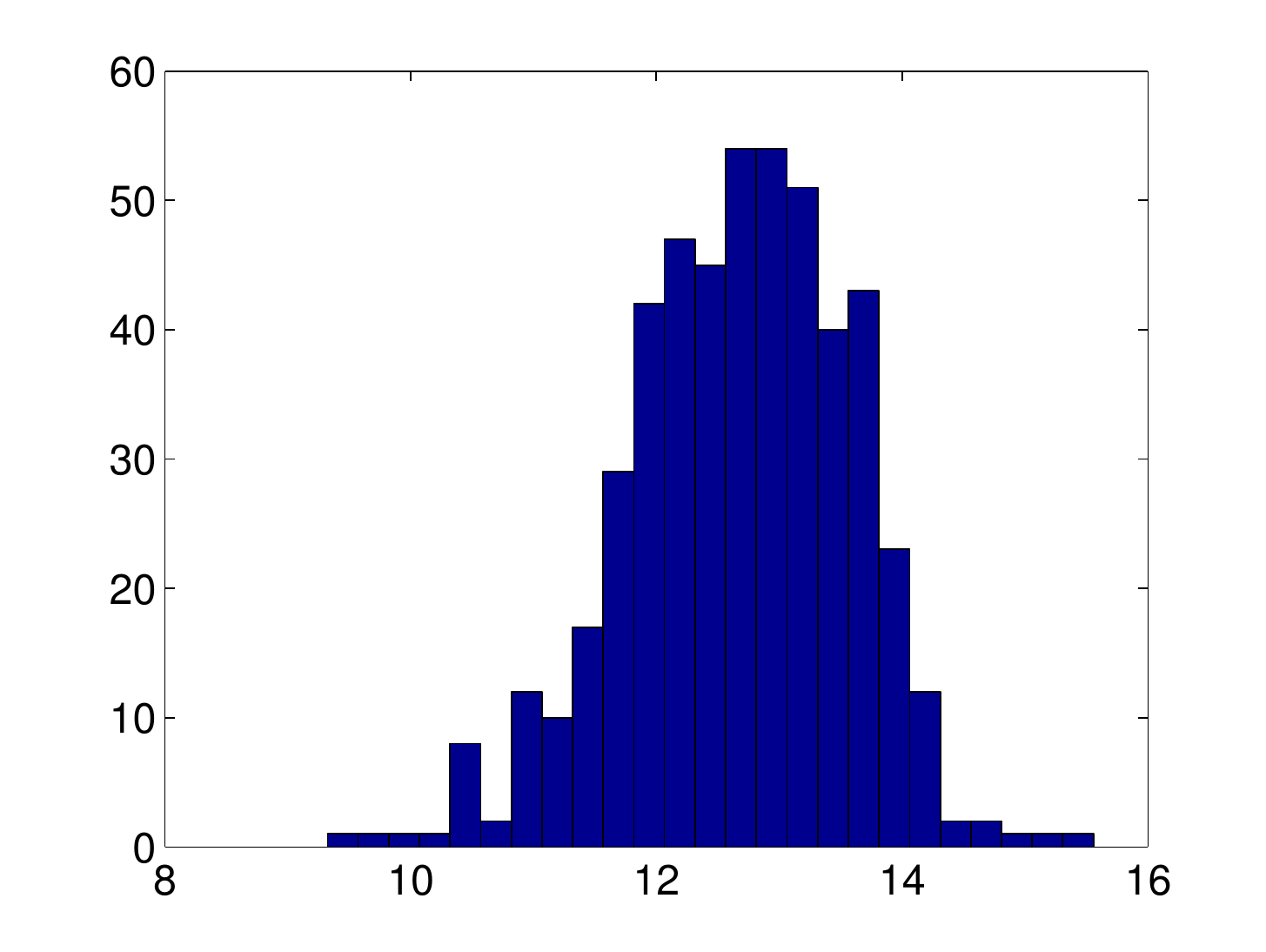}}&%
\resizebox{7.5cm}{!}{%
\includegraphics{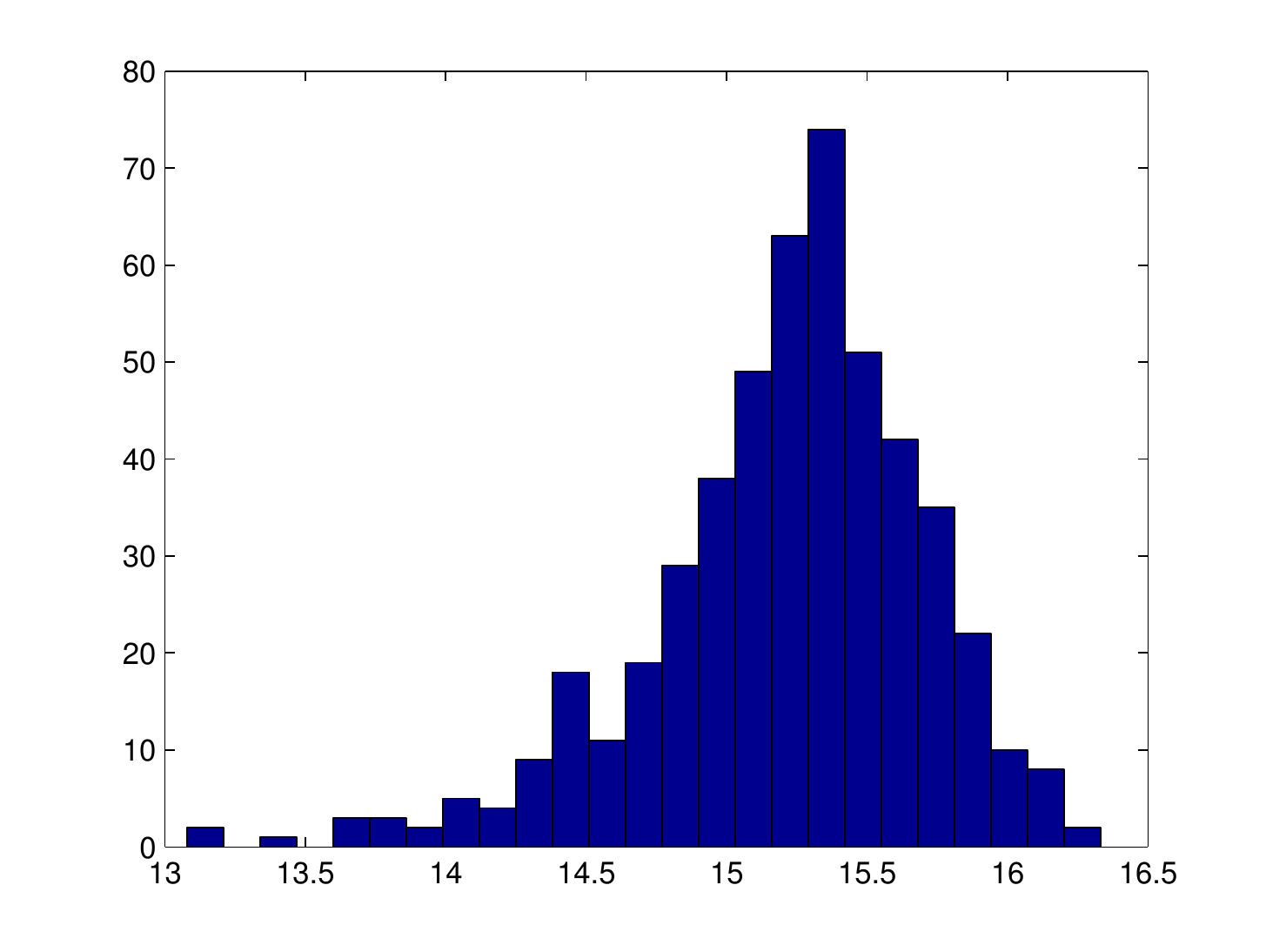}}\\
(c) & (d)\\
\end{tabular}
\caption{\label{fig:bootdist}Bootstrap distributions of the predicted Na$_2$O concentrations by PLS (left column) and PRM (right column), for sample 1 (top row) and sample 50 (bottom row).}
\end{center}
\end{figure}

In Figure \ref{fig:bootdist}, the results are shown for the predictions of two cases from the independent validation set. These are cases 1 and 50 from this set, and noteworthily these are glasses of the potasso-calcic and sodic type, respectively. At first it can be observed that PLS suffers a lot from the presence of the leverage points in the calibration set: the predicted responses vary over a wide range in concentrations and, especially for case 1, appears to be trimodal. For PRM, the distributions are more narrow, but nevertheless, some predictions seem to fall into bins far away from the centre (on the right hand side for case 1 and on the left hand side of the centre for case 50). This is exactly the effect of up-concentration of the outliers: some bootstrap data matrices did contain in practice far more leverage points than the original calibration set, such that even PRM yields erroneous predictions for these few bootstrap samples. It is thus clear that the bootstrap needs a slight modification in order to be applicable to data containing outliers. 

The bootstrap can be modified in two ways: either controlling the number of outliers in the bootstrap sample or limiting the influence on the outcome by using a robust estimator of scale $S$ (in step 5 of the basic algorithm, Section \ref{sec:bootalg}). The first option is more cumbersome: it implies outlier identification and is hard to apply if many borderline cases are present (can they be samples in the bootstrap or not?). Hence, the second option is most often recommended: use a robust scale estimator on the empirical distribution of the estimator (such as the distributions shown in Figure \ref{fig:bootdist}, right column). It is best to use a robust scale estimator which is consistent with the robust estimator which one wants to bootstrap, e.g. if one uses RCR with a 20\% trimmed scale to estimate the concentrations, it is preferable to use a 20\% trimmed scale here as well. 

\subsubsection{Computational efficiency}\label{sec:compeffboot}

Robust estimators are nearly always harder to compute than classical estimators. They frequently involve complex iterative schemes; even some of the most simple robust estimators (in the computational sense), require far more floating point operations than their classical counterparts. For instance, many M-estimators can be computed by an iterative reweighting algorithm. If the classical estimator has on average to be iterated $\kappa$ times, then the overall computation time for the M-estimator is slightly more than $\kappa$ times the computation time of the classical estimator. 

As robust estimators tend to be rather slow in terms of computation, bootstrapping them may be a very slow procedure: the robust estimator may have to be computed 500 or, even worse, 2000 times. Thus, straightforward implementation of the basic algorithm given in Section \ref{sec:bootalg} is only feasible for the fastest robust estimators. For the most common chemometric tools, this implies the methods based on the spatial sign (spatial sign based PCA \cite{Locantore} and PLS \cite{SDV}), iterative reweighting (PRM \cite{SCFV}) or the methods based on ROBPCA (ROBPCA \cite{Hubert2} and RSIMPLS \cite{Hubert}). 

For some of the regression estimators described in Section \ref{sec:robreg}, a specially designed fast bootstrap procedure exists. These {\em fast and robust bootstrap} procedures basically consist of the following: the estimating equations of the estimator are bootstrapped, such that not in every loop the whole estimator has to be constructed. Fast and robust bootstrap exists (up to our knowledge) for S-estimators \cite{WVA}, and least trimmed squares \cite{WVA2}. 

\subsection{Model selection}\label{sec:ModSel}

The last important issue concerning robust estimation, is model selection. For the latent variables based robust estimators, the question is: how many latent variables are optimally used to model the data at hand? The most popular technique to obtain an answer to this question is cross-validation. Cross-validation exists in various flavours; in the context of PLS recent studies favour a so-called {\em full or Monte Carlo} cross-validation, where for a large number of iterations, the data are randomly split up into a model and a test set (now named the CV model and CV test sets, respectively). For each of these iterations, calibration is done based on the CV model set for all possible complexities and the responses are predicted for the CV test with using these models of different complexities. From these predicted responses a root mean-squared error of cross-validation is computed. The complexity leading to the minimum in RMSECV is considered to be optimal. For a thorough simulation study concerning different types of cross-validation, see \cite{BAV,BVA}.

When outliers are present in the data, the same problems may also affect cross-validation. At first, it is possible that due to random selection of the cases which belong to the CV model and test sets, the CV model set will contain a higher fraction of outliers than the original data set (cf. the bootstrap, Section \ref{sec:robboot}). Hence, it is possible that the robust estimator breaks down for the CV model set whereas it would not for the original data set, leading to misleading root mean squared error of cross validation curves. Secondly, it is also probable that the CV test set contains outliers. The robust estimator is supposed to fit the clean data well and to give good predictions for regular cases. However, the robust estimator should not fit the outliers well, and should thus also not provide good predictions for outlying cases. Hence, the outliers which are possibly part of the CV test set, will be predicted badly, giving rise to an artificially high root mean squared error of cross validation. 

In order to tackle these problems, two solutions can be presented, analogous to the bootstrap (see Section \ref{sec:robboot}). On the one hand, one can do an outlier detection prior to cross-validation, exclude these outliers from the CV test sets and keep their fraction in the CV model sets quasi constant. Howbeit, this procedure give rise to some questions, e.g. if you are doing robust PCA in order to detect better the outliers, and you do cross validation to know the complexity of this model, in fact the outliers have not yet been detected such that they cannot be monitored in cross validation. Thus also for cross validation, the more straightforward strategy is to compute a robust variant of the root mean squared error of cross validation. This is simply done by selecting a percentage of trimming corresponding to the fraction of outliers that is likely to arise. This modus operandi has been implemented in the TOMCAT toolbox \cite{TOMCAT}. 

Finally, in the computational sense it holds again that cross validation is only applicable to the fastest robust methods (cf. the bootstrap, Section \ref{sec:compeffboot}). A specifically designed fast and robust cross validation method has, up to our knowledge, only been tailored for the ROBPCA method \cite{Hubert3}.

\section*{Suggestions for further reading}

In contrast to the research areas covered in several other chapters of this reference work, construction and application of robust methods for chemometrics is still very much a research topic, such that there does not, up to our knowledge, exist a textbook which solely addresses this subject. A good textbook on chemometrics that covers some robustness aspects, is {\em Introduction to Multivariate Statistical Analysis in Chemometrics} by  K. Varmuza and P. Filzmoser, Taylor \& Francis --- CRC Press, Boca Raton, 2009. Beyond this book, we refer the interested reader to the several textbooks on chemometric multivariate methods, to be found in the suggestions for further reading provided in several of the chapters preceding the current one. In addition to these works, we advise to consult recent textbooks on robust statistics. A particularly nice, recently updated and thorough introduction to robust statistics can be found in: {\em Robust Statistics: Theory and Methods (with R), 2$^\mathit{nd}$ Edition} by Ricardo A. Maronna, Douglas R. Martin, Victor J. Yohai and Matt\'{i}as Salibi\'{a}n-Barrera, Wiley, 2019. 
We also recommend the book {\em Robust Methods in Biostatistics} by
Heritier, Cantoni, Copt, and Victoria-Feser, Wiley, 2009.
A somewhat older, nicely written textbook is {\em Robust Estimation and Testing}
by Robert G. Staudte, Simon J. Sheather, Wiley, 1990. An old textbook which is specific to robust regression is: {\em Robust Regression and Outlier Detection} by Peter J. Rousseeuw and Annick M. Leroy, Wiley, 1987. For material on specific robust methods for chemometrics, as no textbook on this subject exists, the reader is referred to articles, i.e. the corresponding references in the list given below.

\end{document}